\documentclass[11pt,a4paper,oneside,openright]{book}
\usepackage[latin1]{inputenc}
\usepackage{amsmath}
\usepackage{amsfonts}
\usepackage{amssymb}

\usepackage[bookmarks, hidelinks]{hyperref}
\usepackage{mathtools}
\usepackage[onehalfspacing]{setspace}
\usepackage{cite}
\usepackage{tocloft}

\usepackage{rotating}
\usepackage{lscape}
\usepackage{sidecap}
\usepackage[left = 1.5in]{geometry}

\newcommand{\ud}{\mathrm{d}}

\DeclareMathOperator{\diag}{diag}

\usepackage{fancyhdr}
\makeatletter
\let\ps@plain\ps@empty
\makeatother

\begin{document}

\pagestyle{empty}

\begin{titlepage}
\vfill
\begin{center}
   \baselineskip=16pt
   {\LARGE \bf Solution-generating transformations in duality-invariant theories and the fluid/gravity correspondence}
   \vskip 2cm
    {\large\bf  Joel Alan Fitzhardinge-Berkeley}
       \vskip 1.6cm
       {\it Submitted in partial fulfilment of the requirements \\ of the Degree of Doctor of Philosophy}
       \vskip 3cm
             {\it Centre for Research in String Theory, \\
             School of Physics and Astronomy,\\
             Queen Mary University of London, \\
            Mile End Road, London, E1 4NS, UK} 
\end{center}
\vfill
\setcounter{footnote}{0}
\end{titlepage}

\setcounter{page}{2}

\pagestyle{empty}

\begin{center} 
\textbf{Declaration}
\end{center} 
\begin{quote}
I, Joel Alan Fitzhardinge-Berkeley, confirm that the research included within this thesis is my own work or that where it has been carried out in collaboration with, or supported by others, that this is duly acknowledged below and my contribution indicated. Previously published material is also acknowledged below.

I attest that I have exercised reasonable care to ensure that the work is original, and does not to the best of my knowledge break any UK law, infringe any third party's copyright or other Intellectual Property Right, or contain any confidential material.

I accept that the College has the right to use plagiarism detection software to check the electronic version of the thesis.

I confirm that this thesis has not been previously submitted for the award of a degree by this or any other university.

The copyright of this thesis rests with the author and no quotation from it or information derived from it may be published without the prior written consent of the author.

\vspace{5mm}
Signature:
\vspace{5mm}

Date: 
\vspace{5mm}

Details of collaboration and publications:

The material presented further describes results of the publications \cite{Berkeley:2012kz} (collaboration with David Berman), and \cite{Berkeley:2014nza} (collaboration with David Berman and Felix Rudolph).
\end{quote}

\newpage
\begin{center} 
\textbf{Acknowledgements}
\end{center} 
\begin{quote}
This PhD thesis has been financially supported by an STFC grant.

My many thanks are to my supervisor Professor David Berman, who has been a knowledgeable and inspiring collaborator, and an understanding and patient guide. I would also like to thank Felix Rudolph for a fruitful, informative and enjoyable collaboration, and Edward Hughes for discussions on differential forms.

I have been incredibly fortunate to have had my family's presence, support, and love, in its countless forms throughout this process. I thank from my heart my dear friend and teacher Maitreyabandhu, who has been there with attentiveness, sense, and the warmest compassion. I also thank Dorcas, Jess, Swati, Zack, Amalavajra, Manjusiha, and the QMUL counselling service, who found me when I lost my way.

There are countless others who have made this work possible whom I have not managed to mentioned here. I am grateful for their contributions nonetheless.
\end{quote}

\newpage
\begin{center} 
\textbf{Abstract}
\end{center} 
\begin{quote}
We explore dualities and solution-generating transformations in various contexts. Our focus is on the T-duality invariant form of supergravity known as double field theory, the $SL(5)$-invariant M-theory extended geometry, and metrics dual under the fluid/gravity correspondence to an incompressible Navier-Stokes fluid.

In double field theory (DFT), a wave solution is shown to embed both the F1 string and the pp-wave. For the former, the Goldstone mode dynamics reproduce the duality symmetric string introduced by Tseytlin.

We consider solution-generating techniques in DFT in the presence of an isometry, firstly via Buscher-like transformations in the DFT string $\sigma$-model, and secondly via the DFT equations of motion.

In the $SL(5)$-invariant geometry, we provide a chain rule derivation of the covariant equations of motion, and present a wave solution embedding the M2 brane.

Lastly, solution-generating transformations for metrics with an isometry are considered in the context of the fluid/gravity correspondence. Our focus is on the vacuum Rindler metric dual to a codimension one Navier-Stokes fluid. In particular, when there is a radially directed Killing vector, the dual fluid is found to exhibit an energy scaling invariance valid to all orders in the hydrodynamic expansion.
\end{quote}

\newpage

\tableofcontents

\listoftables
\listoffigures

\chapter{Introduction}
\label{ch:Introduction}
\pagestyle{fancy}
\fancyheadoffset{\displaywidth}
\renewcommand{\headrulewidth}{0.4pt}
\lhead{}
\chead{\textsc{introduction}}
\rhead{\thepage}
\lfoot{}
\cfoot{}
\rfoot{}

\begin{quote}
It is perhaps because all phenomena play out in one continuous, inseparable experience, that we look to find a theory which describes all physical phenomena on a unified level, that we search for a unified ``theory of everything".
\end{quote}
Two great edifices of modern physics, the standard model of particle physics and Einstein's general theory of relativity, both describe in fantastic detail and accuracy their respective phenomena. The standard model is a quantum field theory (QFT) which describes the states and interactions of half-integer spin fermions and integer spin bosons, which are quite naturally unified in the supersymmetric standard model. Supersymmetry additionally removes shortcomings in the standard model such as the hierarchy problem, and provides a dark matter candidate. Meanwhile, Einstein's theory of gravitation describes the interplay between space-time and energy/matter. These two theories, supersymmetric QFT and general relativity can be unified in what is known as supergravity. However, supergravity is non-renormalisable, and so can only provide a low energy effective unified theory.

One candidate for a unified theory valid at all energy scales is superstring theory, a supersymmetric quantum theory whose fundamental objects are one-dimensional rather than the point-like particles of standard QFT. There exist five different forms of string theory: Heterotic $SO(32)$, Heterotic $E_8 \times E_8$, and Types I, IIA and IIB. In their low energy limits, each reproduces a supergravity theory composed of a fermionic sector, and a bosonic sector of two parts: the RR and NS-NS sectors composed respectively of tensor products of fermions and bosons. The NS-NS sector is common to the low energy limit of all string theories.

For superstring theories to preserve Lorentz invariance, they must have a critical spacetime dimension of ten. Clearly these ten dimensions needs to be reconciled with our apparent four-dimensional world. Analogous to rolling a piece of paper into a thin, infinitely long tube, whose circumference is now much smaller than the wavelength of any common measuring device, and is thus largely invisible, dimensions in string theory are wrapped at distance scales too small for low energy objects to probe. This process is called compactification, and using it one can reduce a spacetime theory from ten to four macroscopic dimensions.

In this process, certain connections are revealed between the different string theories, which rely on the string's extended nature to form closed cycles round compact dimensions. One finds that a Type IIA string compactified on an $S^1$ of radius $R$ is equivalent to a Type IIB string compactified on an $S^1$ with radius $\alpha'/R$, where $1/2\pi\alpha'$ is the string tension. The same holds for the two heterotic string theories. This equivalence is dubbed T-duality, and is not alone in the string dualities. There also exists a duality between theories at strong and weak coupling, called S-duality, under which Heterotic $SO(32)$ and Type I are dual, and Type IIB is dual to itself.

If one goes to the strong coupling limit of the Type IIA theory, one arrives at an eleven-dimensional theory \cite{Horava:1995qa} whose fundamental objects are higher-dimensional branes, and whose low energy effective field theory is eleven-dimensional supergravity \cite{Cremmer:1978km}. The full eleven-dimensional theory, dubbed M-theory, is largely unknown, but yields in its various compactification limits all five string theories, while T-duality and S-duality become united as components of a larger U-duality web which relates all the string theories \cite{Hull:1994ys}.

\section{Duality-invariant theories}
\label{sec:Duality-invariant theories}

We could view this thesis as an exploration of dualities both in themselves, and as the basis for solution-generating techniques in the low energy limit of string theory, and the long-wavelength limit of holography. That this is so was not in fact entirely intentional. However, that we have returned to dualities in various forms serves to illustrate the ubiquitous nature of the duality in modern theoretical physics. We have already encountered the T-, S- and U-dualities of string theory. We look here to illustrate a duality arising in electromagnetism, and from it motivate the \emph{duality-invariant} approach.

Consider Maxwell's unification of the electric and magnetic fields $\mathbf E$ and $\mathbf B$ in the absence of sources
\begin{align}
\nabla \cdot \mathbf E		& = 0
&
\nabla \cdot \mathbf B		& = 0
\\
\nabla \times \mathbf E		& = - \frac{\partial \mathbf B}{\partial t}
&
\nabla \times \mathbf B		& = \frac{1}{c^2} \frac{\partial \mathbf E}{\partial t}		.
\end{align}
where $c$ is the speed of light. A quick check verifies that the full set of equations are invariant under a simple form of S-duality:
\begin{equation}
(\mathbf E/c, \mathbf B) \rightarrow (\mathbf B, - \mathbf E/c)		.
\label{EM S-duality}
\end{equation}
A system with electric field $\mathbf E_0$ and magnetic field $\mathbf B_0$ is \emph{dual} to another system with electric field $c\mathbf B_0$ and magnetic field $-\mathbf E_0/c$. They are non-trivially distinguishable from each other, yet exist as dual solutions to Maxwell's equations. The system is also invariant under Lorentz rotations.

We can define a Lorentz-covariant tensor in Minkowski space with coordinates $x^\mu = (ct, x^i)$ by
\begin{equation}
F^{i0} = E^i/c
\qquad
F^{ij} = \epsilon_{ijk} B^k		,
\end{equation}
such that the Maxwell equations can now be expressed in a Lorentz-covariant form
\begin{equation}
\nabla_\mu F^{\mu\nu} = 0
\qquad
\nabla_{\mu} \left( \epsilon^{\mu\nu\sigma\rho} F_{\sigma\rho} \right) = 0	,
\end{equation}
where $\epsilon$ is the antisymmetric symbol (see app. \ref{app:Alternating symbol and tensor}). In addition, noting that \eqref{EM S-duality} can be generated by
\begin{equation}
F_{\mu\nu}		\rightarrow		\epsilon_{\mu\nu\sigma\rho} F^{\sigma \rho}	,
\end{equation}
we illustrate the possibility of promoting duality (in this case both S-duality and Lorentz duality) to manifest symmetry.

\subsection{Double field theory}

This process is analogous the the promotion of T-duality to a manifest symmetry of \emph{generalised geometry} by Hitchin \cite{2002math......9099H} and Gaultieri \cite{2004math......1221G}, which extends the tangent space $TM$ of a $d$-dimensional theory, whose sections are vectors parametrising diffeomorphisms of a symmetric $g_{\mu\nu}$, to $TM\oplus T^*M$, such that sections additionally include one-forms parametrising gauge transformations of a two-form field $B_{\mu\nu}$. In particular, there exists in this generalised geometry a natural $O(d,d,\mathbb R)$-invariant inner product. In \emph{double field theory} (DFT), with origins in \cite{Siegel:1993th, Siegel:1993xq, Duff:1989tf, Tseytlin:1990nb, Tseytlin:1990va} (see \cite{Berman:2013eva, Hohm:2013bwa, Aldazabal:2013sca} for reviews and \cite{Berman:2007yf, Berman:2007xn, Berman:2007vi, Hull:2009mi, Hull:2009zb, Kwak:2010ew, Thompson:2010sr, Hohm:2010jy, Hohm:2010pp, Hohm:2011ex, Copland:2011yh, Jeon:2011kp, Hohm:2011cp, Albertsson:2011ux, Jeon:2011cn, Kan:2011vg, Kan:2012nf, Hohm:2012gk, Andriot:2012wx, Andriot:2012an, Musaev:2013kpa, Park:2013mpa, Hohm:2013nja, Hohm:2013jaa, Blair:2013noa, Hohm:2014xsa, Berman:2014jba, Papadopoulos:2014ifa, Ma:2014kia, Cederwall:2014opa, Naseer:2015tia, Blumenhagen:2015zma, Betz:2014aia, Hohm:2011zr, Hohm:2011nu, Jeon:2011vx, Hohm:2011dv, Jeon:2011sq, Jeon:2012hp, Jeon:2012kd, Cho:2015lha, Lee:2015kba, Hull:2004in, Copland:2011wx, Lee:2013hma, DeAngelis:2013wba, Blumenhagen:2013zpa, Blumenhagen:2014gva, Blumenhagen:2014iua, Ma:2014vqm, Pezzella:2015hfa, Ma:2015yma, Polyakov:2015wna, Hohm:2010xe, Jeon:2010rw, Hohm:2011si, Hohm:2012mf, Berman:2013uda, Cederwall:2014kxa, Berman:2011kg, Ma:2014ala, Wu:2013sha, Aldazabal:2011nj, Geissbuhler:2011mx, Grana:2012rr, Dibitetto:2012rk, Geissbuhler:2013uka, Berman:2013cli, Hassler:2014sba} for further literature) one additionally doubles the coordinate space itself, and allows dependence of the fields on the the full set of coordinates. The extended coordinates are reduced to a physical subset by a \emph{weak constraint} arising from the level matching condition of string theory, which can be supplemented by a \emph{strong constraint} which ensures closure of gauge transformations on the extended space. Various solutions to these constraints yield NS-NS supergravity backgrounds related under T-duality, in addition to gauged supergravities with non-geometric fluxes recovered via Scherk-Schwarz reductions \cite{Aldazabal:2011nj, Geissbuhler:2011mx, Grana:2012rr, Dibitetto:2012rk, Geissbuhler:2013uka, Berman:2013cli, Hassler:2014sba, Lee:2015xga}.

The supergravity fields (including the dilaton) are collected in an \\$O(d,d,\mathbb R)/O(d,\mathbb R) \times O(d, \mathbb R)$ coset representative generalised metric and a DFT dilaton. These objects' dynamics are governed by an action whose minima provide the equations of motion. When directions are compactified, T-duality transformations appear as coordinate transformations on the doubled space.

\subsection{U-duality invariant M-theory}

In M-theory's low energy limit of eleven-dimensional supergravity, compactification on $T^n$ reveals an $E_n$ symmetry originating from the M-theory U-duality group \cite{Julia:1981}. Analogous to how $O(d,d,\mathbb R)$ is promoted to a manifest symmetry in generalised geometry and DFT, one can promote U-duality to a manifest symmetry, constructing $E_d$ duality-invariant theories without requiring compactified directions. This is achieved in various frameworks: $d$-dimensional generalised geometry where the tangent bundle is extended to allow for membrane charges \cite{Coimbra:2011nw, Coimbra:2012yy, Hillmann:2009ci, Coimbra:2011ky, Coimbra:2012af, Godazgar:2013dma, Lee:2014mla}; extended geometry where the $d$ spacetime dimensions are supplemented with $w$ ``wrapping" coordinates dual to membrane charges, and fields are allowed to depend on all coordinates \cite{malek2012u, Berman:2010is, Berman:2011cg, Berman:2011jh, Musaev:2013kpa, Park:2013gaj, Blair:2013gqa, Berman:2011pe, Aldazabal:2013mya, Berman:2012vc, Berman:2011kg, Berman:2014jsa}. From the perspective of the parent eleven-dimensional theory, the $d$-dimensional internal space is decoupled from the $(11-d)$-dimensional external space; exceptional field theory where the full eleven dimensional supergravity is supplemented by $w$ wrapping coordinates, and the internal and eternal spaces are no longer decoupled \cite{Hohm:2013uia, Hohm:2013vpa, Hohm:2013pua, Blair:2014zba, Godazgar:2014nqa, Musaev:2014lna, Hohm:2014fxa, Cederwall:2015ica, Wang:2015hca, Musaev:2015pla, Hohm:2015xna, Abzalov:2015ega, Malek:2013sp, Godazgar:2014sla, Berman:2014hna, Baguet:2015xha}. Without a level matching constraint equivalent for the membrane, the dependence on the additional coordinates appearing in extended and exceptional geometry is constrained by closure of gauge transformations on the extended space. Again, gauged supergravities can be recovered from exceptional and extended geometry via Scherk-Schwarz reductions \cite{Berman:2012uy, Musaev:2013rq, Hohm:2014qga, Lee:2015xga}.

\section{The solution-generating transformation}
\label{sec:The solution-generating transformation}

Many sets of equations we encounter in physics are intractable as they stand, examples of particular relevance to this work being the Einstein field equations of general relativity and the incompressible Navier-Stokes equations of hydrodynamics. A natural question with solving such notoriously difficult equations, is whether there is some sensible constraint we can place on solutions, such as time independence, which will reduce the equations to a more tractable system. Not only would the more tractable system be easier to solve, it may possess symmetries not present in the full, unconstrained set of equations. Such symmetries would manifest as dualities between solutions. In particular, simple solutions may be dual to more complex ones, and we can use the symmetry groups to generate these additional solutions perhaps more easily than trying to solve any of the equations directly. Let us offer an example from $d$-dimensional static vacuum Einstein gravity, originally derived by Buchdahl \cite{BUCHDAHL01011954}. We will present a derivation using a similar method to that we will employ in \S\ref{sec:Solution-generating symmetries from an effective action}. Consider a spacetime static with respect to a coordinate $x^0$. Defining coordinates $x^\mu = (x^0, x^a)$, the metric can be written
\begin{equation}
	g_{\mu \nu}
	=
		\begin{pmatrix}
			F			&		0
			\\
			0			&		g_{ab}
		\end{pmatrix}
	=	
		\begin{pmatrix}
			F			&		0
			\\
			0			&		F^{-1/(d-3)} \hat \gamma_{ab}
		\end{pmatrix},
	\qquad
	\frac{\partial}{\partial x^0} g_{\mu\nu}	=	0,
\end{equation}
where
\begin{equation}
\hat \gamma_{\mu \nu} = F^{1/(d-3)} ( g_{\mu \nu}		-		F^{-1} g_{0\mu} g_{0\nu} )
\end{equation}
is a metric on the space orthogonal to the isometry $\partial_0$. The vacuum field equations $R_{\mu\nu} = 0$ reduce to
\begin{equation}
(^{\hat \gamma} R)_{\mu \nu}		=	  \frac{d-2}{4(d-3)} \frac{1}{F^2} \hat D_\mu F \hat D_\nu F
\qquad
(\hat\gamma^{-1})^{\mu\nu} \hat D_\mu \left( \frac{1}{F} \hat D_\nu F \right)					= 0		,
\label{Hypersurface EFE (static, d-dimensional)}
\end{equation}
where $\hat D_\mu$ is the covariant derivative, and $(^{\hat \gamma} R)_{\mu \nu}$ the Ricci tensor, with respect to $\hat \gamma_{\mu \nu}$. For fixed $\hat \gamma_{\mu \nu}$, equations \eqref{Hypersurface EFE (static, d-dimensional)} are invariant under
\begin{equation}
F		\rightarrow		\frac{1}{F}		,
\end{equation}
and thus we can generate the vacuum solution
\begin{equation}
	g'_{\mu \nu}
	=
		\begin{pmatrix}
			F^{-1}		&		0
			\\
			0			&		F^{1/(d-3)} \hat \gamma_{ab}
		\end{pmatrix}
	=
		\begin{pmatrix}
			F^{-1}		&		0
			\\
			0			&		F^{2/(d-3)} g_{ab}
		\end{pmatrix}		.
\end{equation}
We will see throughout this thesis how we can apply generalisations of this technique to the NS-NS fields of supergravity, the generalised metric of DFT, and to the hydrodynamics holographically dual to gravity, to which we now turn our attention.

\section{The fluid/gravity correspondence}
\label{sec:The fluid gravity correspondence}

The limit of long wavelengths and long time scales of any interacting field theory at finite temperature is governed by hydrodynamics. This fact, leading to the great plethora of such phenomena, from the atmosphere to the oceans, from plasmas in the RHIC to the sun itself, make the study of hydrodynamics not only fascinating, but virtually unavoidable.

This area was perhaps somewhat unexpectedly connected with string theory by Maldacena's observation \cite{Maldacena:1997re} that the dynamics on a codimension one hypersurface in a general relativistic anti-de-Sitter bulk spacetime is dual to a conformal field theory. This is an example of holography, where a gravitational theory is dual to a codimension one quantum field theory. It was realised in \cite{Policastro:2001yc, Policastro:2002se}, that if we scale to the hydrodynamic limit of this duality, we find that the long wavelength, long time scale limit of general relativity is holographically dual to hydrodynamics. This is the fluid/gravity correspondence, which has been explored in \cite{Policastro:2002tn, Kovtun:2004de, Baier:2007ix, Fouxon:2008ik, Bhattacharyya:2008kq, Iqbal:2008by, Gubser:2008sz, Policastro:2008cx, Hansen:2008tq, Hansen:2009xe, Eling:2009sj, Bredberg:2010ky, Brattan:2011my, Kuperstein:2011fn, Leigh:2011au, Huang:2011he, Kolekar:2011gg, Huang:2011kj, chirco2011thermodynamic, Buchel:2011wx, Adams:2012th, Bai:2012ci, Caldarelli:2012cm, Berkeley:2012kz, Cai:2012mg, Zhang:2012uy, Compere:2012mt, Caldarelli:2013aaa, Zou:2013fua, Green:2013zba, Ashok:2013jda, brattan2013aspects, Wu:2013mda, Kuperstein:2013hqa, Cai:2013uye, Wu:2013kqa, Armas:2013hsa, Klemm:2014nka, Bu:2014ena, Arean:2015wea, Sadeghi:2015noa, Chesler:2015lsa, Bhattacharyya:2015dva, Emparan:2013ila, Puletti:2015gwa, Hao:2015zxa, Crossley:2015tka, deBoer:2015ija, Blake:2015aa, Donos:2015gia} (see \cite{Rangamani:2009xk, Hubeny:2011hd} for reviews). One finds that under a hydrodynamic expansion governed by this limit, whilst demanding regularity on the future horizon, the Goldstone mode dynamics of metrics in general relativity are governed by the equations of motion of a codimension one fluid \cite{Bhattacharyya:2008jc, Bhattacharyya:2008mz, Eling:2012ni, Pinzani-Fokeeva:2014cka, Bredberg:2011jq, Compere:2011dx, Cai:2011xv}. The correspondence has been shown to hold for higher derivative gravity corrections \cite{Niu:2011gu, Hu:2011ze, Cai:2012vr, Zou:2013ix, Cai:2014sua, Bu:2015bwa}. By considering various bulk spacetime backgrounds and stress-energy content, one finds dual hydrodynamics on curved backgrounds \cite{Anninos:2011zn, Bredberg:2011xw} and in the presence of forcing terms responsible for e.g.\ magnetohydrodynamics \cite{Lysov:2013jsa}.

\section{Structure of the thesis}

We will begin in chapter \ref{ch:Double Field Theory} with an exploration of DFT, starting with an introduction in \S\ref{sec:NS-NS supergravity} to the NS-NS sector of supergravity and in \S\ref{sec:T-duality} to T-duality via Buscher's procedure of the string $\sigma$-model. Section \ref{sec:The doubled space} continues with an introduction to $O(d,d,\mathbb R)$-invariant double field theory, adapted largely from \cite{Giveon:1994fu, Aldazabal:2013sca, Berman:2013eva}. In \S\ref{Action and equations of motion (DFT)}, a novel derivation of the full DFT equations of motion is presented and, using this, we offer in \S\ref{sec:A DFT wave solution recovers the string} a pp-wave type DFT solution which embeds both the fundamental string and the spacetime pp-wave for two different solutions to the section condition. The Goldstone modes of this solution are analysed in \S\ref{sec:Goldstone Modes of the Wave Solution (DFT)} for the string case, where the dynamics are governed by a self-duality relation in agreement with the duality-symmetric string of \cite{Duff:1989tf, Tseytlin:1990nb, Tseytlin:1990va}. We proceed in \S\ref{sec:Buscher-type transformation in the DFT sigma-model} with an adaptation of the Buscher procedure for the string $\sigma$-model to $\sigma$-model actions in DFT.

We move in chapter \ref{ch:U-duality invariant M-theory} to an introduction to $E_d$ invariant geometries, in particular the $SL(5)$-invariant extended geometry in \S\ref{sec:The $SL(5)$-invariant theory}, including in \S\ref{sec:Action and equations of motion (SL(5))} a full chain rule derivation of the equations of motion, and in \S\ref{sec:An SL(5) wave solution recovers the M2-brane} a wave solution which corresponds to the M2-brane from the spacetime perspective.

Chapter \ref{ch:Solution-generating transformations in NS-NS supergravity and DFT} discusses solution generating techniques in general relativity, supergravity and DFT in the presence of an isometry. We detail in \S\ref{sec:Solution-generating symmetries from an effective action} a $(d-1)+1$ split of $d$-dimensional spacetime in the presence of one Killing vector. We introduce in \S\ref{sec:Solution-generating symmetries from an effective action} a formalism due to \cite{ANDP:ANDP19694790108} for determining solution generating symmetries of general relativity with an isometry in the case that the dynamics can be encoded in an effective action of scalar potentials. In \S\ref{sec:Application to bosonic NS-NS supergravity} this is applied to two sectors of static NS-NS supergravity. Lastly, we briefly explore analogous solution-generating techniques in DFT in \S\ref{sec:Extension to double field theory}.

In chapter \ref{ch:Solution-generating transformation in the fluid/gravity correspondence} we look to extend these solution-generating techniques in the presence of an isometry to the fluid/gravity correspondence. Section \ref{sec:Hydrodynamics} begins with an introduction, largely adapted from \cite{Rangamani:2009xk}, to relativistic hydrodynamics and the non-relativistic limit governed by the incompressible Navier-Stokes equations (INS). Section \ref{sec:The fluid/gravity correspondence} continues with a brief introduction to the fluid/gravity correspondence, leading into a summary of the derivation in \cite{Compere:2011dx} of a vacuum metric dual to a codimension one incompressible Navier-Stokes fluid. We proceed in \S\ref{sec:The generalised Ehlers transformation in the fluid/gravity correspondence} to apply the generalised Ehlers group of \cite{Mars:2001gd} in the presence of a Killing vector to this vacuum metric to derive transformations of the Navier-Stokes fluid, and find in \S\ref{sec:Deriving the transformation on the fluid} and \S\ref{sec:Fixed viscosity mathbb Z_2 transformations} a set of invariance transformations of the fluid. We discuss how these transformations act from the bulk spacetime perspective in \S\ref{sec:Are the metric transformations vacuum to vacuum?}. Section \S\ref{sec:Extension to magnetohydrodynamics} takes a brief look at similar solution-generating transformations for a metric dual to magnetohydrodynamics.

Finally, chapter \ref{ch:Conclusion and outlook} concludes with a discussion of our results and their implications, and a look at possible directions of further research. The appendices include notations and useful formulae for differential forms.

\newpage
\chapter{Double field theory}
\label{ch:Double Field Theory}
\chead{\textsc{double field theory}}

\section{NS-NS supergravity}
\label{sec:NS-NS supergravity}

In the low energy limit of the five string theories one obtains various supergravity theories, each with fermionic and bosonic contributions. These supergravity theories share a common bosonic NS-NS sector, whose fields derive from purely bosonic modes on the string worldsheet. This sector forms a consistent supersymmetric theory on its own---supersymmetry generators transform the NS-NS fields among themselves. On this basis, it is quite reasonable, for simplicity, to restrict our analysis to the bosonic NS-NS sector of supergravity.

The degrees of freedom are the spacetime metric $g_{\mu\nu}$, two-form Kalb-Ramond field $B_{\mu\nu}$, and dilaton $\phi$. These vary as functions of the coordinates $x^\mu$, and the action is given by
\begin{equation}
S		=		\frac{1}{2 \kappa^2} \int \, \ud^d x \sqrt{-\det (g)} \, e^{-2\phi}
				\left(
					R		+		4 \partial_\mu \phi \partial^\mu \phi		-		\frac{1}{12} H_{\mu \sigma \rho} H^{\mu \sigma \rho}
				\right)		,
\label{NS-NS supergravity action in d dimensions}
\end{equation}
where $\kappa$ is a constant. The three-form field strength of $B$
\begin{equation}
H		=		\ud B	
\end{equation}
additionally satisfies the Bianchi identity
\begin{equation}
\ud H		=		0		.
\label{Bianchi identity for H}
\end{equation}
The minima of \eqref{NS-NS supergravity action in d dimensions} are given by the equations of motion
\begin{subequations}
\begin{align}
R_{\mu \nu}		-		\frac{1}{4} H_\mu{}^{\sigma \rho} H_{\nu \sigma \rho}		+		2 \nabla_\mu \nabla_\nu \phi		& =	0
\label{NS-NS SUGRA eom: R_mu nu - 1/4 H^2 _mu nu + 2 nabla_mu nabla_nu phi}
\\
\nabla_\mu (e^{-2\phi} H^{\mu \nu \rho}	)	& =		0
\\
\frac{1}{6} H^{\mu \sigma \rho} H_{\mu \sigma \rho}		+		2 \nabla^\mu \nabla_\mu \phi		-		4 \nabla^\mu \phi \nabla_\mu \phi		& =		0		.
\end{align}
\label{NS-NS SUGRA equations of motion}%
\end{subequations}

The equations of motion and action are invariant under gauge transformations of the two-form, along with diffeomorphisms whose infinitesimal action is given by the Lie derivative. For one-forms $\ell$ and infinitesimal vectors $k$, these are
\begin{equation}
			g	 \rightarrow	g	+ \mathcal{L}_k g
		\qquad
			B		 \rightarrow	B		+ \mathcal{L}_k B		+		\ud\ell
		\qquad
			\phi			 \rightarrow	\phi			+ \mathcal{L}_k \phi		,
\label{infinitesimal diffeos and gauge transformations in sugra}
\end{equation}
where the Lie derivative of (zero weight) tensors $T^{\mu_1 \ldots \mu_p}{}_{\nu_1 \ldots \nu_q}$ with respect to vectors $a$ is
\begin{equation}
(\mathcal L_a T)^{\mu_1 \ldots \mu_p}{}_{\nu_1 \ldots \nu_q}		=
\begin{aligned}[t]
	&	a^\rho \partial_\rho T^{\mu_1 \ldots \mu_p}{}_{\nu_1 \ldots \nu_q}
\\	&	- \sum_{s=1}^p T^{\mu_1 \ldots \mu_{s-1} \rho \mu_{s+1} \ldots \mu_p}{}_{\nu_1 \ldots \nu_q} \partial_\rho a^{\mu_s}
\\	&	+ \sum_{r=1}^q T^{\mu_1 \ldots \mu_p}{}_{\nu_1 \ldots \nu_{r-1} \rho \nu_{r+1} \ldots \nu_q} \partial_{\nu_r} a^\rho .
\end{aligned}
\label{Lie derivative definition}
\end{equation}
The action exhibits no further symmetries as is, but when the theory is compactified on $T^r$, the equations of motion (but not the action) are invariant under a solution-generating symmetry group $O(r,r,\mathbb R)$.

\section{T-duality}
\label{sec:T-duality}

\subsubsection{The Buscher rules}
\label{sec:The Buscher rules}

The extended nature of the string allows it to wrap round non-contractible compact dimensions, resulting in ``winding mode" contributions to its dynamics in addition to the momentum mode contributions familiar from particle dynamics. We will present here a type of duality that relates these two modes, called T-duality, which can be traced back to work by \cite{Kikkawa:1984cp, Sakai:1985cs}. A thorough review of the subject is given in \cite{Giveon:1994fu}. The particular presentation given here was derived by Buscher \cite{buscher1987symmetry, Buscher:1987qj} from the string $\sigma$-model, whose action is
\begin{equation}
S		=		\frac{ 1 }{ 4 \pi \alpha' }	\int \, \ud ^2 \sigma \sqrt{ \det h } \, [	h^{a b} g_{\mu\nu} \partial_a x^\mu \partial_b x^\nu		+		i \epsilon^{a b} B_{\mu\nu} \partial_a x^\mu \partial_b x^\nu		+		\alpha' R^{ (2) } \phi (x)		],
\label{sigma-model action}
\end{equation}
where the target space (coordinates $x^\mu$) has metric $g_{\mu\nu}$, two-form $B_{\mu\nu}$, and dilaton $\phi$, and the worldsheet (coordinates $\sigma^a$) has metric $h_{a b}$ with scalar curvature $R^{(2)}$, and alternating symbol $\epsilon_{ab}$ (see app. \ref{app:Alternating symbol and tensor}). The worldsheet is two-dimensional and thus the metric $h_{ab}$ can be brought (locally on the worldsheet) to Minkowski form via conformal rescaling and diffeomorphisms.

If one imposes an abelian isometry $\partial/\partial x^0 \cdot = 0$ on the target space, where $x^\mu = ( x^0 , x^m )$, then the action
\begin{equation}
S'		=		
			\begin{aligned}[t]
				&		\frac{ 1 }{ 4 \pi \alpha' }	\int \, \ud^2 \sigma \sqrt{ \det h } \, [	 h^{a b} ( g_{0 0} V_a V_b								+		2 g_{0 n} V_a \partial_b x^n		+		g_{mn} \partial_a x^m \partial_b x^n )
				\\
				&		+		i \epsilon^{a b} ( 2 B_{0n} V_a \partial_b x^n		+		B_{mn} \partial_a x^m \partial_b x^n	)		+		2 i \hat x^0 \epsilon^{a b} \partial_a V_b		+		\alpha' R^{ (2) } \phi (x)		]
			\end{aligned}
\end{equation}
provides equation of motion for Lagrange multiplier $\hat x^0$
\begin{equation}
\epsilon^{a b} \partial_a V_b			=			0,
\end{equation}
which on topologically trivial worldsheets fixes $V_a = \partial_a x^0$, which returns the original action \eqref{sigma-model action}. If one instead integrates by parts the term with $\hat x^0$, and substitutes in the resulting equations of motion for $V_a$,
\begin{equation}
V_a		=		-		\frac{ 1 }{ g_{00} }
							(		g_{0 n} \partial_a x^n		+		i h_{a b} \epsilon^{b c} (B_{0 n} \partial_c x^n		+		\partial_c \hat x^0		)		),
\end{equation}
one recovers the dual action
\begin{equation}
S'		=		\frac{ 1 }{ 4 \pi \alpha' }	\int \, \ud ^2 \sigma \sqrt{ \det h } \, [	h^{a b} g'_{\mu\nu} \partial_a \hat x^\mu \partial_b \hat x^\nu		+		i \epsilon^{a b} B'_{\mu\nu} \partial_a \hat x^\mu \partial_b \hat x^\nu		+		\alpha' R^{ (2) } \phi' (\hat x)		],
\end{equation}
with dual coordinates $\hat x^\mu = (\hat x^0, x^m)$, background
\begin{equation}
\begin{gathered}
g'_{0 0}				=		\frac{ 1 }{ g_{0 0} }
\qquad
g'_{0n}					=		\frac{ B_{0n} }{ g_{0 0} }
\qquad
g'_{mn}					=		g_{mn}		-		\frac{ g_{0 m} g_{0 n}		-		B_{0 m} B_{0 n} }{ g_{0 0} }
\\
B'_{0m}					=		\frac{ g_{0m} }{ g_{0 0} }
\qquad
B'_{mn}					=		B_{mn}		-		\frac{ g_{0m} B_{0n}		-		B_{0m} g_{0n} }{ g_{0 0} }		
\end{gathered}
\label{Buscher rules}
\end{equation}
and dilaton
\begin{equation}
\phi'		=	\phi		- \frac{1}{2} \ln g_{00}		.
\end{equation}
The dilaton shift is found by a careful regularisation of determinants \cite{Tseytlin:1990va, tseytlin1991duality, schwarz1993dilation}.

This is called T-duality, and can be generalised to to the case of toroidal $T^r$ compactifications defined by $r$ compact bosonic directions identified as $X^i \sim X^i+2\pi$, where the full T-duality group is given by $O(r,r,\mathbb Z)$ acting on the background matrix
\begin{equation}
E_{ij}		=		g_{ij}		+ B_{ij}
\label{background matrix E = g+B}
\end{equation}
and dilaton in the non-linear form
\begin{equation}
E'		= \frac{a E + b}{c E + d}
\qquad
\phi'	= \phi + \frac{1}{4} \ln \left[ \frac{\det(g')}{\det(g)} \right]
\qquad
\begin{pmatrix}
	a	&	b	\\
	c	&	d
\end{pmatrix}
\in O(r,r,\mathbb Z)		.
\label{O(r,r,Z) transformation of background matrix}
\end{equation}
The $O(r,r,\mathbb Z)$ group is generated by three elements:
\begin{itemize}
\item	Large diffeomorphisms of $T^r$ preserving periodicities which produce a $GL(r,\mathbb Z)$ basis change
		\begin{equation}
			O_A			=		\begin{pmatrix}
									A^T		&		0		\\
									0		&		A^{-1}
								\end{pmatrix}
								\qquad
								\text{where}
								\qquad
								A \in GL(r,\mathbb Z)
		\end{equation}
\item	Integer shifts in the $B$ field which enact a shift $2\pi\mathbb Z$ in the action and thus do not change the path integral
		\begin{equation}
			O_\Omega	=		\begin{pmatrix}
									1_r		&		\Omega		\\
									0			&		1_r
								\end{pmatrix}
								\qquad
								\text{where}
								\qquad
								\Omega_{ij} = - \Omega_{ji} \in GL(r,\mathbb Z)
		\end{equation}
\item	Factorised dualities, which in even dimensions correspond to the generalisation of the radial inversion $g_{00} \rightarrow 1/g_{00}$ in \eqref{Buscher rules}
		\begin{equation}
			O_{e_s}		=		\begin{pmatrix}
									1_r-e_s	&		e_s			\\
									e_s		&		1_r-e_s
								\end{pmatrix}
								\qquad
								\text{where}
								\qquad
								(e_s)_{ij}		=		\delta_{sj}\delta_{sk}
		\end{equation}
		for some $s\in [1,r]$.
\end{itemize}
In the low energy limit, this symmetry becomes the $O(r,r,\mathbb R)$ of NS-NS supergravity compactified on $T^r$.

\section{The doubled space}
\label{sec:The doubled space}

There has been a long-standing interest \cite{Duff:1989tf, Tseytlin:1990nb, Tseytlin:1990va} in formulating string dynamics in a manifestly T-duality-invariant manner, where T-duality acts linearly on fields. Exactly how to do this was greatly aided by the generalised geometry construction of Hitchin \cite{2002math......9099H} and Gaultieri \cite{2004math......1221G}, who studied structures on the tangent space $TM\oplus T^* M$, where sections consist of the sum of vectors and one-forms. This sum can then parametrise gauge transformations of a symmetric tensor $g$ and two-form $B$. Of particular interest here, is that the tangent space has a natural $O(d,d,\mathbb R)$ inner product which, in local coordinates such that $A = A^\mu\partial_\mu$ and $a = a_\mu \ud x^\mu$, becomes
\begin{equation}
\langle A + a, B + b \rangle		=		\frac{1}{2} ( A^\mu b_\mu  +  B^\mu a_\mu )		.
\end{equation}
In double field theory (DFT), one doubles not just the tangent space, but the coordinates space itself. We shall follow the expositions in \cite{Berman:2013eva, Aldazabal:2013sca, Hohm:2013bwa} for this short introduction to DFT. The dimension of the coordinate space is given by the dimension $2d$ of the fundamental of $O(d,d,\mathbb R)$, and the doubled coordinates
\begin{equation}
X^M		=		( x^\mu, \tilde x_\mu )
\end{equation}
then transform linearly as a group representative
\begin{equation}
X^M		\rightarrow		O^M{}_N X^N	,
\end{equation}
where $O$ is an $O(d,d,\mathbb R)$ matrix defined to preserve the inner product
\begin{equation}
O_M{}^P \eta_{PQ} O_N{}^Q		=	\eta_{MN}		,
\qquad
\eta_{MN}	=	\begin{pmatrix}
						0					&		\delta^\mu_\nu
				\\		\delta_\mu^\nu		&		0
				\end{pmatrix}		
\end{equation}
and we have introduced the $O(d,d,\mathbb R)$ metric $\eta$. This $O(d,d,\mathbb R)$ is a global symmetry of DFT.

The supergravity fields $g$, $B$ and $\phi$ will now depend on the $2d$ coordinates of this doubled space. This is clearly undesired from a supergravity perspective. Before we formally introduce how this is remedied, let us illustrate the connection with the $O(d,d,\mathbb Z)$ of T-duality. Consider a particular case where the fields depend only on the directions $x^{\mu\neq d}$ (and not the $\tilde x_\mu$), as is the case in supergravity with an isometry $\partial/\partial x^d \cdot = 0$. Then we are quite free to demand that the physical coordinates of spacetime are the $x^{\mu\neq d}$, plus one of the pair $(x^d, \tilde x_d)$. This freedom corresponds to the $O(1,1,\mathbb Z) = \mathbb Z_2$ T-duality transformation. In particular, the choice $x^d$ corresponds to the usual spacetime coordinate, whilst the choice $\tilde x_d$ corresponds to the T-dual coordinate. In this way, in compactification on tori, one can see the $\tilde x_\mu$ as ``winding" coordinates, Fourier dual to the winding momenta in the string context, though this is not their interpretation in general. One can thus see the connection with the T-duality of closed string theory (for a more detailed discussion, see \cite{Cederwall:2014opa}).

In fact, one finds on formulating the level matching constraint in closed string field theory in a $O(d,d,\mathbb R)$ covariant manner, the condition needed to restrict the dependence of the fields on the number of coordinates \cite{Hull:2009mi}. It is termed the \emph{weak constraint} and requires that
\begin{equation}
Y^{MN}{}_{PQ} \partial_M \partial_N \theta = 0
\label{weak constraint}
\end{equation}
annihilates single fields $\theta$, where the DFT Y-tensor\footnote{We write the weak constraint explicitly in terms of the Y-tensor as this form is also applicable to the $E_d$-invariant extended geometries we will discuss in chapter \ref{ch:U-duality invariant M-theory} (where the form of the Y-tensor depends on the relevant duality group $E_d$). In DFT, it is frequently written in the literature in the simpler yet equivalent form $\eta^{MN}\partial_M \partial_N \theta = 0$, for single fields $\theta$.}
\begin{equation}
Y^{MN}{}_{PQ}		=	\eta^{MN} \eta_{PQ}		.
\end{equation}

One can define local gauge transformations on this space in terms of a generalised Lie derivative with respect to a generalised vector $U^M$, the most general form of which, acting on a tensor $A_M{}^N$ with weight $w(A)$, is
\begin{multline}
(L_U A)_M{}^N		= U^P\partial_P A_M{}^N		+ A_P{}^N \partial_M U^P		-	A_M{}^P \partial_P U^N \\		+ A _M{}^P Y^{QN}{}_{PR} \partial_Q U^R		- A _P{}^N Y^{QP}{}_{RM} \partial_Q U^R		+		w(A) \partial_P U^P A_M{}^N.
\label{Generalised lie derivative}
\end{multline}
It is important to note that these generalised coordinate transformations are not conventional coordinate transformations on a doubled space---one cannot form \eqref{Generalised lie derivative} from the conventional diffeomorphism group on the $2d$ space with the additional condition that diffeomorphisms respect the $O(d,d,\mathbb R)$ metric $\eta_{MN}$. To ensure that the generalised Lie derivative sends tensors into tensors requires its closure
\begin{equation}
[L_{U_1} , L_{U_2}]	V^M	=	L_{[U_1,U_2]_C} V^M
\end{equation}
on to the \emph{C-bracket}
\begin{equation}
[U_1,U_2]_C		\equiv		\frac{1}{2}\left( L_{U_1} U_2 - L_{U_2} U_1 \right),
\end{equation}
which, for weight zero vectors $U_1$, $U_2$, $V$, requires
\begin{equation}
Y^{RS}{}_{PQ} \left( U_1^P \partial_R U_2^Q \partial_S V^M		+		 2 \partial_R U_1^P \partial_S U_2^M V^Q \right)		-		U_1 \leftrightarrow U_2		=	0		.
\label{closure of generalised Lie derivative}
\end{equation}
One solution to this is the \emph{strong constraint} or \emph{section condition}, that in addition to the weak constraint \eqref{weak constraint},
\begin{equation}
Y^{MN}{}_{PQ} \partial_M \theta_1 \partial_N \theta_2= 0
\label{strong constraint or section condition}
\end{equation}
annihilates products of fields. In \cite{Betz:2014aia} the authors discuss the CFT origin of this constraint. It can be shown that all solutions to the strong constraint are $O(d,d,\mathbb R)$ rotations of $\tilde \partial^\mu \theta = 0$ for any field $\theta$. We will refer to the frame where all objects have no dependence on the $\tilde x_\mu$ coordinates as the \emph{supergravity frame}. We note that \eqref{strong constraint or section condition} is not the most general solution to \eqref{closure of generalised Lie derivative}. More general solutions are possible, and are pivotal in the Scherk-Schwarz reduction of DFT to gauged supergravities \cite{Aldazabal:2011nj, Geissbuhler:2011mx, Grana:2012rr, Berman:2012uy, Berman:2013cli, Hassler:2014sba, Lee:2015xga}.

The supergravity fields may be cast into representations of $O(d,d,\mathbb R)$. One can unify the spacetime metric and two-form into a $O(d,d,\mathbb R)/O(d,\mathbb R)\times O(d,\mathbb R)$ coset representative\footnote{The $O(d,\mathbb R)\times O(d,\mathbb R)$ forms a local symmetry of the theory, analogous to the Lorentz symmetry of relativity.} \emph{generalised metric}
\begin{equation}
H_{MN}		=		\begin{pmatrix}
							g_{\mu\nu} - B_{\mu\rho} g^{\rho\sigma} B_{\sigma\nu}		&		B_{\mu\rho} g^{\rho\nu}												\\
							- g^{\mu\sigma} B_{\sigma\nu}											&		g^{\mu\nu}
						\end{pmatrix}		,
\label{DFT metric}
\end{equation}
transforming as an $O(d,d,\mathbb R)$ tensor:
\begin{equation}
H_{MN}		\rightarrow		O_M{}^P H_{PQ} O_N{}^Q		.
\end{equation}
The coset form ensures the metric has inverse $H^{MN} = \eta^{MP} H_{PQ} \eta^{QN}$. Meanwhile, the supergravity dilaton is shifted to define an $O(d,d,\mathbb R)$ scalar density DFT dilaton $d$ defined by\footnote{To align with the literature, we use $d$ to mean both the spacetime dimension and the DFT dilaton field, though it should be clear which is implied in each instance.}
\begin{equation}
e^{-2d}		=		\sqrt{ -\det(g) } \, e^{-2\phi}		.
\label{DFT dilaton: definition in terms of sugra fields}
\end{equation}
Infinitesimal gauge transformations of the generalised metric and dilaton are given by the generalised Lie derivative with respect to an infinitesimal generalised vector $(k^\mu, \ell_\mu)$. In the supergravity frame, these generate the spacetime diffeomorphisms and gauge transformations \eqref{infinitesimal diffeos and gauge transformations in sugra} of the supergravity fields.

\subsection{Action and equations of motion}
\label{Action and equations of motion (DFT)}

One can form an action in the DFT metric and dilaton. The supergravity Lagrangian contains terms up to second order in derivatives, so one also expects two derivatives on the metric $H$ and dilaton $d$ in the DFT Lagrangian\footnote{Higher derivative corrections have been considered in \cite{Hohm:2013jaa, Hohm:2014xsa, Lee:2015kba}.}. All possible $O(d,d,\mathbb R)$-covariant terms can be linearly summed over, and the relevant constants fixed on requiring the Lagrangian transforms covariantly under DFT gauge transformations. Meanwhile, the DFT dilaton forms the measure in the action. The action is thus \cite{Hohm:2010jy, Hohm:2010pp}
\begin{equation}
S		=		\int \ud^{2d}X	e^{-2d} R		,
\label{DFT action}
\end{equation}
where the $O(d,d,\mathbb R)$ scalar
\begin{equation}
R		=
\begin{aligned}[t]
	&	\frac{1}{8} H^{MN} \partial_M H^{KL} \partial_N H_{KL} 
		- \frac{1}{2} H^{MN}\partial_M H^{KL}\partial_K H_{NL} \\
	&	+ 4 H^{MN}\partial_M\partial_N d - \partial_M\partial_N H^{MN}
		- 4 H^{MN}\partial_M d \partial_N d + 4\partial_M H^{MN}\partial_N d		.
\end{aligned}
\label{O(d,d,R) scalar R in DFT action}
\end{equation}
We have neglected an additional term (see e.g. \cite{Grana:2012rr}) whose contribution vanishes under the strong constraint here and in all further dynamics, but is necessary for the Scherk-Schwarz reduction to gauged supergravity. A treatment of boundary contributions for the DFT action can be found in \cite{Berman:2011kg}. It is non-trivial that the action \eqref{DFT action} reproduces the NS-NS supergravity action \eqref{NS-NS supergravity action in d dimensions} in the supergravity frame.

We now introduce a novel derivation of the DFT equations of motion. The degrees of freedom are encoded within the metric $H$ and dilaton $d$. Varying the action with respect to the latter produces
\begin{equation}
\delta_d S = \int \ud^{2d} X \left( -2 e^{-2d} R \delta d	\right)	+ \text{boundary terms}		,
\end{equation}
yielding the dilaton equation of motion
\begin{equation}
R		=		0.
\label{DFT dilaton eom}
\end{equation}
Meanwhile, varying the action with respect to the metric yields
\begin{equation}
\delta_H S = \int \ud^{2d} X e^{-2d} \mathcal{K}_{MN} \delta H^{MN}		+ \text{boundary terms}		,
\end{equation}
where
\begin{equation}
\mathcal K_{M N}		=
\begin{aligned}[t]
&		\frac{1}{8} \partial_M H^{K L} \partial_N H_{K L}
		- \frac{1}{2} H^{P K} H^{Q L} \partial_L H_{P M} \partial_K H_{N Q}
\\
&		+ \frac{1}{4} H^{P Q} H^{K L} \partial_P H_{K M} \partial_Q H_{N L}
		+ 2 \partial_M \partial_N d
\\		
&		+ ( \partial_L		- 2 (\partial_L d) )
		\left[
			H^{L K} \left(  \partial_{ (M } H_{ N) K}		- \frac{1}{4} \partial_K H_{M N} \right)
		\right].
\end{aligned}
\end{equation}
However, this does not yield the equations of motion $\mathcal K_{MN} = 0$, since $\delta_H S$ is required to vanish only for those $\delta H^{MN}$ constrained by the $O(d,d,\mathbb R)/O(d,\mathbb R)\times O(d,\mathbb R)$ coset form of the generalised metric. This was realised by the authors of \cite{Hohm:2010pp}, who derived the full equations of motion by ensuring that the coset form was satisfied through the condition $\eta_{MP} H^{PQ} \eta_{QN} = H_{MN}$. We will take a different route, which while considerably more cumbersome, may be adapted quite naturally to the more demanding case of $E_d$-covariant extended geometries, where the full equations of motion are largely unknown. We will
\begin{itemize}
\item	Demand that the metric assumes the coset form \eqref{DFT metric}
\item	Expand variations of the generalised metric using the chain rule in terms of variations of the field content: the spacetime metric $g$ and two-form $B$
\item	Re-express the result in a manifestly duality-covariant form
\item	Determine the duality-covariant equations of motion which contains precisely the information of the equations of motion for the fields $g$ and $B$
\end{itemize}
That is,
\begin{subequations}
\begin{align}
\mathcal K_{MN} \delta H^{MN}
& =		\mathcal K_{MN} \left( \frac{\partial H^{MN}}{\partial g_{\mu \nu}} \delta g_{\mu \nu}		+ \frac{\partial H^{MN}}{\partial B_{\mu \nu}} \delta B_{\mu \nu} \right)
\\
& =		
\begin{aligned}[t]
	&	\begin{aligned}[t]
		\Big[
		&	- \mathcal K_{\mu\nu}g^{\mu\rho}g^{\sigma\nu}
			+ 2\mathcal K_\mu{}^\nu g^{\mu\rho}g^{\sigma\tau}B_{\tau\nu}
			\\
		&	+ \mathcal K^{\mu\nu}
			\left(
				\delta_\mu^\rho \delta_\nu^\sigma
				+ B_{\mu\tau}g^{\tau\rho}g^{\sigma\lambda}B_{\lambda\nu}
			\right)
		\Big]
		\delta g_{\rho\sigma}
		\end{aligned}
	\\
	&	+ \left[
			- 2\mathcal K_\mu{}^\nu g^{\mu\tau}\delta_\tau^\rho \delta_\nu^\sigma
			- 2\mathcal K^{\mu\nu} B_{\mu\tau} g^{\tau\lambda} \delta_\lambda^\rho\delta_\nu^\sigma
		\right]
		\delta B_{\rho\sigma}
\end{aligned}
\\
& =
\begin{aligned}[t]
	&	\begin{aligned}[t]
		\big[
		&	- \mathcal K_{\mu\nu} H^{\mu\rho} H^{\sigma\nu}
			+ 2\mathcal K_\mu{}^\nu H^{\mu\rho}H^{\sigma}{}_\nu 
		\\
		&	+ \mathcal K^{\mu\nu}
			\left(
				\delta_\mu^\rho \delta_\nu^\sigma
				- H_\mu{}^\rho H^\sigma{}_\nu
			\right)
		\big]
		\delta g_{\rho\sigma}
		\end{aligned}
	\\
	& 	-
		\left(
			\mathcal K_\mu{}^\nu H^{\mu\tau}
			+ \mathcal K^{\mu\nu} H_\mu{}^\tau
		\right)
		\left(
			\delta_\tau^\rho \delta_\nu^\sigma 
			- \delta_\tau^\sigma \delta_\nu^\rho
		\right)
		\delta B_{\rho\sigma}
\end{aligned}
\\
& =
\begin{aligned}[t]
	&	\mathcal K_{KL}
		\left(
			\eta^{K\rho}\eta^{\sigma L} - H^{K\rho} H^{\sigma L}
		\right) 
		\delta g_{\rho\sigma}
	\\
	&	- \mathcal K_{KL}
		\left(
			H^{KP}\eta_{PM}\eta^{LN} - H^{KP}\delta_P^N \delta_M^L
		\right)
		\eta^{M\rho} \delta^\sigma_N \delta B_{\rho\sigma}
\end{aligned}
\\
& =		P_{MN}{}^{KL} \mathcal K_{KL}
		\left(
			\eta^{M\rho}\eta^{\sigma N}\delta g_{\rho\sigma}
			+ \eta^{M\rho} H^{\sigma N}\delta B_{\rho\sigma}
		\right)		,
\end{align}
\label{DFT variation of metric x mathcal K in terms of g, B}%
\end{subequations}
where
\begin{equation}
P_{MN}{}^{KL}		= \delta_M^K \delta_N^L 		- H_{MP}\eta^{PK}\eta_{NQ} H^{QL}		.
\label{DFT projector}
\end{equation}
Thus,
\begin{equation}
P_{MN}{}^{KL} \mathcal K_{KL}		= 0
\label{DFT metric eoms}
\end{equation}
are \emph{sufficient} covariant equations of motion which correspond to minima of the action \eqref{DFT action}. That they are \emph{necessary} (i.e.\ that they do not over-constrain $g$ and $B$), can be seen from a simple degree of freedom counting. We do this for vanishing two-form field and a Minkowski metric. We find $P_{MN}{}^{KL}$ has a kernel on symmetric objects $J_{KL}$ of dimension $d(d+1)$ which, of the $2d(2d+1)/2$ degrees of freedom in the symmetric $\mathcal K_{MN}$, leaves $d^2$ degrees of freedom, i.e.\ \eqref{DFT metric eoms} contains $d^2$ equations of motion. Alongside the 1 dilaton equation of motion \eqref{DFT dilaton eom}, this corresponds to the $d(d+1)/2+d(d-1)/2+1$ degrees of freedom in the supergravity fields\footnote{We note that this degrees of freedom counting, while highly suggestive, is not strictly a full proof that \eqref{DFT metric eoms} are the covariant DFT metric equations of motion. Specifically, we have not shown that the variation of the spacetime metric and two-form do not lie within the kernel of the projector.} $(g, B, \phi)$. Indeed, the equations of motion \eqref{DFT dilaton eom}, \eqref{DFT metric eoms} are in agreement with those calculated in \cite{Hohm:2010pp} up to contributions which vanish under the section condition.

\section{A double field theory wave solution recovers the string}
\label{sec:A DFT wave solution recovers the string}

We seek a generalised metric solution for the DFT equations of motion corresponding to a null wave whose momentum is pointing the $\tilde x_d$ direction. The ansatz will be analogous to that of a pp-wave in general relativity \cite{aichelburg1971gravitational}. That this is a solution of the DFT equations of motion does not follow obviously from the existence of the spacetime pp-wave. As we have seen, the equations of motion of the generalised metric in DFT are certainly not the same as the equations of motion of the metric in relativity. Let us remove any source of confusion. The pp-wave as a solution for $g$ may of course, by construction, be embedded as a solution in DFT by simply inserting the pp-wave solution for $g$ and $B$ into $H$. However, here we will consider a pp-wave analogue (that is the usual pp-wave ansatz of \cite{aichelburg1971gravitational}) not for $g$ but for the DFT metric $H$ itself and then determine its interpretation in terms of the usual metric $g$ and two-form $B$.

With generalised coordinates written
\begin{equation}
X^M
	= (x^\mu,\tilde x_\mu )
	= (t,y^m,z,\tilde t,\tilde y_m,\tilde z)	,
\end{equation}
the following metric
\begin{equation}
H_{MN}\ud X^M \ud X^N	=
\begin{aligned}[t]
	 	& (K-2)(\ud t^2 - \ud z^2) + \delta_{mn}\ud y^m\ud y^n
\\
		& + 2(K-1) (\ud t\ud\tilde z + \ud\tilde t\ud z )
\\
		& - K (\ud\tilde t^2 - \ud\tilde z^2 ) + \delta^{mn}\ud\tilde y_m\ud\tilde y_n
\end{aligned}
\label{DFT wave solution}
\end{equation}
and constant dilaton $d$ constitute a solution to the DFT equations of motion \eqref{DFT dilaton eom}, \eqref{DFT metric eoms} provided $K$ satisfies certain conditions we will shortly discuss.

Since it is of the same form as the usual pp-wave solution, the natural interpretation is of a pp-wave in the doubled geometry. One can therefore imagine it propagates with momentum in the $\tilde z$ direction. To determine whether it truly carries momentum would require the construction of conserved charges in DFT, which is yet to be done. It would be useful to consider objects like generalised Komar integrals, and other ways one defines charges in general relativity, but now for DFT. Nevertheless, we shall proceed with our interpretation.

From a DFT perspective, it is required (at least naively) that $K$ satisfies the section condition. We take $K$ to be a harmonic function of the usual transverse coordinates $y^m$ (but not of their duals $\tilde y_m$):
\begin{equation}
K = K(y^m),
\qquad
{}^y\Box K	=	0,
\qquad
\text{where}
\quad
{}^y\Box \equiv \delta^{mn}\partial_m\partial_n.
\label{wave equation in K (DFT case)}
\end{equation}
This is sufficient for the solution to satisfy the equations of motion. For illustrative purposes, we will adopt the explicit form
\begin{equation}
K		=		1		+		\frac{k_0}{|y|^{d-4}},		\qquad		|y|^2 = y^m y^n\delta_{mn} ,
\end{equation}
where $k_0$ is a constant. In particular, this gives an asymptotically flat background metric. We note that the precise form of $K$ does not affect the Goldstone mode calculations appearing later in \S\ref{sec:Goldstone Modes of the Wave Solution (DFT)}.
That $H$ depends only on the $y^m$ and not their duals implies that the wave solution is smeared in the $\tilde y_m$ directions. One can think of it as a plane wave front extending along the dual $\tilde y_m$ directions but with momentum in the $\tilde z$ direction.

The generalised metric as \eqref{DFT metric} corresponds to a Kaluza-Klein ansatz from which we can determine the supergravity metric and two-form:
\begin{equation}
H_{MN} \ud X^M\ud X^N	=
		(g_{\mu\nu} - B_{\mu\rho}g^{\rho\sigma}B_{\sigma\nu})\ud x^\mu \ud x^\nu
		+ 2B_{\mu\rho}g^{\rho\nu}\ud x^\mu \ud\tilde x_\nu + g^{\mu\nu}\ud\tilde x_\mu\ud\tilde x_\nu 		,
\label{KK reduction of DFT wave}
\end{equation}
while we can determine the dilaton $\phi$ from \eqref{DFT dilaton: definition in terms of sugra fields}. If we choose the spacetime coordinates to be the subset $x^\mu = (t,y^m,\tilde z)$, the DFT wave \eqref{DFT wave solution} reduces via \eqref{KK reduction of DFT wave} to the \emph{spacetime} pp-wave solution
\begin{equation}
\begin{gathered}
g_{\mu\nu} \ud x^\mu x^\nu = (K-2)\ud t^2		+ 2(K-1)\ud t\ud \tilde z		+ K \ud \tilde z^2		+ \delta_{mn} \ud y^m \ud y^n
\\
B_{\mu\nu}		= 0
\qquad
e^{-2\phi}		= e^{-2d}		,
\end{gathered}
\end{equation}
with momentum along the $\tilde z$ direction. Alternatively, with spacetime coordinates $x^\mu=(t,y^m,z)$, we find
\begin{equation}
\begin{gathered}
g_{\mu \nu} \ud x^\mu \ud x^\nu = - K^{-1}(\ud t^2-\ud z^2)		+ \delta_{mn}\ud y^m\ud y^n
\qquad
e^{-2(\phi-d)} = K
\\
B_{tz} =		1 - K^{-1}
\qquad
B_{\mu n} = 0		,
\end{gathered}
\label{fundamental string solution}
\end{equation}
which is the fundamental string solution extended along the $z$ direction \cite{Dabholkar:1990yf}. We have thus shown that the solution \eqref{DFT wave solution} which carries momentum in the $\tilde z$ direction in the doubled space corresponds in the supergravity frame to a string extended along the $z$ direction from the spacetime perspective, and a spacetime pp-wave with momentum in the $\tilde z$ direction in a T-dual frame.

Of course this is no surprise from the point of view of T-duality. Momentum and string winding exchange under T-duality. It is precisely as expected that momentum in the dual direction corresponds to a string. What is more surprising is when one views this from the true DFT perspective. There are null wave solutions that can point in any direction. When we analyse these null waves from the reduced theory we see them as fundamental strings in the supergravity frame, or as pp-waves from a dual direction. It is a simple rotation of direction of propagation that takes one solution into the other. This is duality from the DFT perspective.

Moreover, in the doubled formalism the solution is a massless wave with \\$\mathcal{P}_M \mathcal{P}_N H^{MN}=0$ (where the $\mathcal{P}^M$ are some generalised momenta), while from the spacetime perspective in the supergravity frame the string has a tension $T$ and charge $q$ which are given by the momenta in the dual directions with a resulting BPS equation $T = |q|$.

\subsection{Goldstone modes of the wave solution}
\label{sec:Goldstone Modes of the Wave Solution (DFT)}

In the previous section we presented a solution to the equations of motion of DFT which reduces in the supergravity frame to the fundamental string. It will be interesting to analyse the Goldstone modes of this DFT solution in the supergravity frame. Especially since the advent of M-theory, it was understood that branes are dynamical objects and that when one finds a solution of the low energy effective action one can learn about the theory by examining the dynamics of the Goldstone modes. For D-branes in string theory this was done in \cite{Adawi:1998ta} and for the membrane and fivebrane in M-theory, where such an analysis was really the only way of describing brane dynamics, this was done in \cite{Kaplan:1995cp, Adawi:1998ta}. We will follow the excellent exposition and the method described in \cite{Adawi:1998ta} as closely as possible.

We will consider small perturbations $h_{MN}$ and $\lambda$ of the generalised metric and dilaton respectively, each generated by the generalised Lie derivative \eqref{Generalised lie derivative} with respect to a generalised vector $\zeta^M$,
\begin{align}
h_{MN}		& =		(L_\zeta H)_{MN}
\label{h as lie derivative of H wrt zeta}
\\
\lambda		& =		L_\zeta d			=	\zeta^M \partial_M d - \frac{1}{2}\partial_{M}\zeta^M	.
\label{lambda as lie derivative of d wrt zeta}
\end{align}
Note that \eqref{lambda as lie derivative of d wrt zeta} contains a weight term for the dilaton (whose exponential $e^{-2d}$ has weight 1).

The wave solutions are extended objects and therefore sweep out a worldvolume in space, spanned by the coordinates $(t,z)$. The solution clearly breaks translation symmetry and so one naturally expects worldvolume-scalar zero-modes. One immediate puzzle would be to ask about the number of degrees of freedom of the Goldstone modes. Given that the space is now doubled one might imagine that any solution which may be interpreted as a string would have $2d-2$ degrees of freedom rather than the expected $d-2$. We will answer this question and show how the Goldstone modes have the correct number of degrees of freedom despite the solution living in a $2d$ dimensional space. The projected form of the equations of motion are crucial in making this work out.

To carry out the analysis it will be useful to split up the space into parts longitudinal and transverse to the string. We will do this using an alternative coordinate notation for this section $X^M = (x^\mu, \tilde x^{\bar\mu})$, so as not to confuse the inverse metric and dual coordinates. We collect the coordinates into $x^a = (t,z)$, $\tilde x^{\bar a} = ( \tilde t,\tilde z )$ such that the generalised coordinates are $X^M=(x^a,y^m,\tilde x^{\bar a},\tilde y^{\bar m})$. We will refer to the ($\tilde x^{\bar a}$) $x^a$ as the (dual) worldsheet coordinates, and the ($\tilde y^{\bar m}$) $y^m$ as the (dual) transverse coordinates. In this notation, the non-zero components of the metric and its inverse are
\begin{equation}
\begin{aligned}
	H_{ab}				& = (2-K)\mathbb I_{ab} 			&		H^{ab}				& =		K \mathbb I^{ab}
\\
	H_{\bar a\bar b}	& = K \mathbb I_{\bar a\bar b} 		&		H^{\bar a\bar b}	& =		(2-K)\mathbb I^{\bar a\bar b}
\\
	H_{a\bar b}			& = (K-1)\mathbb J_{a\bar b}		&		H^{a\bar b} 		& =		(K-1)\mathbb J^{a\bar b}
\\
	H_{mn} 				& = \delta_{mn}						&		H^{mn}				& = \delta^{mn}
\\
	H_{\bar m\bar n}	& =		\delta_{\bar m\bar n}		&		H^{\bar m\bar n}	& = \delta^{\bar m\bar n}
\end{aligned}
\end{equation}
where $\mathbb I_{ab}=\mathbb I^{ab}$ ($\mathbb I_{\bar{a}\bar{b}}=\mathbb I^{\bar{a}\bar{b}}$) is the Minkowski space metric $\mathbb I = \diag (-1,1)$ in the (dual) worldsheet coordinates, and the $2\times 2$ matrix
\begin{equation}
\mathbb J = \begin{pmatrix} 0 & 1 \\ 1 & 0 \end{pmatrix}		.
\end{equation}

We now choose a transformation parameter $\zeta^M$ with components along the transverse directions (and their duals),
\begin{equation}
\zeta^M = (0,K^\alpha \varphi^m, 0, K^\beta \tilde \varphi^{\bar m})		.
\end{equation}
Here, $\varphi^m$ and $\tilde\varphi^{\bar m}$ are coordinate-independent vectors that will become the Goldstone modes once we allow them to depend on the worldsheet coordinates, and $\alpha$ and $\beta$ are constants that will be determined by demanding that the Goldstone modes become normalisable. From \eqref{h as lie derivative of H wrt zeta}, the \emph{non-zero} components of the metric perturbation are
\begin{equation}
\begin{aligned}
h_{ab}					& =		-\varphi^m K^\alpha\partial_m K  \mathbb I_{ab}
	& h_{mn}			& =		2 \varphi^q \delta_{q(m} \delta_{n)}{}^p \partial_p K^\alpha
\\
h_{\bar a\bar b}		& =		\varphi^m K^\alpha\partial_m K \mathbb I_{\bar a\bar b}
	& h_{\bar m\bar n}	& =		- 2 \varphi^q \delta_{q(\bar m} \delta_{\bar n)}{}^p\partial_p K^\alpha
\\
h_{a\bar b} 			& =		\varphi^m K^\alpha\partial_m K \mathbb J_{a\bar b}
	& h_{m\bar n} 		& =		- 2\tilde\varphi^{\bar q} \delta_{\bar q [m} \delta_{\bar n]}{}^p \partial_p K^\beta,
\end{aligned}
\label{DFT metric perturbation h with constant varphi, bar varphi}
\end{equation}
while the dilaton perturbation \eqref{lambda as lie derivative of d wrt zeta} is
\begin{equation}
\lambda =  - \frac{1}{2}\varphi^m\partial_{m} K^\alpha  .
\label{DFT dilaton perturbation lambda with constant varphi, bar varphi}
\end{equation}

We now promote the moduli to depend on the worldsheet coordinates (but not their duals, so as to satisfy the section condition),
\begin{equation}
\varphi^m		\rightarrow		\varphi^m(x^a)		,
\qquad
\tilde \varphi^{\bar m}		\rightarrow		\tilde \varphi^{\bar m}(x^a)		.
\label{DFT zero modes}
\end{equation}
These are the Goldstone modes, which are really the normalisable modes corresponding to broken gauge symmetry. Whereas for gravity gauge transformations are ordinary diffeomorphisms, in the case of DFT they are generated by the generalised Lie derivative.

The equations of motion of the modes are determined by those of the perturbed DFT metric $H+h$ and dilaton $d+\lambda$. We will determine them to linear order in $h$ and $\lambda$ (it would certainly be interesting to move beyond this expansion and compare with a Nambu-Goto type action but we will not do so here). In approaching the equations of motion to linear order, we will simplify our calculation with the fact that both the background and gauge transformations of the background (which were calculated for \emph{constant} $\varphi^m$, $\tilde\varphi^{\bar m}$) will satisfy the equations of motion. Thus, terms will only contribute to the equations of motion when there are $x^a$ derivatives acting on modes.

For the metric equations of motion, we note that the background satisfies $\mathcal K= 0$. Therefore, in the linear order equations of motion, $\mathcal K$ will only contain terms linear in the modes, while $P$ will contain no contribution from the modes---it will be given purely by the background metric. With the projector only in terms of the background metric, the equations of motion automatically satisfy
\begin{equation}
(P \mathcal K)_{MN} = - H_{MK} Y^{KP}{}_{NL} H^{LQ} (P \mathcal K)_{PQ},
\label{relation for independent metric eoms for modes}
\end{equation}
which we can use to determine the independent components, which are as follows.

We begin with the transverse-worldsheet cross components (and their duals), $(P\mathcal K)_{ma}$ and $(P\mathcal K)_{m\bar a}$, where
\begin{equation}
(P\mathcal K)_{ma}		+		\mathbb I_{ab} \mathbb J^{b\bar a} (P\mathcal K)_{m\bar a}
	=	\begin{multlined}[t]
		- \frac{1}{2} \left( \delta_m^p \delta_q^r - \delta_{mq} \delta^{pr} \right) \\ \times \left[ \partial_a\varphi^q \partial_r\partial_p K^\alpha - \mathbb I_{ab} \epsilon^{bc} \partial_c \tilde \varphi^{\bar q} \delta_{\bar q}^q \partial_r\partial_p K^\beta \right] ,
		\end{multlined}
\end{equation}
with antisymmetric symbol $\epsilon^{ab}$ (see app. \ref{app:Alternating symbol and tensor}). So as to obtain an equation of motion for the modes independent of the background function $K$, we choose $\alpha = \beta$, where the above vanishes if the modes satisfy a duality relation
\begin{equation}
\partial_a\varphi^q = \mathbb I_{ab} \epsilon^{bc} \partial_c \tilde \varphi^{\bar q} \delta_{\bar q}^q		,
\label{duality-symmetric string relation}
\end{equation}
which in turn implies the wave equations
\begin{equation}
\Box \varphi^q = \Box \varphi^{\bar q}\delta_{\bar q}^q = 0,		\qquad \text{where} \qquad \Box \equiv \mathbb I^{ab} \partial_a \partial_b.
\label{DFT Goldstone mode wave equations}
\end{equation}
Employing $\alpha = \beta$ and the duality relation \eqref{duality-symmetric string relation}, we also have the contribution
\begin{equation}
\mathbb I_{ab} \mathbb J^{b\bar a} (P\mathcal K)_{m\bar a}
\begin{aligned}[t]
&	=		\frac{1}{2} \left(\delta_m^p \delta_q^r - \delta_{mq} \delta^{pr} \right) \left[ \partial_a \varphi^q \partial_r(K^\alpha\partial_p K) + \mathbb I_{ab} \epsilon^{bc} \partial_c \tilde \varphi^{\bar q} \delta_{\bar q}^q \partial_r (K\partial_p K^\beta) \right]
\\
&	=		\frac{1}{2} (\alpha+1) \left( \delta_m^p \delta_q^r - \delta_{mq} \delta^{pr} \right) \partial_a \varphi^q \partial_r(K^\alpha\partial_p K),
\end{aligned}
\end{equation}
which vanishes for $\alpha = -1$. Indeed, this is the value derived in \cite{Adawi:1998ta} for the Goldstone dynamics of branes. In particular, for the D3, M2 and M5 brane, $\alpha = -1$ provided normalisable modes. The independent transverse and worldsheet components, along with the scalar $R$ of the dilaton equation \eqref{DFT dilaton eom} are
\begin{subequations}
\begin{align}
(P\mathcal K)_{mn}			& =		-K \Box \varphi^q \delta_{q(m} \partial_{n)} K^\alpha
\\
(P\mathcal K)_{m\bar n}		& =		K \Box \tilde\varphi^{\bar q} \delta_{\bar q[m} \delta_{\bar n]}^p \partial_p K^\beta		
\\
(P\mathcal K)_{ab}				& =		f_1(K,\alpha) \mathbb I_{ab} \Box \varphi^q \partial_q K		+		(\alpha + 1)(K-2) K^\alpha \partial_a\partial_b \varphi^q \partial_q K
\\
\mathbb I_{bc} \mathbb J^{c\bar b}(P\mathcal K)_{a\bar b}
										& =		f_2(K,\alpha) \mathbb I_{ab} \Box \varphi^q \partial_q K		+		(\alpha + 1)(1-K) K^\alpha \partial_a\partial_b \varphi^q \partial_q K
\\
R										& =		-	(2\alpha + 1) \Box\varphi^q K^\alpha\partial_q K		,
\end{align}
\end{subequations}
where $f_1$ and $f_2$ are functions whose form is unimportant for this analysis. These all vanish by means of the wave equations \eqref{DFT Goldstone mode wave equations} and $\alpha = -1$.

Returning to the duality relation \eqref{duality-symmetric string relation}, we can combine the Goldstone modes into a generalised vector $\Phi^M=(0,\varphi^m,0,\tilde\varphi^{\bar m})$, where we find
\begin{equation}
H_{MN}\partial_a\Phi^M = \eta_{MN}\mathbb I_{ab}\epsilon^{bc} \partial_c \Phi^N ,
\end{equation}
in agreement with Duff's result in \cite{Duff:1989tf} for the duality symmetric string. Alternatively, expressed in terms of the (anti-)chiral combinations
\begin{equation}
\psi_\pm^m 		= \varphi^m		\pm \tilde\varphi^{\bar{n}}\delta_{\bar{n}}^m		,
\end{equation}
equation \eqref{duality-symmetric string relation} describes the familiar (anti-)self-dual left- and right-movers
\begin{equation}
\partial_a \psi_\pm^m		= \pm \mathbb I_{ab}\epsilon^{bc} \partial_c \psi_\pm^m
\end{equation}
of the Tseytlin-string \cite{Tseytlin:1990nb, Tseytlin:1990va}. Thus the dynamics of the Goldstone modes of the wave solution reproduce the duality symmetric string in doubled space. The number of physical degrees of freedom are not doubled but just become rearranged in terms of chiral and anti-chiral modes on the worldsheet.

\subsection{Comparison with the \texorpdfstring{$\sigma$}{sigma}-model evaluated in the string or wave background}

The equations of motion derived in the previous section recover the equations of motion of the Tseytlin string. A natural question would be to ask what background is the string in? Is the target space of the doubled solution the combination of the fundamental string with the spacetime pp-wave background? The answer to this question can be seen immediately from the Goldstone mode analysis which gives the equations of motion of the free string i.e.\ that of the $\sigma$-model in a flat background. 

To understand this it is worth understanding what the Goldstone mode analysis provides us with in other cases where this has been carried out in a more conventional setting. In \cite{Adawi:1998ta} the Goldstone mode analysis for the D3, M2 and M5 branes was completed and used to determine the effective equations of motion for each of those objects. In each case the analysis gave the description of the brane in a flat background---as one would expect---the Goldstone mode analysis must give the equations of motion of the brane in a flat background since the solution for which one is determining the moduli is that of the brane in a flat background. The same argument applies here---the string dynamics must be those of the string in a flat background.

A string solution in the background of other strings i.e.\ a string $\sigma$-model in a string background would be a different solution and as such obey a different set of equations of motion. Describing this more technically, to find the $\sigma$-model in a non-trivial background one must find the back-reacted wave solution not for asymptotically flat space but for one with asymptotically switched on NS-fluxes and then determine its moduli and their equations of motion. Once one has determined the effective equations of motion through a Goldstone mode analysis, one can then proceed to covariantize these equations (in terms of the geometry of moduli space) to determine the general equations of motion. In terms of the doubled string above, this would imply just replacing the flat target space generalised metric with the generalised metric of an arbitrary background.

\section[\texorpdfstring{Buscher-type transformation in the double field theory $\sigma$-model}{Buscher-type transformation in the double field theory sigma-model}]{Buscher-type transformation in the double field \\theory $\sigma$-model}
\label{sec:Buscher-type transformation in the DFT sigma-model}

We have seen in \S\ref{sec:The Buscher rules} how Buscher recovered T-duality transformations from the string $\sigma$-model with an isometry in the target space. We have also seen in \S\ref{sec:A DFT wave solution recovers the string} how string dynamics can be approached in an $O(d,d,\mathbb R)$-covariant manner via DFT. In this section, we look at manifestly $O(d,d,\mathbb R)$-covariant string $\sigma$-models in DFT, and apply a Buscher procedure to these. A Floreanini-Jackiw-style \cite{Floreanini:1987as} string $\sigma$-model was derived for $O(d,d,\mathbb R)$-invariant field theory initially in \cite{Tseytlin:1990nb, Tseytlin:1990va}, which takes the form
\begin{equation}
S		=		\frac{ 1 }{ 2 }		\int \, \ud ^2 \sigma		(		- H_{M N} \partial_1 X^M \partial_1 X^N		+		\eta_{M N} \partial_1 X^M \partial_0 X^N		),
\label{Non-covariant doubled sigma model action}
\end{equation}
where indices on the partial derivatives correspond to worldsheet directions (as opposed to the doubled target space indices $M, N \ldots $). This is not, however, manifestly worldsheet covariant---covariance is recovered on-shell. A manifestly worldsheet-covariant action was proposed by Hull \cite{Hull:2004in} (which has been shown to be equivalent to \eqref{Non-covariant doubled sigma model action} \cite{Pezzella:2015hfa}, see also \cite{DeAngelis:2013wba, Polyakov:2015wna}), which can be gauged by a derivative-index-valued connection $A_M \sim \psi\partial_M \vartheta$, to produce the gauged action of \cite{Lee:2013hma}
\begin{equation}
S		=		\int\ud^2 \sigma \frac{1}{2} H_{MN} D_a X^M D_b X^N h^{ab}		-		\epsilon^{ab} D_a X^M A_{bM}		.
\label{gauged covariant doubled sigma model action}
\end{equation}
Here, $D_a X^M		=		\partial_a X^M		-		A_a^M$, and target space indices are here raised and lowered with $\eta$. We will apply the Buscher procedure with one Lagrange multiplier to the \emph{ungauged} version of the covariant action \eqref{gauged covariant doubled sigma model action}, and with two Lagrange multipliers to the non-covariant action \eqref{Non-covariant doubled sigma model action}. For both cases, we will split the coordinates as $\mu = (0,m)$, where $x^0$ need not be the time coordinate.

\subsubsection{A single Buscher transformation}

The action
\begin{equation}
S'		=
\begin{multlined}[t]
\frac{ 1 }{ 2 }		\int \, \ud ^2 \sigma		\left[ (		H_{00} C_a C_b			+ 2H_{0m} C_a \partial_b X^m		+ H_{mn} \partial_a X^m \partial_b X^n		\right.
\\
\left. + 2H_0{}^\nu C_a \partial_b X_\nu		+ 2H_m{}^\nu \partial_a X^m \partial_b X_\nu		+ H^{\mu\nu} \partial_a X_\mu \partial_b X_\nu		) h^{ab}		+		\alpha \partial_a C_b \epsilon^{ab}	\right]
\end{multlined}
\label{dual action: one LM}
\end{equation}
has equations of motion for the Lagrange multiplier $\alpha$
\begin{equation}
\partial_{[a} C_{b]}		=		0		,
\end{equation}
which on identifying $C_a = \partial_a X^0$ recovers the covariant action \eqref{gauged covariant doubled sigma model action} for $A_M = 0$. If instead we integrate by parts the Lagrange multiplier term, we find (up to boundary contributions)
\begin{equation}
S'		=
\begin{multlined}[t]
\frac{ 1 }{ 2 }		\int \, \ud ^2 \sigma		\left[ (		H_{00} C_a C_b			+ 2H_{0m} C_a \partial_b X^m		+ H_{mn} \partial_a X^m \partial_b X^n		\right.
\\
\left. + 2H_0{}^\nu C_a \partial_b X_\nu		+ 2H_m{}^\nu \partial_a X^m \partial_b X_\nu		+ H^{\mu\nu} \partial_a X_\mu \partial_b X_\nu		) h^{ab}		-		\partial_a \alpha C_b \epsilon^{ab}	\right]		,
\end{multlined}
\label{dual action integrated by parts: one LM}
\end{equation}
which has equations of motion
\begin{equation}
h^{ab} \left( 2 H_{00} C_b		+ 2H_{0m} \partial_b X^m		+ 2 H_0{}^\nu \partial_b X_\nu \right)		-		\epsilon^{ab} \partial_b \alpha		=		0		.
\end{equation}
If we insert this into the action \eqref{dual action integrated by parts: one LM}, we find
\begin{equation}
S'		=
		\int \, \ud ^2 \sigma		\frac{ 1 }{ 2 } H'_{MN} \partial_a (X')^M	\partial_b (X')^N h^{ab}
+  \epsilon^{ab} \partial_a (X')^M \partial_b (X')^0 \frac{H_{0M}}{ H_{00} },
\label{"dual" DFT sigma-model on LM}
\end{equation}
where $(X')^M = ( \alpha/2, X^m, X_\mu)$ and
\begin{equation}
\begin{aligned}
H'_{00}			& =		\frac{1}{ H_{00} }
&
H'_{mn}			& =		H_{mn}		-		\frac{ H_{0m} H_{0n} }{ H_{00} }
\\
H'_{0n}			& =		0
&
H'_m{}^\nu		& =		H_m{}^\nu		-		\frac{ H_{0m} H_0{}^\nu }{ H_{00} }
\\
H'_0{}^\mu		& =		0
&
H'^{\mu\nu}		& =		H^{\mu\nu}		-		\frac{ H_0{}^\mu H_0{}^\nu }{ H_{00} }.
\end{aligned}
\end{equation} 
While it may appear that \eqref{"dual" DFT sigma-model on LM} takes the form of the gauged $\sigma$-model \eqref{gauged covariant doubled sigma model action}, we recall that $A_M$ is required to be derivative-index-valued. With this in mind, we do not find that \eqref{"dual" DFT sigma-model on LM} corresponds to a gauged DFT $\sigma$-model. That is, we find no solution-generating transformations for one Lagrange multiplier.

\subsubsection{A double Buscher transformation}

We now extend the analysis to include an additional Lagrange multiplier term in the dual coordinates, but here for the action \eqref{Non-covariant doubled sigma model action}. Consider the action
\begin{equation}
S'		=		
\begin{multlined}[t]
\frac{ 1 }{ 2 }		\int \, \ud ^2 \sigma		(		- H_{00} A_1^2			- 2H_{0m} A_1 \partial_1 X^m		- 2H_0{}^0 A_1 B_1		- 2H_0{}^m A_1 \partial_1 X_m
\\
- H_{mn} \partial_1 X^m \partial_1 X^n		- 2 H_m{}^0 \partial_1 X^m B_1		- H^{mn} \partial_1 X_m \partial_1 X_n
\\
- 2 H_m{}^n \partial_1 X^m \partial_1 X_n		- H^{00} B_1^2		- 2 H^{0m} B_1 \partial_1 X_m		+ A_1 B_0		+ A_0 B_1
\\
+ \partial_1 X^m \partial_0 X_m		+ \partial_1 X_m \partial_0 X^m		- \alpha(\partial_1 A_0 - \partial_0 A_1)		- \beta (\partial_0 B_1 - \partial_1 B_0)		) .
\end{multlined}
\label{Equivalent Floreanini-Jackiw action in A, B, alpha, beta}
\end{equation}
The Lagrange multipliers $\alpha$ and $\beta$ have resulting equations of motion
\begin{equation}
\partial_1 A_0		- \partial_0 A_1		= 0
\qquad
\partial_0 B_1	- \partial_1 B_0		= 0,
\end{equation}
solutions of which are $A_a = \partial_a C$, $B_a = \partial_a D$. When we reinsert these solutions in to the action \eqref{Equivalent Floreanini-Jackiw action in A, B, alpha, beta} we recover the action \eqref{Non-covariant doubled sigma model action} with the identification $C = X^0$, $D = X_0$.

If instead we integrate \eqref{Equivalent Floreanini-Jackiw action in A, B, alpha, beta} by parts so that the derivatives act on the Lagrange multipliers, we find, up to total derivatives,
\begin{equation}
S'		=		
\begin{multlined}[t]
\frac{ 1 }{ 2 }		\int \, \ud ^2 \sigma		(		- H_{00} A_1^2			- 2H_{0m} A_1 \partial_1 X^m		- 2H_0{}^0 A_1 B_1		- 2H_0{}^m A_1 \partial_1 X_m
\\
- H_{mn} \partial_1 X^m \partial_1 X^n		- 2 H_m{}^0 \partial_1 X^m B_1		- H^{mn} \partial_1 X_m \partial_1 X_n
\\
- 2 H_m{}^n \partial_1 X^m \partial_1 X^n		- H^{00} B_1^2		- 2 H^{0m} B_1 \partial_1 X_m		+ A_1 B_0		+ A_0 B_1	
\\
+ \partial_1 X^m \partial_1 X_m		+ \partial_1 X_m \partial_0 X^m		+ A_0 \partial_1 \alpha		- A_1 \partial_0 \alpha		- B_1 \partial_0 \beta		+ B_0 \partial_1 \beta		) .
\end{multlined}
\label{Equivalent Floreanini-Jackiw action in A, B, alpha, beta after integration by parts}
\end{equation}
The equations of motion for $A_a$ and $B_a$ are now
\begin{subequations}
\begin{align}
A_1		+ \partial_1 \beta		& =	0
\\
- 2 H_0{}^0 A_1		- 2H_m{}^0 \partial_1 X^m		- 2H^{00} B_1		- 2H^{0m} \partial_1 X_m		+ A_0		- \partial_0 \beta		& =	0
\\
B_1		+ \partial_1 \alpha		& =	0
\\
- 2 H_{00} A_1		- 2H_{0m} \partial_1 X^m		- 2H_0{}^0 B_1		- 2H_0{}^m \partial_1 X_m		+ B_0		- \partial_0 \alpha		& =	0		.
\end{align}
\end{subequations}
Inserting these into \eqref{Equivalent Floreanini-Jackiw action in A, B, alpha, beta after integration by parts}, we recover the action
\begin{equation}
\begin{gathered}
S'		=		\frac{ 1 }{ 2 }		\int \, \ud ^2 \sigma		(		- H_{M N} \partial_1 (X')^M \partial_1 (X')^N		+		\eta_{M N} \partial_1 (X')^M \partial_0 (X')^N		),
\\
\text{where}
\qquad
(X')^M			=		(-\beta, X^m, -\alpha, X_m).
\end{gathered}
\end{equation}
We observe that the Lagrange multiplier $\alpha$ for the $A$ equation of motion (which was used to replace $X^0$) becomes the dual coordinate $X_0$, likewise for $\beta$ and $B$. That is, the transformation sends $x^0 \leftrightarrow \tilde x_0$. One can see this is the T-duality invariance of the theory manifesting if instead we equivalently write
\begin{equation}
\begin{gathered}
S'		=		\frac{ 1 }{ 2 }		\int \, \ud ^2 \sigma		(		- H'_{M N} \partial_1 (X')^M \partial_1 (X')^N		+		\eta_{M N} \partial_1 (X')^M \partial_0 (X')^N		),
\\
\text{where}
\qquad
(X')^M			=		(-\alpha, X^m, -\beta, X_m)
\end{gathered}
\end{equation}
where we find the corresponding spacetime metric and two-form field transformations are given by the Buscher rules \eqref{Buscher rules}.

\chapter{U-duality invariant M-theory}
\label{ch:U-duality invariant M-theory}
\chead{\textsc{u-duality invariant m-theory}}

The strong coupling limit of type IIA string theory is an eleven dimensional theory, termed M-theory, which has as its fundamental objects not strings, but membranes and five-branes. Various compactification limits of M-theory yield all five string theories, while T-duality and S-duality are united as part of a larger web of U-dualities. Though our understanding of the full theory remains rather incomplete, we can look to the low energy limit, eleven dimensional supergravity, discovered by Cremmer, Julia, and Scherk \cite{Cremmer:1978km}, and whose action in the bosonic sector is given by
\begin{equation}
S_{11}		=		\frac{1}{2\kappa^2} \int_{M_{11}} \sqrt{\det(g)} \left( R		- \frac{1}{48} F_4^2 \right)		- \frac{1}{4} F_4 \wedge F_4 \wedge C_3		,
\label{11-d sugra action}
\end{equation}
where $F_4 = \ud C_3$ is the field strength of the three-form gauge field.

In 1980, Julia \cite{Julia:1981} demonstrated that compactifications of eleven dimensional supergravity on $T^n$ exhibit the $E_n$ symmetries listed in table \ref{table: symmetry groups of 11-d sugra compactified on tori}.
\begin{table}
\begin{center}
\begin{tabular}{cccc}
	$11-n$	&		$n$	&		$G$							&		$H$
\\	\hline
	10		&		1		&		$SO(1,1)$					&		1					
\\	9		&		2		&		$SL(2)$						&		$SO(2)$
\\	8		&		3		&		$SL(3)\times SL(2)$	&		$SO(3)\times SO(2)$
\\	7		&		4		&		$SL(5)$						&		$SO(5)$
\\	6		&		5		&		$SO(5,5)$					&		$SO(5)\times SO(5)$
\\	5		&		6		&		$E_6	$						&		$USp(8)$
\\	4		&		7		&		$E_7	$						&		$SU(8)$
\\	3		&		8		&		$E_8$						&		$SO(16)$
\end{tabular}
\end{center}
\caption{The symmetry groups $G$ and their maximally compact subgroups $H$ of 11-dimensional supergravity compactified on $T^n$.}
\label{table: symmetry groups of 11-d sugra compactified on tori}
\end{table}
These symmetries are inherited from the U-duality group of M-theory. We will here on use the name U-duality interchangeably to mean both the continuous hidden symmetry group of supergravity and the parent M-theory duality group.

It has been proposed that these U-duality groups are not only present on compactification, but are symmetries of the full M-theory \cite{deWit:1986mz}---when compactifying on $T^n$, an $E_n$ subset of the full $E_{11}$ symmetry group is revealed, acting on the compactified directions. In the supergravity theory, the symmetries act on the metric and gauge fields on equal footing, and so one may ask whether there is some way of unifying these objects in a manifestly duality-invariant form. However, since the symmetry is not present in the supergravity theory as it stands, some kind of extension is required. This is the role of the exceptional geometries.

Recall that in $O(d,d,\mathbb R)$-invariant generalised geometry, the duality group of compactified NS-NS supergravity is made manifest without compactification by extending the tangent space to include one-forms which can parametrise gauge transformations of the two-form field. Analogously, Hull \cite{Hull:2007zu}, and later Pacheco and Waldram \cite{Pacheco:2008ps} made manifest the $E_d$ duality group in \emph{exceptional generalised geometry} by extending the tangent space to include forms which can parametrise gauge transformations of the form field content inherited from the parent membrane theory. The tangent spaces are given for $d=4$--7 in table \ref{table:Tangent space and dimensions of duality-invariant M-theory}. One can think of the tangent space contributions as arising with three-form gauge transformations ($\in\wedge^2 T^*M$), fivebrane-modes ($\in\wedge^5 T^*M$) for $d\geq 5$, and six-form (dual to $C_3$) gauge transformations ($\in\wedge^6 T^*M$) for $d\geq 7$.
\begin{table}
\begin{center}
\begin{tabular}{cccc}
$n$ & $G= E_d$ & Tangent space & $D$ \\ 
\hline 
4 & $SL(5)$ & $TM\oplus \wedge^2 T^*M$ & 10 \\ 
5 & $Spin(5,5)$ & $TM\oplus \wedge^2 T^*M \oplus \wedge^5 T^*M$ & 16 \\ 
6 & $E_6$ & $TM\oplus \wedge^2 T^* M \oplus \wedge^5 T^* M$ & 27 \\ 
7 & $E_7$ & $TM\oplus \wedge^2 T^* M \oplus \wedge^5 T^* M \oplus \wedge^6 T^* M$ & 56 \\ 
\end{tabular}
\end{center}
\caption{The tangent space and coordinate representation $D$ of $E_d$ duality-invariant extended geometries}
\label{table:Tangent space and dimensions of duality-invariant M-theory}
\end{table}
Meanwhile, the form fields and metric are unified into a $G/H$ coset representative which encodes precisely the
\begin{equation}
d(d+1)/2		+ \binom{d}{3}		+ \binom{d}{6}		=	\dim (G/H)
\end{equation}
degrees of freedom in $g$, $C_3$ and $C_6$ (these have been constructed for $d=3,4$ \cite{malek2012u}, $d=5$--7 \cite{Berman:2011jh} and $d = 8$ \cite{Godazgar:2013rja}).

As DFT is to generalised geometry, where one extends not just the tangent space, but the coordinate space itself, \cite{malek2012u, Berman:2010is, Berman:2011cg, Berman:2011jh, Musaev:2013kpa, Park:2013gaj, Blair:2013gqa, Berman:2011pe, Aldazabal:2013mya, Berman:2012vc, Berman:2011kg, Berman:2014jsa} constructed \emph{extended geometries} where the $d$ spacetime dimensions are supplemented with $w$ ``wrapping" coordinates which in the supergravity picture correspond to membrane charges. The $E_d$ duality group now acts as a manifest symmetry of a $(d+w)$-dimensional space. Again, there has been no compactification, unlike in the standard torus compactification of eleven-dimensional supergravity. That $E_d$ can act on the space is precisely due to the presence of the additional wrapping coordinates\footnote{More accurately, the extended geometries exhibit manifest $E_d(\mathbb R)$ covariance, while compactification on $T^r$ reveals an $E_r(\mathbb Z)$ symmetry group, analogous to how $O(d,d,\mathbb R)$-covariant DFT compactified on $T^r$ reveals an $O(r,r,\mathbb Z)$ T-duality group.}. The tangent spaces of the extended geometries are the same as those of their corresponding exceptional generalised geometry, given in table \ref{table:Tangent space and dimensions of duality-invariant M-theory}. Again, the crucial difference between generalised and extended geometries is that, in the latter, one allows dependence of all the fields on the additional coordinates, while imposing a section condition which determines the physical subset. Different solutions to this section condition are related by $E_d$ rotations.

Extended geometry is in fact a truncation of \emph{exceptional field theory} \cite{Hohm:2013uia, Hohm:2013vpa, Hohm:2013pua, Blair:2014zba, Godazgar:2014nqa, Musaev:2014lna, Hohm:2014fxa, Cederwall:2015ica, Wang:2015hca, Musaev:2015pla, Hohm:2015xna, Abzalov:2015ega, Malek:2013sp, Godazgar:2014sla, Berman:2014hna, Baguet:2015xha}. There, the eleven-dimensional supergravity coordinates are split $11 = (11-d) + d$  without dimensional reduction, and then supplemented with the $w$ wrapping coordinates. While the extended geometries can be seen as (a truncation of) an eleven-dimensional theory which exhibits manifest $E_d\times GL(11-d,\mathbb R)$ invariance, and where the internal and external sectors are decoupled, in exceptional field theory these sectors are no longer decoupled.

In this chapter, we focus on the four-dimensional $SL(5)$-invariant extended geometry constructed in \cite{Berman:2010is, Berman:2011cg, malek2012u, Park:2013gaj, Blair:2013gqa}, whose corresponding four-dimensional supergravity has action \eqref{11-d sugra action} (short of the Chern-Simons terms which can only be defined in eleven dimensions). The picture we illustrate in the next section is similar to that in $d=5$--7.

\section{The \texorpdfstring{$SL(5)$}{SL(5)}-invariant theory}
\label{sec:The $SL(5)$-invariant theory}

In $SL(5)$-invariant M-theory, one supplements the four-dimensional spacetime coordinates with the six coordinates dual to wrapping configurations of M2-branes in four dimensions. These can be combined into an $SL(5)$ coordinate on the full space via the decomposition\footnote{Note that with our summation convention, we have \[A_M\ud x^M = A_{\mu 5} \ud X^{\mu 5} + A_{5\mu} \ud X^{5\mu} + A_{\mu\nu} \ud X^{\mu\nu} = 2A_{\mu 5} \ud x^\mu + \frac{1}{2} A_{\mu\nu} \epsilon^{\mu\nu\sigma\rho} \ud y_{\sigma\rho}.\] In particular, we note the factor of 2 for the $\ud x^\mu$ contribution.}
\begin{equation}
X^M			=		X^{[mm']}
				=		\begin{cases}
							X^{\mu 5}		= x^\mu		\\
							X^{\mu \nu}	=	\frac{1}{2} \epsilon^{\mu\nu\rho\sigma 5} y_{\rho\sigma}
						\end{cases}
\label{SL(5) coordinates}
\end{equation}
where $\mu \in\{1, \ldots ,4\}$. Here, capital Roman indices run $M \in\{1,\ldots,10\}$ and correspond to an antisymmetric pair of later-alphabet lower Roman indices $m,\ldots ,z\in\{1,\ldots, 5\}$. Fields are allowed to depend on all $X^M$.

For the $SL(5)$ geometry, the Y-tensor takes the form
\begin{equation}
Y^{MN}{}_{KL}		=	\frac{1}{4} \epsilon^{aMN} \epsilon_{aKL}		,
\label{SL(5) Y tensor}
\end{equation}
where $\epsilon^{aMN} = \epsilon^{amm'nn'}$ is the five-dimensional antisymmetric symbol. In terms of this Y-tensor, the reduction to the physical set of coordinates is once again achieved by a section condition of the form \eqref{weak constraint} with \eqref{strong constraint or section condition}. This additionally guarantees closure of gauge transformations on the extended space, which are given by the corresponding generalised Lie derivative \eqref{Generalised lie derivative} (in this comparison, indices $M,N \ldots $ should of course be understood here as those of the $SL(5)$ representation \eqref{SL(5) coordinates}). Solving the section condition by demanding independence from the dual coordinates $y_{\rho\sigma}$ recovers the supergravity frame, while other solutions are $SL(5)$ rotations of this. One may relax the section condition to allow Scherk-Schwarz reductions to gauged supergravities \cite{Berman:2012uy, Musaev:2013rq, Lee:2015xga}.

The metric and form field $C_3$ are unified into the $10\times 10$ symmetric coset generalised metric
\begin{equation}
M_{M N}		=		g^{1/5} \begin{pmatrix}
								g_{\mu \nu} + \frac{1}{2} C_\mu{}^{\rho \sigma} C_{\nu \rho \sigma}		&		- \frac{1}{2 \sqrt 2} C_\mu{}^{\rho \sigma} \epsilon_{\rho \sigma \lambda \lambda'}	\\
								- \frac{1}{2 \sqrt 2} C_\nu{}^{\rho \sigma} \epsilon_{\rho \sigma \eta \eta'}	&		(\det g)^{-1} g_{\eta \eta', \lambda \lambda'}
								\end{pmatrix},
\label{SL(5) metric}
\end{equation}
where $g_{\eta \eta', \lambda \lambda'} \equiv g_{\eta [\lambda} g_{\lambda'] \eta'}$ (also $g^{\eta \eta', \lambda \lambda'} \equiv g^{\eta [\lambda} g^{\lambda'] \eta'}$) and $C = C_3$ is the three-form gauge field. The inverse\footnote{
Note that the identity here takes the form
\begin{equation*}
\delta_M^N
						=		\begin{pmatrix}
										\delta_{ [\mu }^\nu	\delta_{ 5] }^5
								&		\delta_{ [\mu }^\nu	\delta_{ \mu'] }^5
								\\		\delta_{ [\mu }^\nu	\delta_{ 5] }^\nu
								&		\delta_{ [\mu }^\nu	\delta_{ \mu'] }^{ \nu' }
								\end{pmatrix}
						=		\begin{pmatrix}
										\frac{1}{2} \delta_\mu^\nu
								&		0
								\\		0
								&		\delta_{ [\mu }^\nu	\delta_{ \mu'] }^{ \nu' }
								\end{pmatrix}
\end{equation*}
} metric is
\begin{equation}
M^{M N}		=		g^{-1/5} \begin{pmatrix}
										\frac{1}{4} g^{\mu \nu}
								&		\frac{1}{4 \sqrt 2} g^{\mu \kappa} C_{\kappa \omega \omega'} \epsilon^{ \omega \omega' \lambda \lambda' } 
								\\
										\frac{1}{4 \sqrt 2} g^{\nu \kappa} C_{\kappa \omega \omega'} \epsilon^{ \omega \omega' \sigma \sigma' }
								&		(\det g) g^{\sigma \sigma', \lambda \lambda'}		+		\frac{1}{8} g^{\alpha \beta} C_{\alpha \omega \omega'} \epsilon^{\omega \omega' \sigma \sigma'} C_{\beta \chi \chi'} \epsilon^{\chi \chi' \lambda \lambda'}
								\end{pmatrix}.
\end{equation}

There is further field content which can be seen to arise from the truncation of the eleven dimensional theory to four dimensions. In this process, the $E_{11}$ decomposes as $SL(5)\times GL(7)$, and the eleven-dimensional metric as $g_{11} = \diag(g,g_7)$, where $g_7$ is the metric on the seven-dimensional space. Extending the four-dimensional spacetime with six wrapping coordinates, the resulting metric on the $(4+6+7)$-dimensional space is \cite{malek2012u}
\begin{equation}
\mathcal M		=
	|\det(g_{11})|^{-1/2}
	\begin{pmatrix}
		\det(g)^{-1/5} M		&	0		\\
		0							&	g_7
	\end{pmatrix}.
\end{equation}
If we assume the diagonal form\footnote{As would arise on compactification on $T^7$ where the tori have equal radii $\det(g_7)^{1/7}$.} $g_7 = \det(g_7)^{1/7} \delta_7$, we can use $\det(g_{11}) = \det(g) \det(g_7)$ to find
\begin{equation}
\mathcal M		=
	\begin{pmatrix}
		e^{-\Delta} M		&	0						\\
		0						&	e^{-5\Delta/7} \delta_7
	\end{pmatrix}		,
\end{equation}
where
\begin{equation}
e^\Delta = \det(g_7)^{1/2} \det (g)^{7/10}
\end{equation}
defines the additional degree of freedom, the \emph{volume factor} scalar density $\Delta$.

\subsection{Action and equations of motion}
\label{sec:Action and equations of motion (SL(5))}

We can now apply the method used in \S\ref{Action and equations of motion (DFT)} to derive the full equations of motion for $SL(5)$-invariant M-theory. The action (up to boundary contributions, treated in \cite{Berman:2011kg}) is
\begin{equation}
S		= \int \ud^{10} X e^\Delta R	,
\end{equation}
where the $SL(5)$ scalar
\begin{equation}
R		=	
		\begin{aligned}[t]
		&	\frac{1}{12} M^{MN} \partial_M M^{KL} \partial_N M_{KL}
			- \frac{1}{2} M^{MN} \partial_M M^{KL} \partial_L M_{KN}
			- \partial_M \partial_N M^{MN}
		\\
		&	- \frac{1}{7} M^{MN} \partial_M \Delta \partial_N \Delta
			- \frac{2}{7} \partial_M M^{MN} \partial_N \Delta
			- \frac{2}{7} M^{MN} \partial_M \partial_N \Delta		.
		\end{aligned}
\label{SL(5) scalar R}
\end{equation}
The first three terms reproduce the four-dimensional supergravity action in the supergravity frame, whilst the latter three are kinetic terms for $\Delta$.

Varying the action with respect to the volume factor yields the equation of motion
\begin{equation}
R = 0	,
\label{volume factor eom}
\end{equation}
while varying the action with respect to the metric yields
\begin{equation}
\delta_M S = \int \ud^{10} X e^\Delta \mathcal K_{MN} \delta M^{MN},
\end{equation}
where
\begin{equation}
\mathcal K_{M N}		=
\begin{aligned}[t]
&		\frac{1}{12} \partial_M M^{K L} \partial_N M_{K L}
		- \frac{1}{2} M^{P K} M^{Q L} \partial_L M_{P M} \partial_K M_{Q N}
\\
&		+ \frac{1}{6} M^{P Q} M^{K L} \partial_P M_{K M} \partial_Q M_{L N}
		- \partial_M \partial_N \Delta		
		- \frac{6}{7} \partial_M \Delta \partial_N \Delta
\\
&		+ ( \partial_L		+ (\partial_L \Delta) )
		\left[
			M^{L K} \left( \partial_{ (M } M_{ N) K}		- \frac{1}{6} \partial_K M_{M N} \right)
		\right].
\end{aligned}
\end{equation}
Again, this does not require $\mathcal K_{MN} = 0$, since $\delta M^{MN}$ is constrained by its $SL(5)/SO(5)$ coset structure \eqref{SL(5) metric}. As shown in \cite{Park:2013gaj}, this particular coset structure allows one to write the metric and volume factor in full generality in terms of a ``little metric" $m_{ab}$ where $M_{ab,cd} = m_{a[c}m_{d]b}$. From this, equations of motion were derived via construction of a semi-covariant derivative and curvature tensor analogous to the Riemann tensor. We will determine the metric equations of motion by the chain rule method of \S\ref{Action and equations of motion (DFT)}. For this purpose, it is convenient to define a rescaled metric $\tilde M^{MN} = g^{1/5} M^{MN}$. Thus,
\begin{subequations}
\begin{align}
g^{1/5} & \mathcal K_{MN} \delta M^{MN}
\\
& =		g^{1/5} \mathcal K_{MN}	\left(	g^{-1/5} \delta \tilde M^{MN}		+ \delta g^{-1/5} \tilde M^{MN} \right)
\\
& =		\mathcal K_{MN}
		\left[
			\frac{\partial \tilde M^{MN}}{\partial g_{\mu\nu}}\delta g_{\mu\nu}
			+ \frac{\partial \tilde M^{MN}}{\partial C_{\mu\nu\rho}} \delta C_{\mu\nu\rho}
			+ \delta \ln g^{-1/5} \tilde M^{MN}
		\right]
\\
& =		
\begin{aligned}[t]
	&	\begin{aligned}[t]
			\bigg[
			&	- g^{\mu\sigma}g^{\nu\rho} \mathcal K_{\mu 5\nu 5}
				- \frac{1}{\sqrt 2} g^{\mu\sigma} g^{\rho\kappa}C_{\kappa\omega\omega'}\epsilon^{\omega\omega'\nu\nu'} \mathcal K_{\mu 5\nu\nu'}
		\\
			&	+
				\bigg(
					g g^{\sigma \rho} g^{\mu\mu',\nu\nu'}
					- g g^{\mu\sigma} g^{\rho[\nu}g^{\nu']\mu'}
					- g g^{\mu[\nu} g^{\nu']\sigma} g^{\rho\mu'}
		\\
			& 		- \frac{1}{8} g^{\sigma\alpha} g^{\rho\beta} C_{\alpha\omega\omega'}\epsilon^{\omega\omega'\mu\mu'} C_{\beta\chi\chi'}\epsilon^{\chi\chi'\nu\nu'}
				\bigg)
				\mathcal K_{\mu\mu'\nu\nu'}
				- \frac{1}{5} g^{\rho\sigma} \mathcal K_{MN} \tilde M^{MN}
			\bigg]
			\delta g_{\sigma\rho}
		\end{aligned}
	\\
	&	+ \left[
			\frac{1}{\sqrt 2} g^{\mu\alpha} \epsilon^{\beta\gamma\nu\nu'} \mathcal K_{\mu 5\nu\nu'}
			+ \frac{1}{4} g^{\sigma\alpha} \epsilon^{\beta\gamma\mu\mu'}C_{\sigma\chi\chi'} \epsilon^{\chi\chi'\nu\nu'} \mathcal K_{\mu\mu'\nu\nu'}
		\right]
		\delta C_{\alpha\beta\gamma}
\end{aligned}
\\
& =
\begin{aligned}[t]
		\bigg[
		&	- 4 \tilde M^{\sigma 5 M} \tilde M^{\rho 5 N} \mathcal K_{MN}
			+ \tilde M^{\rho 5 N} \epsilon^{\mu \sigma 5 P} \epsilon_{\mu N R} \tilde M^{RQ} \mathcal K_{PQ}
		\\	
		&	- \frac{4}{5} \tilde M^{\sigma 5 \rho 5} \mathcal K_{MN} \tilde M^{MN}
		\bigg]
		\delta g_{\rho\sigma}
		+ \sqrt 2 M^{M\alpha 5} \epsilon^{\beta\gamma\nu\nu'} \mathcal K_{M\nu\nu'} \delta C_{\alpha\beta\gamma}
\end{aligned}
\\
& =
\begin{aligned}[t]
	&	4 \tilde M^{\sigma 5 M} \tilde M^{\rho 5 N}
		\left(
			- \delta_M^P \delta_N^Q
			+ M_{ML} Y^{LP}{}_{NR} M^{RQ}
			- \frac{1}{5} M_{MN} M^{PQ}
		\right)
		\mathcal K_{PQ} \delta g_{\rho\sigma}
	\\
	&	\begin{multlined}
			+ \frac{\sqrt{2}}{1+2x} M^{M\sigma 5}
			\left(
				\delta_M^P \delta_N^Q		- x M_{ML} Y^{LP}{}_{NR} M^{RQ}		- y M_{MN} M^{PQ}
			\right)
		\\
			\times \mathcal K_{PQ} \delta C_{\sigma\omega\omega'} \epsilon^{\omega\omega'N5},
		\end{multlined}
\end{aligned}
\end{align}
\end{subequations}
where the constants $x\neq -1/2$ and $y$ are otherwise arbitrary. \emph{Sufficient} equations of motion are thus
\begin{equation}
P_{MN}{}^{PQ} \mathcal K_{PQ}		=		0,
\label{SL(5) metric equations of motion}
\end{equation}
where the projector
\begin{equation}
\begin{split}
P_{MN}{}^{PQ}
& =
	\delta_M^P \delta_N^Q
	- M_{ML} Y^{LP}{}_{NR} M^{RQ}
	+ \frac{1}{5} M_{MN} M^{PQ}
\\
& =
	\left(
		\delta_M^R \delta_N^S
		- \frac{1}{10} M_{MN} M^{RS}
	\right)
	\left(
		\delta_R^P \delta_S^Q
		- M_{RK} Y^{KP}{}_{SL} M^{LQ}
	\right)		.
\end{split}
\label{SL(5) projector}
\end{equation}
As in \S\ref{Action and equations of motion (DFT)}, one can determine if \eqref{SL(5) metric equations of motion} are \emph{necessary} by counting the dimension of the symmetric kernel of $P$ for Minkowski metric and vanishing $C_3$. We find, of the $10(10+1)/2 = 55$ degrees of freedom in $\mathcal K_{MN}$, that 41 are projected out, leaving precisely the $\binom{4}{3}$ and $4(4+1)/2$ degrees of freedom in $C_3$ and $g$ respectively\footnote{Again, we note that this degrees of freedom counting, while highly suggestive, is not strictly a full proof that \eqref{SL(5) metric equations of motion} are the $SL(5)$ covariant metric equations of motion. Specifically, we have not shown that the variation of the spacetime metric and three-form do not lie within the kernel of the projector.}. Further work includes verifying that these equations agree with those of \cite{Park:2013gaj}.

In \eqref{SL(5) projector}, the first bracket projects out terms proportional to the metric. This can be seen to arise from the relation
\begin{equation}
(\mathcal K_{PQ}+M_{PQ})\delta M^{PQ} = \mathcal K_{PQ}\delta M^{PQ} -\delta \ln \det M = \mathcal K_{PQ}\delta M^{PQ}		,
\end{equation}
where we have used the fact that $\det(M) = 1$. Such a term does not explicitly arise in the DFT projector simply because the metric is already projected out by the projector \eqref{DFT projector}. Thus, both projectors are of the form
\begin{equation}
P_{MN}{}^{PQ}		=
\left(
	\delta_M^R \delta_N^S
	- \frac{1}{D} M_{MN} M^{RS}
\right)
\left(
	\delta_R^P \delta_S^Q
	- M_{RK} Y^{KP}{}_{SL} M^{LQ}
\right)		,
\label{conjectured projector for generic extended geometries}
\end{equation}
where $D$ is the dimension of the extended geometry. A natural question is whether the equations of motion in extended geometries in higher dimensions are also given by \eqref{SL(5) metric equations of motion} with projector \eqref{conjectured projector for generic extended geometries}, where metrics/vielbein have been constructed in \cite{malek2012u} ($d=3,4$), \cite{Berman:2011jh} ($d=5$--7), \cite{Godazgar:2013rja} ($d=8$), and the corresponding Y-tensors in \cite{Berman:2012vc} for $d=3$--7.

\subsection{An \texorpdfstring{$SL(5)$}{SL(5)} wave solution recovers the M2-brane}
\label{sec:An SL(5) wave solution recovers the M2-brane}

In this section, we will redefine many of the symbols used in chapter \ref{ch:Double Field Theory} to keep analogy to the DFT case clear.  A wave solution for the $SL(5)$ duality invariant theory is given by the generalised metric $M_{MN}$ with line element
\begin{equation}
M_{MN}\ud X^M \ud X^N	=
\begin{aligned}[t]
	&	(K-2) [(\ud x^1)^2 - (\ud x^2)^2 - (\ud x^3)^2 ]		+ (\ud x^4)^2
\\
	&	+ 2(K-1) [\ud x^1\ud y_{23}		+ \ud x^2\ud y_{13}		- \ud x^3\ud y_{12} ]
\\
	&	- K [(\ud y_{13})^2 + (\ud y_{12})^2 - (\ud y_{23})^2 ]	+ (\ud y_{34})^2  + (\ud y_{24})^2 - (\ud y_{14})^2,
\end{aligned}
\label{eq:SL5ppwave}
\end{equation}
and constant volume factor $\Delta$. It can be interpreted as a pp-wave in the extended geometry which carries momentum in the directions dual to $x^2$ and $x^3$, i.e.\ combinations of $y_{12}, y_{13}$ and $y_{23}$ (though as for the DFT case in \S\ref{sec:A DFT wave solution recovers the string}, we have not constructed the conserved charges). In the pp-wave interpretation, it has no mass or charge and the solution is pure metric, i.e.\ there is no form field it couples to. $K$ is a harmonic function of the transverse radial coordinate $|x^4|$: $K=1+k_0|x^4|$, where $k_0$ is a constant. The wave is smeared in the remaining dual directions.

So that we may interpret this solution from the perspective of objects originating in eleven-dimensional supergravity, we use a Kaluza-Klein reduction from eleven to four dimensions: $\mathcal M_{11} = \mathcal M_4 \times \mathcal M_7$, followed by an augmentation by the relevant winding coordinates to a ten-dimensional extended geometry. We do this assuming the supergravity frame, with spacetime coordinates $x^\mu$. The metric ansatz is thus
\begin{equation}
M_{MN}\ud X^M \ud X^N	=
\begin{aligned}[t]
	&	\left(g_{\mu\nu} + e^{2w} C_\mu{}^{\rho\sigma} C_{\rho\sigma\nu}\right)\ud x^\mu \ud x^\nu \\
	&	+ 2e^{2w} C_\mu{}^{\rho\sigma} \ud x^\mu \ud y_{\rho\sigma}		+ e^{2w} g^{\lambda\tau,\rho\sigma} \ud y_{\lambda\tau} \ud y_{\rho\sigma},
\end{aligned}
\label{eq:KKforSL5}
\end{equation}
where the function $e^{2w(x^\mu)}$ is a scale factor inherited from the KK reduction from eleven to four dimensions. Comparing \eqref{eq:KKforSL5} with \eqref{eq:SL5ppwave}, we find
\begin{equation}
\begin{gathered}
g_{\mu\nu}\ud x^\mu \ud x^\nu = -K^{-1}\left[(\ud x^1)^2 - (\ud x^2)^2 - (\ud x^3)^2 \right] + (\ud x^4)^2
\\
C_{123} = 1-K^{-1}
\qquad
C_{\mu\nu 4} = 0
\end{gathered}
\label{membrane in ? frame}
\end{equation}
with $e^{2w}=K^{-1}$. We now view this solution in the Einstein frame related to \eqref{membrane in ? frame} by
\begin{equation}
\mathfrak g_{\mu\nu} = \sqrt{|\det (e^{2w} g^{\mu\nu,\rho\sigma})|} \, g_{\mu\nu} = K^{3/2} g_{\mu\nu},
\end{equation}
where the determinant is calculated by considering $g^{\mu\nu,\rho\sigma}$ as a $6\times 6$ matrix with rows given by $\{\mu\nu \} = \{ 12,13,14,23,24,34\}$, and similarly for columns. The C-field is not transformed (only its field strength obtains a different factor in the action) and we find
\begin{equation}
\begin{gathered}
\mathfrak g_{\mu \nu} \ud x^\mu \ud x^\nu = -K^{1/2}\left[(\ud x^1)^2-(\ud x^2)^2-(\ud x^3)^2 \right]+K^{3/2}(\ud x^4)^2
\\
C_{123} = 1-K^{-1}
\qquad
C_{\mu\nu 4} = 0 ,
\end{gathered}
\label{membrane in Einstein frame}
\end{equation}
which is the four-dimensional M2-brane in the Einstein frame, extended in the $x^2-x^3$ plane. We have thus shown that the solution \eqref{eq:SL5ppwave} which carries momentum in the directions dual to $x^2$ and $x^3$ in the extended geometry corresponds to a membrane stretched along these directions from a spacetime perspective. By similar arguments as in the string case, the mass and charge of the M2-brane are given by the momenta in the dual directions.

It would be interesting to see a Goldstone mode analysis for the $SL(5)$ wave analogous to that achieved for the DFT wave in \S\ref{sec:Goldstone Modes of the Wave Solution (DFT)}. One would consider perturbations of the metric and volume factor given by the generalised Lie derivative with respect to a generalised vector $\sim K^\alpha \varphi\partial/\partial x^4 + K^\beta \tilde\varphi_\mu \partial/\partial y_{\mu 4}$, which is constant on the worldvolume. Promoting the constants $\varphi$ and $\tilde\varphi_\mu$ to functions of the worldvolume coordinates, one could then determine the linear equations of motion, which we expect would produce the duality relation of Duff \cite{Duff:1990hn} for the membrane.

\newpage
\chapter{Solution-generating transformations in NS-NS supergravity and double field theory}
\label{ch:Solution-generating transformations in NS-NS supergravity and DFT}
\chead{\textsc{solution-generating transformations \\in ns-ns supergravity and double field theory}}

In 1954, Buchdahl \cite{BUCHDAHL01011954} found a symmetry group of the vacuum Einstein equations for static vacuum metrics. Metrics are defined to be static if they possess a Killing vector $\xi$ which is additionally hypersurface orthogonal. This is true if and only if the \emph{twist}, proportional to  $\iota_\xi \star \ud \xi_\flat$, vanishes. In adapted coordinates $\xi = \partial_0$ (where $x^\mu = (x^0, x^a)$), this is equivalent to the statement that the metric takes the block diagonal form $g = \diag(g_{00}, g_{ab})$.

Buchdahl's symmetry group was extended significantly in $(3+1)$ dimensions by the work of Ehlers \cite{Ehlers:1957zz} and Geroch \cite{Geroch:1970nt} to an $SL(2,\mathbb R)$ symmetry group of the vacuum Einstein equations in the presence of a non-null Killing vector\footnote{Mars \cite{Mars:2001gd} showed that their analysis was also valid for null Killing vectors.}. Harrison \cite{harrison1968new} extended their work to generate maps between solutions to Einstein-Maxwell theory and vacuum Einstein gravity. These symmetry groups were found by considering a split of the metric and electromagnetic field into two parts:
\begin{itemize}
\item		Those defining a codimension one \emph{reduced} metric orthogonal to Killing direction, for example $g_{ab}$ when $\xi = \partial_0$. The symmetry groups preserve a conformal transformation of this metric.
\item		All remaining contributions, which are decomposed into a set of scalar functions or \emph{potentials}, whose invariance transformations determine the symmetry group. 
\end{itemize}
The symmetry group was completed by Neugebauer and Kramer \cite{ANDP:ANDP19694790108} in Einstein-Maxwell theory for any non-null Killing vector.

In this chapter, we look at how we might find similar solution-generating symmetries in NS-NS supergravity and DFT (see \cite{buchdahl1959reciprocal, Galtsov:1994pd, Pinkstone:1995wk, Kechkin:1996pg, yazadjiev2001exact, Yazadjiev:2005pf} for similar studies in supergravity). We have seen that NS-NS supergravity with an isometry has an $O(1,1,\mathbb R)$ symmetry group, a remnant of string theory T-duality, and how this is made manifest in DFT. However, we also look for symmetries which lie outside this manifest symmetry group, and how they might appear within the DFT context. We begin our exploration with a reformulation of general relativity with an isometry, decomposing the metric into a reduced metric and $p$-form potentials, and then apply this to NS-NS supergravity before we approach DFT.

\section{Reduction of Einstein gravity with respect to an isometry}
\label{sec:Reduction of Einstein gravity with respect to an isometry}

\subsection*{The space $V_{d-1}$ orthogonal to a Killing vector}

Consider a gravitational theory with arbitrary field content. Define a $d$-dimensional manifold $V_d$ with metric $g_{\mu \nu}$ which exhibits some isometry
\begin{equation}
(\mathcal L_\xi g)_{\mu \nu}		=		2 \nabla_{(\mu} \xi_{\nu)}		=		0		
\label{Killing equation}
\end{equation}
with respect to a Killing vector $\xi$. We can define a $(d-1)$-dimensional hypersurface $V_{d-1}$ in $V_d$ where objects $T_{\mu_1 \ldots \mu_p}{}^{\nu_1\ldots \nu_q}$ are defined to live on $V_{d-1}$ if they satisfy
\begin{subequations}
\begin{align}
(\mathcal L_\xi T)_{\mu_1 \ldots \mu_p}{}^{\nu_1\ldots \nu_q}		& =		0
\\
\xi^{\mu_s} T_{\mu_1 \ldots \mu_s \ldots \mu_p}{}^{\nu_1\ldots \nu_q}		& = 0		\qquad \forall s\in\{1, \ldots, p\}
\\
\xi_{\nu_s} T_{\mu_1 \ldots \mu_p}{}^{\nu_1 \ldots \nu_s \ldots \nu_q}		& = 0		\qquad \forall s\in\{1, \ldots, q\}.
\end{align}
\end{subequations}
The \emph{reduced metric} on $V_{d-1}$ is
\begin{equation}
\gamma_{\mu \nu}		=		g_{\mu \nu}		-		\frac{1}{F} \xi_\mu \xi_\nu
\qquad
F		=		\xi^\mu \xi_\mu	,
\label{metric on V_d-1}
\end{equation}
where the \emph{norm} $F$, which lives on $V_{d-1}$, is required to be non-vanishing throughout the geometry\footnote{The results of this formalism may in fact remain valid for null $\xi$, but we have not verified as such. For example, in \cite{Mars:2001gd}, the author presents the $(3+1)$-dimensional case in a manner valid for null Killing vectors.} ($\xi$ may be spacelike \emph{or} timelike). The reduced metric acts in $V_d$ as a projector on to $V_{d-1}$. The derivative on $V_d$ of the Killing vector can be written in terms of objects on $V_{d-1}$ as
\begin{equation}
\nabla_\mu \xi_\nu		=		\frac{1}{F} \left(
	-\frac{1}{2 (d-3)!} \xi^\rho \hat \epsilon_{\sigma_1 \ldots \sigma_{d-3} \rho \mu \nu} \omega^{\sigma_1 \ldots \sigma_{d-3}}		+		 \xi_{[ \nu} \nabla_{\mu ]} F
	\right),
\label{derivative of lowered killing vector in terms of F and omega}
\end{equation}
where $\hat\epsilon_{\mu_1 \ldots \mu_d}$ is the alternating tensor (see app. \ref{app:Alternating symbol and tensor}), and we have introduced the \emph{twist}
\begin{equation}
\omega^{\sigma_1 \ldots \sigma_{d-3}}		\equiv		\hat \epsilon^{\sigma_1 \ldots \sigma_{d-3} \alpha \beta \gamma} \xi_\alpha \nabla_\beta \xi_\gamma		.
\label{twist: general definition}
\end{equation}
which lives on $V_{d-1}$.

We can define the unique metric-compatible derivative operator on $V_{d-1}$ \cite[p.~257]{Wald1984} by
\begin{equation}
D_\mu T^{\sigma_1 \ldots \sigma_k}{}_{\rho_1 \ldots \rho_l}		=		\gamma^{\sigma_1}_{\alpha_1}	\ldots	\gamma^{\sigma_k}_{\alpha_k} \gamma^{\beta_1}_{\rho_1}	\ldots	\gamma^{\beta_l}_{\rho_l}	\gamma^\nu_\mu	\nabla_\nu T^{\alpha_1 \ldots \alpha_k}{}_{\beta_1 \ldots \beta_l}		,
\label{derivative operator on V_d-1}
\end{equation}
in terms of which the twist and norm satisfy the following relations which will prove useful:
\begin{align}
D_\mu \left( F^{-3/2} \omega^{\sigma_1 \ldots \sigma_{d-4} \mu} \right)		& = 0		
\label{divergence of twist}
\\
\nabla_{ [\nu } \omega_{\mu_1 \ldots \mu_{d-3}] }		& = - 2 \xi^\chi \hat \epsilon_{\chi \nu \mu_1 \ldots \mu_{d-3}  \lambda} R^\lambda{}_\eta \xi^\eta	
\label{exterior derivative of twist}
\\
D^\mu D_\mu F		& =		\frac{1}{2F} D^\mu F D_\mu F		-		\frac{1}{(d-3)! F} \omega^{\sigma_1 \ldots \sigma_{d-3}} \omega_{\sigma_1 \ldots \sigma_{d-3}}		 - 2 \xi^\mu \xi^\nu R_{\mu\nu}.
\label{Laplacian of F on V_(d-1)}
\end{align}
In particular, while the twist is \emph{not} one of the potentials which we use to define our solution-generating transformations, we will find the necessary potential via \eqref{exterior derivative of twist}---for example, consider vacuum spacetimes where $\ud\omega = 0$ implies $\omega = \ud \chi$.

We can define a reduced determinant of the reduced metric $\gamma$ as follows. First, we note that $\gamma$ has a zero eigenvalue and thus zero determinant
\begin{equation}
\gamma_{\mu \nu} \xi^\nu		=		0
\quad
\Rightarrow
\quad
\det( \gamma_{\mu \nu} )			=		0.
\label{det(gamma) = 0}
\end{equation}
Writing the determinant of a $p\times p$ matrix $A_{ab}$ as
\begin{equation}
\det( A_{ab} )			=		\frac{1}{p!} \epsilon^{c_1 \ldots c_p} \epsilon^{d_1 \ldots d_p}	A_{c_1 d_1} \ldots A_{c_p d_p},
\label{general definition of a determinant}
\end{equation}
where lower Roman indices here run 1 to $p$ (note that we use the alternating symbol, rather than the tensor), we can decompose \eqref{det(gamma) = 0} to derive
\begin{equation}
\det( g ) 		=		F \gamma_\xi		,
\label{det(g) in terms of F and det(hat gamma)}
\end{equation}
where we have defined, for an arbitrary $d\times d$ matrix $J_{\mu \nu}$, a \emph{reduced determinant}
\begin{equation}
J_\xi		\equiv		\frac{1}{(d-1)! F^2} \xi_\alpha \epsilon^{\alpha \sigma_1 \ldots \sigma_{d-1}} \xi_\beta \epsilon^{\beta \phi_1 \ldots \phi_{d-1}}	J_{\sigma_1 \phi_1} \ldots J_{\sigma_{d-1} \phi_{d-1}}		,
\label{definition of reduced determinant}
\end{equation}
which is normalised such that for $\xi = \partial_0$, with $x^\mu = ( x^0, x^a )$, we have
\begin{equation}
\gamma_\xi		=		\det( \gamma_{ab} )		.
\end{equation}

It is central to our construction that for a given metric $\gamma$, twist $\omega$, and norm $F$, all living on $V_{d-1}$, and satisfying \eqref{divergence of twist}, in addition to the Killing vector $\xi$ and reduced metric determinant $\gamma_\xi$, we can reconstruct entirely the original metric $g$ (up to diffeomorphisms) as follows. We can rewrite equation \eqref{derivative of lowered killing vector in terms of F and omega} in the form
\begin{equation}
\nabla_{[\mu} (F^{-1} \xi_{\nu]})		=		- \frac{1}{2 (d-3)! F^2} \sqrt{-F\gamma_\xi} \, \xi^\rho \epsilon_{\sigma_1 \ldots \sigma_{d-3} \rho \mu \nu} \omega^{\sigma_1 \ldots \sigma_{d-3}}		,
\label{curl of (F^-1 xi_mu)}
\end{equation}
which has a solution $F^{-1} \xi_\mu$ if the right hand side of \eqref{curl of (F^-1 xi_mu)} has vanishing exterior derivative
\begin{equation}
\nabla_{[\eta} \nabla_\mu (F^{-1} \xi_{\nu]})		=		- \frac{1}{3!(d-4)!} \frac{1}{F^{1/2}} \xi^\delta \hat\epsilon_{\delta \eta\mu\nu \sigma_1 \ldots \sigma_{d-4}} D_\lambda \left( F^{-3/2} \omega^{\sigma_1 \ldots \sigma_{d-4} \lambda} \right)		.
\label{integrability of d(xi/F)}
\end{equation}
By contracting with $\iota_\xi \star$, we find that the RHS of \eqref{integrability of d(xi/F)} vanishes if and only if \eqref{divergence of twist} holds. The physical contribution to the gauge freedom in $F^{-1}\xi_\mu$ from \eqref{curl of (F^-1 xi_mu)} is fully constrained by the definition of the norm $\xi^\mu F^{-1} \xi_\mu = 1$. The remaining gauge freedom
\begin{equation}
F^{-1}\xi_\mu		\rightarrow F^{-1} \xi_\mu		+ \partial_\mu z(x^\nu)
\qquad
\xi^\mu \partial_\mu z = 0
\end{equation}
corresponds merely to a coordinate transformation which, in the coordinate frame where $\xi = \partial_0$ ($x^0$ need not be the time coordinate) is $x^0 \rightarrow x^0 + z(x^{\mu \neq 0})$.

\subsection*{Curvature and conformal transformations on $V_{d-1}$}

The Riemann tensor on $V_{d-1}$ is defined in terms of a one-form $u$ on $V_{d-1}$ by
\begin{equation}
({}^\gamma R)_{\mu \nu \rho}{}^\sigma u_\sigma		=		(D_\mu D_\nu	-	D_\nu D_\mu ) u_\rho		.
\end{equation}
One can show that the Riemann tensors on $V_d$ and $V_{d-1}$ are related by
\begin{equation}
({}^\gamma R)_{\mu \nu \rho}{}^\eta			=		\gamma^\phi_{[\mu } \gamma^\sigma_{ \nu]} \gamma^\beta_\rho \gamma_\lambda^\eta
\left[
	R_{\phi \sigma \beta}{}^\lambda		+		\frac{2}{F} \gamma^{\lambda \chi} ( \nabla_\phi \xi_\beta \nabla_\sigma \xi_\chi		+		\nabla_\phi \xi_\sigma \nabla_\beta \xi_\chi	)
\right]		,
\label{Riemann on V_d-1 in terms of Riemann on V_d}
\end{equation}
where we have used the relation \cite[p.~100]{Kramer1980}
\begin{equation}
\nabla_\mu \nabla_\nu \xi_\sigma		=		R_{\sigma \nu \mu \lambda} \xi^\lambda		.
\label{nabla^2 xi = Riemann xi}
\end{equation}

Substituting the expression \eqref{derivative of lowered killing vector in terms of F and omega} for the derivative of the Killing vector in to the Riemann curvature \eqref{Riemann on V_d-1 in terms of Riemann on V_d} on $V_{d-1}$, the Ricci tensor on $V_{d-1}$ is then
\begin{equation}
({}^\gamma R)_{\mu \rho}	=		({}^\gamma R)_{\mu \nu \rho}{}^\nu		=
	\begin{aligned}[t]
	&	\frac{1}{2F} D_\mu D_\rho F
		-	\frac{1}{4F^2} D_\rho F D_\mu F		
	\\
	&	+ \frac{1}{2 F^2} \frac{1}{(d-3)!} [ (d-3) \omega^{\sigma_1 \ldots \sigma_{d-4}}{}_\rho \omega_{\sigma_1 \ldots \sigma_{d-4} \mu}
	\\
	&	-  \gamma_{\mu \rho} \omega^{\sigma_1 \ldots \sigma_{d-3}} \omega_{\sigma_1 \ldots \sigma_{d-3}} ]
		+	\gamma^\sigma_\mu \gamma^\nu_\rho R_{\sigma \nu}		.
	\end{aligned}
\label{ricci tensor on V_d-1 (1)}
\end{equation}
We note at this point that all components of the Ricci tensor on $V_d$ are encoded in the equations \eqref{exterior derivative of twist}, \eqref{Laplacian of F on V_(d-1)}, \eqref{ricci tensor on V_d-1 (1)}. With these equations interpreted as equations of motion for $F$, $\omega$ and $\gamma$, one can then reconstruct $g$ from these objects provided the integrability condition \eqref{divergence of twist} is satisfied. That is, these four equations fully determine the dynamics of the metric $g$---they are the full metric equations of motion.

It is useful for solution-generating transformations to consider a conformal rescaling of $V_{d-1}$ with respect to a scalar $\kappa$ on $V_{d-1}$. Define the space $\hat V_{d-1}$ with metric and ``inverse"
\begin{equation}
\hat \gamma_{\mu \nu}		=		\kappa \gamma_{\mu \nu}
\qquad
(\hat \gamma^{-1})^{\mu\nu}		=		\kappa^{-1} \gamma^{\mu\nu}
\qquad
\mathcal L_\xi \kappa		= 0.
\end{equation}
where the inverse is defined such that for tensors $T$ on $\hat V_{d-1}$ (equivalently on $V_{d-1}$), $\hat\gamma^{-1}\hat\gamma T = T$ (in particular, $\hat\gamma^{\mu\nu} = \kappa^2 (\hat\gamma^{-1})^{\mu\nu}$). If we write the covariant derivatives on $V_{d-1}$ and $\hat V_{d-1}$ (acting on a objects on these spaces) as
\begin{align}
D_\mu u_\lambda		& =		\partial_\mu u_\lambda		-	({}^\gamma \Gamma)^\rho{}_{\mu \lambda} u_\rho
\\
\hat D_\mu u_\lambda		& =		\partial_\mu u_\lambda		-	({}^{\hat\gamma} \Gamma)^\rho{}_{\mu \lambda} u_\rho
\end{align}
respectively, we can use metric compatibility on $V_{d-1}$: $D_\mu \gamma_{\nu \rho} =	0$, and demand metric compatibility on $\hat V_{d-1}$: $\hat D_\mu \hat \gamma_{\nu \rho}	=	0$ to determine
\begin{equation}
\left( ({}^{\hat\gamma}\Gamma)^\sigma{}_{\rho \nu} - ({}^\gamma \Gamma)^\sigma{}_{\rho \nu} \right) \gamma^\lambda_\sigma	
		=	 - \frac{1}{2} \gamma^{\lambda \mu} \left( \gamma_{\nu\rho} \partial_\mu \ln \kappa		- \gamma_{\nu\mu} \partial_\rho \ln \kappa		- \gamma_{\mu\rho} \partial_\nu \ln \kappa \right).
\end{equation}
This is sufficient to relate the Ricci tensors on $\hat V_{d-1}$ and $V_{d-1}$ by
\begin{equation}
({}^{\hat \gamma} R)_{\mu \rho}		=
\begin{aligned}[t]
	&	({}^\gamma R)_{\mu \rho}		-		\frac{1}{2} \gamma_{\mu \rho} \gamma^{\sigma \nu}
		\left[ \hat D_\sigma \hat D_\nu \ln \kappa		-		\frac{1}{2} (d-3) \hat D_\sigma \ln \kappa \hat D_\nu \ln \kappa
		\right]
	\\
	&	- \frac{1}{2} (d-3) \hat D_\mu \hat D_\rho \ln \kappa
		- \frac{1}{4} (d-3) \hat D_\mu \ln \kappa \hat D_\rho \ln \kappa		.
\end{aligned}
\end{equation}

\section[Solution-generating symmetries from an effective action]{Solution-generating symmetries from an effective \\action}
\label{sec:Solution-generating symmetries from an effective action}

We look now to construct solution-generating transformations deriving from isometries, which preserve the conformally rescaled reduced metric $\hat\gamma$. Transformed quantities will be denoted with primes. Suppose the full equations of motion of the theory can be recovered by variation of an action
\begin{equation}
\int_{\hat V_{d-1}} \sqrt{|\hat\gamma_\xi|} \left( {}^{\hat\gamma} R		+		L_p \right)		\qquad\text{where}\qquad		\delta \hat\gamma_\xi		=	\hat\gamma_{\mu\nu} \delta  (\hat\gamma^{-1})^{\mu\nu} 	,
\label{action for eoms on V_{d-1}}
\end{equation}
we define ${}^{\hat\gamma} R= \left(\hat \gamma^{-1}\right)^{\mu\nu} \left({}^{\hat\gamma} R\right)_{\mu\nu}$, and $L_p$ denotes all contributions from remaining fields, both metric and non-metric.

That $L_p$ is the only part of the action that transforms is seen as follows. Firstly, the scalar curvature ${}^{\hat\gamma} R$ is conserved. That the reduced determinant $\hat\gamma_\xi$ is invariant is shown as follows. Expanding $\hat\gamma' = \hat\gamma$, the transformed metric $g'$ is
\begin{equation}
g'_{\mu \nu}		=		\frac{\kappa}{\kappa'} g_{\mu \nu}		-		\frac{\kappa}{\kappa'} \frac{1}{F} \xi_\mu \xi_\nu		+		\frac{1}{F'} \xi'_\mu \xi'_\nu,
\label{metric transformation g' = ... from hat gamma' = hat gamma}
\end{equation}
where $\xi'_\mu \equiv \xi^\nu g'_{\mu \nu}$. We can then use the following equation valid for any invertible $d\times d$ symmetric matrix $M_{ij}$ \cite{Mars:2001gd}
\begin{gather}
\det(a_1 M_{ij}		+		a_2 b_i b_j		+		a_3 c_i c_j)
		=		a_1^{d-2} \det(M_{ij})
				\left[
					(a_1		+		a_2 b^2) (a_1		+		a_3 c^2)		-		a_2 a_3 (b\cdot c)^2
				\right]
\nonumber
\\
b^2					\equiv		b_i b_j (M^{-1})^{ij}
\qquad
c^2					\equiv		c_i c_j (M^{-1})^{ij}
\qquad
b\cdot c			\equiv		b_i c_j (M^{-1})^{ij},
\label{formula for determinant of matrix M shifted by column vectors b and c}
\end{gather}
where $a_1\neq 0$, $a_2$ and $a_3$ are all real constants and $b_i$, $c_i$ are arbitrary $d$-column vectors. Applying \eqref{formula for determinant of matrix M shifted by column vectors b and c} to \eqref{metric transformation g' = ... from hat gamma' = hat gamma} we find
\begin{equation}
\frac{(\kappa')^{d-1}}{F'} \det (g')		=		\frac{\kappa^{d-1}}{F} \det (g)			,
\end{equation}
or, using \eqref{det(g) in terms of F and det(hat gamma)} and $\hat\gamma_\xi = \kappa^{d-1}\gamma_\xi$ from \eqref{definition of reduced determinant}, we arrive at our result
\begin{equation}
\hat\gamma'_\xi		=		\hat\gamma_\xi		.
\label{preserved det(hat gamma)}
\end{equation}
Invariance transformations of $L_p$ can in simple cases be found by observation or comparison with previous known results. However, in the case $L_p$ forms a scalar sigma model, \cite{ANDP:ANDP19694790108} developed a methodical procedure to find all resulting transformations.

Suppose all potentials are given by $N$ scalars $\varphi^A$, $A = 1, \ldots,N$, and their contribution to the action \eqref{action for eoms on V_{d-1}} can be written
\begin{equation}
L_p = G_{AB} (\varphi^C) (\hat\gamma^{-1})^{\mu\nu} \partial_\mu \varphi^A \partial_\nu \varphi^B,
\end{equation}
where $\det(G_{AB}) \neq 0$. One can consider the object $G_{AB}$ as a metric on a \emph{potential space} with coordinates $\varphi^A$, whose line element is
\begin{equation}
G_{AB} (\varphi^C) \ud \varphi^A \ud \varphi^B.
\end{equation}
Infinitesimal invariance transformations of $L_p$ are then generated by $m$ Killing vectors $\Phi^{(m)} = \Phi^{(m) A} \partial/\partial \varphi_A$ on this potential space satisfying
\begin{equation}
(\mathcal L_\Phi G)_{AB}		=		0,
\end{equation}
as
\begin{equation}
\varphi^A		\rightarrow		\varphi^A		+		\lambda^{(m)} \Phi^{(m) A},
\end{equation}
for infinitesimal parameters $\lambda^{(m)}$. Finite transformations can be found by exponentiation, while discrete transformations can be derived by taking appropriate limits of the finite continuous transformations.

\subsection{An example: \texorpdfstring{$(3+1)$}{(3+1)}-d vacuum Einstein gravity with an isometry}
\label{sec:An example: (3+1)-d vacuum Einstein gravity with an isometry}

It will be instructive for our exploration of solution-generating transformations in supergravity to briefly review the corresponding transformations in $(3+1)$-dimensional vacuum Einstein gravity. The Einstein equations
\begin{equation}
R_{\mu\nu}		=		0
\end{equation}
can be reduced in the presence of a Killing vector $\xi$ in terms of the metric $F\gamma$ (i.e.\ $\kappa = F$) and the scalars $F$ and $\chi$, where $\partial_\mu \chi= \omega_\mu$ and the existence of this scalar is guaranteed by \eqref{exterior derivative of twist} in vacuum. The full equations of motion are
\begin{subequations}
\begin{align}
({}^{\hat \gamma} R)_{\mu \rho}		& =		\frac{1}{2F^2} \hat D_\mu F \hat D_\rho F		+ \frac{1}{2F^2} \hat D_\rho \chi \hat D_\mu \chi
\label{ricci tensor on V_d-1: (3+1)-d vacuum}
\\
\hat D^\mu \hat D_\mu \chi	& = 		\frac{2}{F} \hat D^\mu F \hat D_\mu \chi
\label{divergence of twist: (3+1)-d vacuum}
\\
\hat D^\mu \hat D_\mu F		& =	 \frac{1}{F} \hat D^\mu F \hat D_\mu F		- \frac{1}{F} \hat D^\mu \chi \hat D_\mu \chi
\end{align}
\label{EFE with one Killing vector in (3+1)-d vacuum}%
\end{subequations}
where indices are raised with $\hat \gamma^{-1}$, and can be found by variation of the action
\begin{equation}
\int_{\hat V_{d-1}}\sqrt{-\hat\gamma_\xi} \, \left[ {}^{\hat\gamma} R		- \frac{1}{2F^2} \left( \hat D^\mu F \hat D_\mu F	+ \hat D^\mu \chi \hat D_\mu \chi\right)	\right].
\end{equation}
Following \S\ref{sec:Solution-generating symmetries from an effective action}, one can then form the potential space line element
\begin{equation}
\ud S^2		= 		\frac{1}{2F^2} ( \ud F^2 + \ud \chi^2),
\end{equation}
whose Killing vectors generate the transformations with infinitesimal real parameters $k$, $p$, $q$,
\begin{equation}
F		\rightarrow		F+ kF\chi + pF
\qquad
\chi		\rightarrow		\chi		+ \frac{1}{2} k (\chi^2	- F^2)		+ p\chi +q	.
\end{equation}
The finite counterparts are the finite invariance transformations of the $(3+1)$-dimensional vacuum Einstein equations in the presence of an isometry, which can be combined into the M\"obius map
\begin{equation}
\sigma = F+i\chi
\qquad
\sigma	\rightarrow		\frac{\alpha \sigma 	- i\beta}{i\gamma\sigma + \delta}
\qquad
\begin{pmatrix}
\alpha	&		\beta		\\
\gamma &		\delta
\end{pmatrix}
\in SL(2,\mathbb R)		
\end{equation}
for constant $\alpha$, $\beta$, $\gamma$, $\delta$.

\section{Application to NS-NS supergravity}
\label{sec:Application to bosonic NS-NS supergravity}

We now apply our analysis to $d$-dimensional NS-NS supergravity with equations of motion
\begin{subequations}
\begin{align}
R_{\mu \nu}		-		\frac{1}{4} H_\mu{}^{\sigma \rho} H_{\nu \sigma \rho}		+		2 \nabla_\mu \nabla_\nu \phi		& =	0
\label{NS-NS SUGRA metric eom (2)}
\\
\ud\star (e^{-2\phi} H)	& =		0
\label{NS-NS SUGRA 2-form eom (2)}
\\
H\wedge \star H		- e^{2\phi} \ud\star\ud e^{-2\phi}	& =		0		,
\label{NS-NS SUGRA dilaton eom (2)}
\end{align}
\label{NS-NS SUGRA equations of motion (2)}%
\end{subequations}
and Bianchi identity 
\begin{equation}
\ud H = 0,
\label{Bianchi identity for H (2)}
\end{equation}
for the case that the field content satisfies\footnote{Owing to the gauge freedom in $B$, we do not require $\mathcal L_\xi B = 0$.}
\begin{equation}
\mathcal L_\xi H		=		0
\qquad
\mathcal L_\xi \phi		=		0
\qquad
\mathcal L_\xi g			=		0		.
\label{Killing isometry on NSNS sugra fields}
\end{equation}

We begin by constructing the potentials of theory, then proceeding to express the equations of motion in terms of these potentials. The field strength can be expressed under an ``electric-magnetic" decomposition (see app. \ref{app:Differential forms})
\begin{align}
F H
& =		- \left( (-1)^d (\iota_\xi \star)^2		+	(\star \iota_\xi)^2 \right) H		\notag
\\
& =		-\iota_\xi \star \beta		- \star \iota_\xi \star \alpha,
\label{H in terms of alpha and beta}
\end{align}
where the forms
\begin{equation}
\alpha		=		\iota_\xi H
\qquad
\beta 		=		(-1)^d \iota_\xi \star H
\end{equation}
satisfy
\begin{align}
\ud \alpha		& =		\mathcal L_\xi H		- \iota_\xi \ud H
\label{d alpha}
\\
\mathcal L_\xi \alpha		& = \iota_\xi \mathcal L_\xi H
\\
\ud \beta		& =		(-1)^d (\mathcal L_\xi \star H		- \iota_\xi \ud \star H )
\label{d beta (1)}
\\
\mathcal L_\xi \beta		& = (-1)^d \iota_\xi \mathcal L_\xi \star H.
\end{align}
We will refer to $\alpha$ and $\beta$ as the electric and magnetic contributions respectively, for \emph{any} $\xi$. Using the equation of motion \eqref{NS-NS SUGRA 2-form eom (2)}, equation \eqref{d beta (1)} becomes
\begin{equation}
\ud \left( e^{-2\phi} \beta \right)		=		(-1)^d\mathcal L_\xi \left(e^{-2\phi} \star H \right).
\end{equation}
Thus we have, at least locally, for all $(g, H, \phi)$ satisfying the equations of motion and Bianchi identity, and annihilated by $\mathcal L_\xi$, that we can write
\begin{equation}
\alpha	 =		\ud a
\qquad
\beta		=		e^{2\phi} \ud b,
\end{equation}
where $\ud a$ and $\ud b$ live on $\hat V_{d-1}$. Additionally, under isometry the exterior derivative of the twist \eqref{exterior derivative of twist} becomes
\begin{equation}
\ud \left( e^{-2\phi} \omega \right)		=		- (-1)^d 4 \ud (a \wedge \ud b),
\end{equation}
which ensures that we can define a potential $\chi$ by
\begin{equation}
\omega		= e^{2\phi} ( \ud \chi		-		4 (-1)^d  \ud b \wedge a )		.
\end{equation}
Our potentials are thus $(F, \chi, \phi, a, b)$.

We now restrict to the scalar potential space, by demanding that all higher form potentials vanish. There are multiple cases where this is true. These include static Einstein-dilaton theory in arbitrary dimension, and the magnetic sector of five-dimensional static NS-NS supergravity (for the latter, we refer the reader to the closely related analysis \cite{Yazadjiev:2005pf}).

\subsection{Static Einstein-dilaton theory}
\label{sec:Static Einstein-dilaton theory}

We first examine static Einstein-dilaton theory in arbitrary dimension, with vanishing $H$ and $\omega$. The equations of motion are (raising indices with $\hat\gamma^{-1}$)
\begin{subequations}
\begin{align}
({}^{\hat\gamma} R)_{\mu\nu}		& =
\begin{aligned}[t]
&		\frac{1}{2F} \hat D_\mu \hat D_\rho F		- \frac{1}{4F} \hat\gamma_{\mu\rho} \hat D^\sigma \ln\kappa \hat D_\sigma F
		+	\frac{1}{2F} \hat D_{(\mu} \ln\kappa \hat D_{\rho)} F		- \frac{1}{4F^2} \hat D_\mu F\hat D_\rho F
\\
&		-	2 \hat D_\mu \hat D_\rho \phi
		+	\hat\gamma_{\mu\rho} \hat D^\sigma \ln\kappa \hat D_\sigma \phi
		-	2 \hat D_{(\mu} \ln\kappa \hat D_{\rho)} \phi
\\
&		-	\frac{1}{2} \hat\gamma_{\mu\rho}
			\left(
				\hat D^\sigma \hat D_\sigma \ln\kappa
				-	\frac{1}{2} (d-3) \hat D^\sigma \ln\kappa \hat D_\sigma \ln\kappa
			\right)
\\
&		-	\frac{1}{2} (d-3) \hat D_\mu \hat D_\rho \ln\kappa		- \frac{1}{4} (d-3) \hat D_\mu \ln\kappa \hat D_\rho \ln\kappa
\end{aligned}
\label{eoms for static Einstein-dilaton (CR ricci)}
\\
\hat D^\mu \hat D_\mu F		& =		\frac{1}{2F} \hat D^\mu F \hat D_\mu F		+ 2 \hat D^\mu F \hat D_\mu \phi		+ \frac{1}{2} (d-3) \hat D^\mu \ln \kappa \hat D_\mu F		
\label{eoms for static Einstein-dilaton (hat D^2 F)}
\\
\hat D^\mu \hat D_\mu \phi		& = 		2 \hat D^\mu \phi \hat D_\mu \phi		- \frac{1}{F} \hat D^\mu F \hat D_\mu \phi		+ \frac{1}{2} (d-3) \hat D^\mu \ln \kappa \hat D_\mu \phi.
\label{eoms for static Einstein-dilaton (hat D^2 phi)}
\end{align}
\end{subequations}
We employ the conformal rescaling\footnote{This is chosen as it is a simple example---in particular to render the equations of motion first order in derivatives. It would be interesting to find if other rescalings allowed different symmetry groups.}
\begin{equation}
\kappa		=		\left(|F|e^{-4\phi}\right)^{1/(d-3)},
\label{conformal factor for static Einstein-dilaton}
\end{equation}
where the equations of motion simplify to
\begin{subequations}
\begin{align}
({}^{\hat\gamma} R)_{\mu\nu}		& =		\frac{1}{4F^2} \hat D_\mu F \hat D_\rho F		+ \frac{1}{d-3} \left[ \frac{1}{4F^2} \hat D_\mu F\hat D_\rho F		- \frac{2}{F} \hat D_{(\mu} \ln\kappa \hat D_{\rho)} F		- 4 \hat D_\mu \phi\hat D_\rho \phi \right]
\label{eoms for static Einstein-dilaton (CR ricci): chosen kappa}
\\
\hat D^\mu \hat D_\mu F		& =		\frac{1}{F} \hat D^\mu F \hat D_\mu F
\label{eoms for static Einstein-dilaton (hat D^2 F): chosen kappa}
\\
\hat D^\mu \hat D_\mu \phi		& = 		0,
\label{eoms for static Einstein-dilaton (hat D^2 phi): chosen kappa}
\end{align}
\end{subequations}
which can be derived by variation of the following action with potentials $F$, $\kappa$
\begin{equation}
\int_{\hat V_{d-1}} \sqrt{|\hat\gamma_\xi|} \left[ {}^{\hat\gamma} R		-		\left(\hat\gamma^{-1}\right)^{\mu\nu}
				\left(
					\frac{1}{4F^2} \hat D_\mu F \hat D_\nu F		+ \frac{d-3}{4\kappa^2} \hat D_\mu \kappa \hat D_\nu \kappa \right) \right]		.
\label{static einstein-dilaton sigma model}
\end{equation}
This action exhibits a $\mathbb Z_2 \times \mathbb R \times \mathbb Z_2 \times \mathbb R$ invariance group with finite generators
\begin{subequations}
\begin{align}
F		&	\rightarrow		\frac{1}{F}
& \& &&
\phi	& \rightarrow		\phi		- \frac{1}{2} \ln|F|
&
\label{F to 1/F static Einstein dilaton transformation}
\\
F		& \rightarrow		q F
& \& &&
\phi	& \rightarrow		\phi
&
q \in \mathbb R
\label{F to qF static Einstein dilaton transformation}
\\
F		& \rightarrow		F
& \& &&
\phi	& \rightarrow		- \phi		+ \frac{1}{2} \ln|F|
&
\\
F		& \rightarrow		F
& \& &&
\phi	& \rightarrow		\phi		- p
&
p \in \mathbb R
\end{align}
\label{Einstein-dilaton invariance group}%
\end{subequations}
for constant $p$ and $q$.

\subsection{5-d static magnetic NS-NS supergravity}
\label{sec:5-d static magnetic NS-NS supergravity}

Here we consider the magnetic sector of static five-dimensional supergravity, where $\alpha$ and $\omega$ vanish. The equations of motion are
\begin{subequations}
\begin{align}
({}^{\hat\gamma} R)_{\mu\nu}		& =
\begin{aligned}[t]
&		\frac{1}{2F} \hat D_\mu \hat D_\rho F		- \frac{1}{4F} \hat\gamma_{\mu\rho} \hat D^\sigma \ln\kappa \hat D_\sigma F
		+	\frac{1}{2F} \hat D_{(\mu} \ln\kappa \hat D_{\rho)} F		- \frac{1}{4F^2} \hat D_\mu F\hat D_\rho F
\\
&		+	\frac{e^{4\phi}}{F} \left( \hat D_\mu b \hat D_\rho b		- \hat\gamma_{\mu\rho} \hat D^\sigma b \hat D_\sigma b \right)
\\
&		-	2
			\left(
				\hat D_\mu \hat D_\rho \phi
				-	\frac{1}{2}
					\left(
						\hat\gamma_{\mu\rho} \hat D^\sigma \ln\kappa \hat D_\sigma \phi
						-	2 \hat D_{(\mu} \ln\kappa \hat D_{\rho)} \phi
					\right)
			\right)
\\
&		-	\frac{1}{2} \hat\gamma_{\mu\rho}
			\left(
				\hat D^\sigma \hat D_\sigma \ln\kappa
				-	\hat D^\sigma \ln\kappa \hat D_\sigma \ln\kappa
			\right)
\\
&		- \hat D_\mu \hat D_\rho \ln\kappa		- \frac{1}{2} \hat D_\mu \ln\kappa \hat D_\rho \ln\kappa
\end{aligned}
\label{eoms for 5-d NSNS sugra potentials (CR ricci)}
\\
\hat D^\mu \hat D_\mu F		& =		\hat D^\mu \ln \kappa \hat D_\mu F		+ \frac{1}{2F} \hat D^\mu F \hat D_\mu F		+ 2 \hat D_\mu F \hat D^\mu \phi
\label{eoms for 5-d NSNS sugra potentials (hat D^2 F)}
\\
\hat D^\mu \hat D_\mu \phi		& = 		\hat D^\mu \ln \kappa \hat D_\mu \phi		+ 2 \hat D^\mu \phi \hat D_\mu \phi		- \frac{1}{2F} \hat D^\mu F \hat D_\mu \phi		+ \frac{e^{4\phi}}{2F} \hat D^\mu b\hat D_\mu b
\label{eoms for 5-d NSNS sugra potentials (hat D^2 phi)}
\\
\hat D^\mu \hat D_\mu b		&=	\hat D^\mu \ln\kappa \hat D_\mu b		+ \frac{1}{2F} \hat D^\mu F\hat D_\mu b		- 2\hat D^\mu\phi \hat D_\mu b		.
\label{eoms for 5-d NSNS sugra potentials (hat D^2 b)}
\end{align}
\end{subequations} 
We employ the conformal factor \eqref{conformal factor for static Einstein-dilaton}
\begin{equation}
\kappa		=		|F|^{1/2} e^{-2\phi},
\label{conformal factor for 5-d sugra solgen}
\end{equation}
where the equations of motion become
\begin{subequations}
\begin{align}
({}^{\hat\gamma} R)_{\mu\nu}		& =
\begin{aligned}[t]
&		\frac{3}{8F^2} \hat D_\mu F \hat D_\rho F		+ 2 \hat D_\mu \phi\hat D_\rho \phi		- \frac{1}{F} \hat D_{(\mu} \phi \hat D_{\rho)} F		+	\frac{e^{4\phi}}{2F} \hat D_\mu b \hat D_\rho b
\end{aligned}
\\
\hat D^\mu \hat D_\mu F		& =		\frac{1}{F} \hat D^\mu F \hat D_\mu F
\\
\hat D^\mu \hat D_\mu \phi		& = \frac{e^{4\phi}}{2F} \hat D^\mu b\hat D_\mu b
\\
\hat D^\mu \hat D_\mu b		&=	\frac{1}{F} \hat D^\mu F \hat D_\mu b		- 4 \hat D^\mu \phi \hat D_\mu b,
\end{align}
\end{subequations}
which can be derived by variation of an action, here written in terms of $\kappa$ (as \eqref{conformal factor for 5-d sugra solgen}), $F$ and $b$, as
\begin{equation}
\int_{\hat V_{d-1}} \sqrt{|\hat\gamma_\xi|} \left[ {}^{\hat\gamma} R		-		\left(\hat\gamma^{-1}\right)^{\mu\nu}
				\left(
					\frac{1}{4F^2} \hat D_\mu F \hat D_\nu F		+ \frac{1}{2\kappa^2} \hat D_\mu \kappa \hat D_\nu \kappa		+ \frac{1}{\kappa^2} \hat D_\mu b \hat D_\nu b \right) \right]		.
\label{5-d static sugra sigma model alpha = 0}
\end{equation}
This action exhibits an $SL(2,\mathbb R)\times \mathbb Z_2 \times\mathbb R$ invariance group with finite generators
\begin{subequations}
\begin{align}
F		&	\rightarrow		\frac{1}{F}
\qquad
\& \qquad
\phi	\rightarrow		\phi		- \frac{1}{2} \ln|F|
\\
F		& \rightarrow		q F
\qquad
\& \qquad
\phi	\rightarrow		\phi		- \frac{1}{4} \ln|q|
\qquad
q \in \mathbb R
\\
\sigma		& \rightarrow		\frac{\alpha \sigma - i\beta}{i\gamma\sigma+\delta}		\qquad		\begin{pmatrix} \alpha & \beta \\ \gamma & \delta \end{pmatrix} \in SL(2,\mathbb R)
\qquad
\sigma = |F|^{1/2}e^{-2\phi}		+ i b.
\end{align}
\label{finite generators 5-d NSNS sugra alpha = omega = 0}%
\end{subequations}
for constant $q$, $\alpha$, $\beta$, $\gamma$ and $\delta$.

\section{Extension to double field theory}
\label{sec:Extension to double field theory}

In this section, we offer a few thoughts on extensions of the solution-generating techniques presented in this chapter to DFT. We will use the coordinate notation $X^M = (x^\mu, \tilde x^{\bar \mu})$ throughout. To do this, we will take inspiration from our analysis of Einstein-dilaton theory and static magnetic five-dimensional NS-NS supergravity in \S\ref{sec:Application to bosonic NS-NS supergravity}. It is the generalised metric $H$ which is dynamical, and not the metric $\eta$ which defines inner products, so it is the former, along with the DFT dilaton, that we will generate solutions of. We will choose an isometry $\partial_0 = 0$, here with respect to a generalised vector $V = \partial_0$ with Killing equations
\begin{equation}
\partial_0 H_{MN} = 0
\qquad
\partial_0 d			= 0.
\end{equation}
We begin by noting that vanishing $\alpha$ and $\omega$ implies, for a given gauge of $B$,
\begin{equation}
B_{0m}		=		g_{0m}		=	0.
\end{equation}
This can be seen from the DFT perspective as some kind of staticity requirement on the generalised metric with respect to $X^0 = x^0$, that
\begin{equation}
H_{0M} = \delta_M^0 H_{00}		\qquad		H_{\bar 0 M} = \delta_M^{\bar 0} H_{\bar 0 \bar 0}		.
\label{generalised metric static wrt partial_0}
\end{equation}

With regard to solution-generating techniques, one can easily see that preserving a reduced metric orthogonal to $\partial_0$ (as we did in supergravity) would be far too restrictive: all of the metric data is contained within this choice of reduced metric, and so preserving it allows only the dilaton to transform. Which quantity is it then that we wish to preserve? In our analysis of 5-d static magnetic supergravity, which can be found from DFT in the supergravity frame, where one imposes the section condition $\tilde\partial_{\bar\mu} = 0$, we preserved a conformally rescaled metric proportional to the $H_{\bar m\bar n}$ part of the generalised metric. Perhaps a natural generalisation is as follows
\begin{itemize}
\item Solve the section condition by requiring the fields depend on spacetime coordinates $X^s$, where $s$ runs over some $d$ of $(1,\ldots ,2d)$.
\item Determine the metric components along these $X^s$ directions, but excluding the direction of the isometry (and its dual).
\item Preserve a conformal rescaling of these components.
\end{itemize}
For example, we could have
\begin{enumerate}
\item For Killing vector $\partial_0$ and section $\tilde\partial_{\bar\mu} = 0$, we preserve $\propto H_{\bar m\bar n}$ 
\item For Killing vector $\tilde\partial_{\bar 0}$ and section $\partial_\mu = 0$, we preserve $\propto H_{mn}$		.
\end{enumerate}
We note however that since the metric is not an arbitrary symmetric object, but takes a coset form, there will be other metric components which must also be preserved by this transformation. It remains unclear how we determine which other components to preserve in general, and how precisely we do this. However, we circumvent this issue here by studying the simple case of vanishing two-form field (equivalently zero $H_{m\bar n}$), corresponding to $O(d,d,\mathbb R)$-covariant Einstein-dilaton theory, where the procedure becomes quite straightforward.

We will impose the supergravity frame. Recalling that the metric is static (see \eqref{generalised metric static wrt partial_0}) with respect to $x^0$, it takes the diagonal form
\begin{equation}
H = \diag(H_{00}, H_{mn}, H_{\bar 0 \bar 0}, H_{\bar m\bar n}).
\end{equation}
We will look  for transformations which preserve a conformal rescaling of the metric components $H_{mn}$. From the coset form of the metric, we can determine that this must also preserve appropriate rescalings of $H_{\bar m\bar n}$, $H^{mn}$ and $H^{\bar m\bar n}$.

The equations of motion which do not vanish identically are
\begin{subequations}
\begin{align}
R		& =		0
\\
\mathcal K_{00} & = (H^{\bar 0\bar 0})^2 \mathcal K_{\bar 0 \bar 0}
\\
\mathcal K_{mn} & = H_{mr} \eta^{r \bar s} H_{nl} \eta^{l \bar u} \mathcal K_{\bar s \bar u}
\end{align}
\end{subequations}
where $R$ and the non-vanishing components of $\mathcal K$ are written explicitly as
\begin{subequations}
\begin{align}
R		& =
\begin{aligned}[t]
& \frac{1}{4} H^{mn} \partial_m H^{qr} \partial_n H_{qr}		-		\frac{1}{2} H^{mn} \partial_m H^{ql} \partial_q H_{nl}
\\
& + 4H^{mn} \partial_m\partial_n d		-		\partial_m \partial_n H^{mn}		-		4H^{mn}\partial_m d \partial_n d		+		4\partial_m H^{mn} \partial_n d
\\
&	+ \frac{1}{8} H^{mn} \left( \partial_m H^{00} \partial_n H_{00} + \partial_m H^{\bar 0\bar 0} \partial_n H_{\bar 0\bar 0} \right)
\end{aligned}
\\
\mathcal K_{00} & = \frac{1}{4} H^{mn} H^{00} \partial_m H_{00} \partial_n H_{00} - \frac{1}{4} (\partial_l - 2 \partial_l d) (H^{lk} \partial_k H_{00} )
\\
\mathcal K_{\bar 0\bar 0} & = \frac{1}{4} H^{mn} H^{\bar 0 \bar 0} \partial_m H_{\bar 0 \bar 0}\partial_n H_{\bar 0\bar 0} - \frac{1}{4} (\partial_l - 2 \partial_l d) (H^{lk} \partial_k H_{\bar 0\bar 0} )
\\
\mathcal K_{mn} & = 
\begin{aligned}[t]
& \frac{1}{8} \left( \partial_m H^{00} \partial_n H_{00}		+ \partial_m H^{\bar 0\bar 0} \partial_n H_{\bar 0\bar 0} \right) + \frac{1}{8} \partial_m H^{qr} \partial_n H_{qr}		+ \frac{1}{8} \partial_m H^{\bar q \bar r} \partial_n H_{\bar q \bar r}
\\
&	- \frac{1}{2} H^{rs} H^{ql} \partial_l H_{rm} \partial_s H_{nq}		+ \frac{1}{4} H^{rs} H^{ql} \partial_r H_{qm} \partial_s H_{nl}
\\
&	\frac{1}{4} (\partial_l - 2\partial_l d) [ H^{lq}(4\partial_{(m} H_{n) q} - \partial_q H_{mn})] + 2\partial_m \partial_n d
\end{aligned}
\\
\mathcal K_{\bar m \bar n} & = \frac{1}{4} H^{rs} H^{\bar l \bar q} \partial_r H_{\bar l \bar m} \partial_s H_{\bar n \bar q}		-		\frac{1}{4} (\partial_l - 2\partial_l d)(H^{lq}\partial_q H_{\bar m\bar n}).
\end{align}
\end{subequations}
Noting that $H^{\bar 0\bar 0} = 1/H^{00} = H_{00} = 1/H_{\bar 0 \bar 0}$, and fixing $d$ and $H_{mn}$ etc.\ (employing a trivial conformal rescaling of 1), the equations are clearly invariant under the transformations
\begin{subequations}
\begin{align}
H_{00} & \rightarrow 1/H_{00}
\\
H_{00} & \rightarrow qH_{00}		\qquad q\in\mathbb R.
\end{align}
\label{DFT Buchdahl transformation directly from eoms}%
\end{subequations}
So that we may compare this with the Einstein-dilaton analysis of \S\ref{sec:Static Einstein-dilaton theory}, we note the relations
\begin{equation}
\gamma_{mn} = H_{mn},
\qquad
F = H_{00},
\qquad
\kappa^{d-3} = \frac{1}{|\gamma_\xi|} e^{-4d}.
\end{equation}
From these equalities, we see that fixed $d$ and $H_{mn}$ corresponds to fixed $\kappa$, and that \eqref{DFT Buchdahl transformation directly from eoms} are the $\mathbb Z_2 \times \mathbb R$ transformations \eqref{F to 1/F static Einstein dilaton transformation} and \eqref{F to qF static Einstein dilaton transformation}. The first constitutes a T-duality transformation, while the second a trivial coordinate rescaling. The remaining symmetries in \eqref{Einstein-dilaton invariance group} are not manifest, since they do not belong to T-duality. Retrieving these, and any further possible invariance transformations from these equations, are avenues for future research. It would indeed be interesting to see more general constructions for finding solution-generating symmetries in DFT, perhaps utilising the constructions of DFT covariant derivatives and curvature tensors in \cite{Hohm:2010xe, Jeon:2010rw, Hohm:2011si, Hohm:2012mf, Berman:2013uda, Cederwall:2014kxa}.

\newpage
\chapter{Solution-generating transformations in the fluid/gravity correspondence}
\label{ch:Solution-generating transformation in the fluid/gravity correspondence}
\chead{\textsc{solution-generating transformations \\in the fluid/gravity correspondence}}

\section{Hydrodynamics}
\label{sec:Hydrodynamics}

The study of hydrodynamics is fundamental to vast areas of physics and engineering, owing to its origin as the long-wavelength limit of any interacting field theory at finite temperature. Such a limit needs a consistent definition. Consider a quantum field theory where quanta interact with a characteristic length scale $\ell_\text{corr}$, the correlation length. The long-wavelength limit simply requires that fluctuations of the thermodynamic quantities of the system vary with a length scale $L$ much greater than $\ell_\text{corr}$, parametrized by the dimensionless Knudsen number
\begin{equation}
	K_n		=	\frac{ \ell_\text{corr} }{ L }.
\label{Knudsen}
\end{equation}
For a fluid description to be useful in non-equilibrium states, we naturally require that $L$ remain small compared to the size of the system. This is usually satisfied trivially by considering systems of infinite size.

The long-wavelength limit allows the definition of a particle as an element of the macroscopic fluid, infinitesimal with respect to the size of the system, yet containing a sufficiently large number of microscopic quanta. One mole contains an Avogadro's number of molecules, for example. Each particle defines a local patch of the fluid in thermal equilibrium, that is, thermodynamic quantities do not vary within the particle. Away from global equilibrium quantities vary between particles as function of time $\tau$ and spatial coordinates $x^i$, combined as $x^a=(\tau,x^i)$. The evolution of particles in the fluid is parametrized by a relativistic velocity $u^b(x^a)$, which refers to the velocity of the fluid at $x^a$. It is well known \cite{landau1987fluid} that the thermodynamic quantities, such as the temperature $T(x^a)$ and the local energy density $\rho(x^a)$, are determined by the value of any two of them, along with the equation of state. The evolution of the system is then specified by the equations of hydrodynamics in terms of a set of transport coefficients, whose values depend on the fluid in question.

Fluid flow is in general relativistic in that the systems it describes are constrained by local Lorentz invariance, and velocities may take any physical values below the speed of light. Applications at relativistic velocities are multitudinous: the dust clouds in galaxy and star formation; the flow of plasmas and gases in stars supporting fusion; the superfluid cores of neutron stars; the near-horizon dynamics of black holes are all described by hydrodynamics. Modelling black holes (and black branes in M/string theory) with hydrodynamics has now developed into a fundamental correspondence of central importance to our present study, as discussed in \S\ref{sec:The fluid/gravity correspondence}. Quark-gluon plasmas behave as nearly ideal fluids and are expected to have formed after the inflationary epoch of the big bang, and are reproduced in collisions at the RHIC and LHC. Non-relativistic fluids are equally ubiquitous, somewhat more familiar, and constitute an endless list of phenomena from the atmosphere to the oceans.

\subsection{The fluid equations}
\label{sec:Relativistic fluids}

We begin with a discussion, adapted from \cite{Rangamani:2009xk}, of the relativistic fluid described by the stress energy tensor $T^{ab}$ and a set of conserved currents $J^a_I$ where $I$ indexes the corresponding conserved charge. The dynamical equations of the $(d-1)$-spacetime dimensional fluid are
\begin{subequations}
\begin{align}
\nabla_a T^{ab}		& = 0		\\
\nabla_a J^a_I		& = 0		\, .
\label{fluid charge current conservation}
\end{align}
\label{relativistic NS}%
\end{subequations}
For an ideal fluid, with no dissipation, the energy-momentum tensor and currents may be expressed in a local rest frame in the form
\begin{subequations}
\begin{align}
	T^{ab} &=\rho u^au^b+p(g^{ab}+u^au^b)
	\label{Tideal}
\\
	J_I^a &=q_I u^a
	\label{Jideal}
\end{align}
\label{TJideal}%
\end{subequations}
where $p$ is the pressure, $q_I$ are the conserved charges and $g_{ab}$ is the metric of the space on which the fluid propagates. The velocity is normalised to $u^au_a=-1$. The entropy current is given by \eqref{Jideal} with the charge $q$ being given by the local entropy density. The conservation of the entropy current illustrates the non-dissipative nature intrinsic to zero entropy production.

In a dissipative fluid, there are corrections to \eqref{TJideal}. We must first take into account the interrelation between mass and energy to define the velocity field more rigorously. This is achieved by using the Landau gauge, which requires that the velocity be an eigenvector of the stress-energy tensor with eigenvalue the local energy density of the fluid. This is satisfied when dissipative terms are orthogonal to the velocity (it is satisfied for the ideal fluid by the normalisation of the velocity). If the stress energy tensor gains a dissipative term $\Pi^{ab}$, and the current a term $\Upsilon_I^a$, this reads
\begin{equation}
\Pi^{ab}u_a=0\qquad \Upsilon_I^a u_a=0.
\end{equation}
Dissipative corrections to the stress tensor are constructed in a derivative expansion of the velocity field and thermodynamic variables, where derivatives implicitly scale with the infinitesimal Knudsen number \eqref{Knudsen}. Recalling that the equations of motion for the ideal fluid are composed of relations between these gradients, we may express $\Pi^{ab}$ purely in terms of the derivative of the velocity. This can be iterated to all orders in the expansion (except when charges are present in which case this is only true to to first order). Now, the derivative of the velocity may be decomposed using the acceleration $A^a$, divergence $\vartheta$, a symmetric traceless shear $\sigma^{ab}$, and the antisymmetric vorticity $w^{ab}$ into the form
\begin{equation}
\nabla^b u^a = -A^a u^b + \sigma^{ab} + w^{ab} + \frac{1}{d-2}\vartheta P^{ab},
\end{equation}
where
\begin{align}
\vartheta	& =		\nabla_a u^a
\\
A^a			& =		u^b\nabla_b u^a
\\
\sigma^{ab} & =		P^{ac} P^{bd} \nabla_{(c}u_{d)}		-		\frac{1}{d-2} \theta P^{ab}
\\
w^{ab} 		& =		P^{bc} P^{ad} \nabla_{[c}u_{d]}.
\end{align}
and $P^{ab} = g^{ab} + u^a u^b$ is a projection operator on to spatial directions. In the Landau frame, only the divergence and shear can contribute to the first-order stress-energy tensor. A similar analysis for the charge current retains the acceleration, and if one includes the parity-violating pseudo-vector contribution
\begin{equation}
\ell^a =\hat\epsilon_{bcd}{}^a u^b\nabla^c u^d,
\end{equation}
the leading order dissipative equations of motion for a relativistic fluid are \eqref{relativistic NS} with
\begin{subequations}
\begin{align}
T^{ab}	& =		\rho u^au^b		+ pP^{ab}		- 2\eta\sigma^{ab}		-\zeta\vartheta P^{ab}	\label{Tdiss}
\\
J_I^a	& =		q_I u^a	-\chi_{IJ}P^{ab}\nabla_b q_J-\Theta_I\ell^a-\gamma_I P^{ab}\nabla_bT,	\label{Jdiss}
\end{align}
\label{TJdiss}%
\end{subequations}
where $\eta$ and $\zeta$ are the shear\footnote{We use the traditional notation of \cite{Kovtun:2004de, Bhattacharyya:2008kq, landau1987fluid} rather than that of \cite{Bredberg:2011jq, Compere:2011dx}. We introduce the kinematical viscosity $\nu$ in \S\ref{sec:non-rel NS}.} and bulk viscosities respectively, $\chi_{IJ}$ is the matrix of charge diffusion coefficients, $\gamma_I$ indicates the contribution of the temperature gradients and $\Theta_I$ the pseudo-vector transport coefficients. The transport coefficients have been calculated in the weakly coupled QFT in perturbation theory, whereas in the strongly coupled theory, a dual holographic description may be employed, see e.g.\ \cite{Baier:2007ix}.

\subsection{Non-relativistic Navier-Stokes fluids} \label{sec:non-rel NS}

We now take the non-relativistic limit, defined by long distances, long times and low velocity and pressure amplitudes, of the relativistic fluid on a spacetime lightly perturbed in the following manner \cite{Bhattacharyya:2008kq}. Consider an arbitrary perturbation
\begin{equation}
	G_{ab}=g_{ab}+H_{ab}
\end{equation}
of the background metric
\begin{equation}
	g_{ab}\ud x^a\ud x^b=-\ud\tau^2+g_{ij}\ud x^i\ud x^j
\end{equation}
(this form is purely a choice of coordinate system for a large class of metrics), where
\begin{equation}
\begin{split}
	H_{\tau\tau} &=\epsilon^2 h_{\tau\tau}(\epsilon^2\tau,\epsilon x^i)\\
	H_{\tau i} &=\epsilon \hat a_i(\epsilon^2\tau,\epsilon x^i)\\
	H_{ij} &=\epsilon^2 h_{ij}(\epsilon^2\tau,\epsilon x^i),
\end{split}
\end{equation}
for infinitesimal scaling parameter $\epsilon$. Treating the fluid on $G_{ab}$ as an effective forced flow of that on $g_{ab}$, whilst preserving the normalisation $u_au^a=-1$, the velocity
\begin{equation}
	u^a=\frac{1}{\sqrt{1-g_{ij}V^i V^j}}(1,V^i),
\end{equation}
density and pressure may be expanded about an equilibrium configuration $(\rho_0,V^i_0,p_0)$ of a stationary, uniform fluid at rest\footnote{We assume the background allows such stationary, uniform solutions. In particular, one would not expect generic non-static metrics $g_{ab}$ to allow stationary fluid solutions. On the other hand, stationary, uniform solutions will of course exist in flat space, for example.}, as
\begin{equation}
\begin{split}
	V^i &=\epsilon v^i(\epsilon^2\tau,\epsilon x^i)\\
	\frac{p-p_0}{\rho_0+p_0} &=\epsilon^2 p_e(\epsilon^2\tau,\epsilon x^i)		.
\end{split}
\end{equation}
The scaling of the density is fixed by the equation of state. The scaling of time $\tau$ and space $x^i$ are implicit: time and space derivatives acting on the velocity and pressure will draw out, via the chain rule, the corresponding power of $\epsilon$. In this limit, the temporal component of the relativistic Navier-Stokes equations becomes
\begin{equation}
\nabla_a T^{a\tau}=\epsilon^2(\rho_0+p_0)\nabla_i v^i+\mathcal{O}(\epsilon^4)		,
\label{incompressibility scaling}
\end{equation}
where $\nabla_i$ is the covariant derivative with respect to the spatial metric $g_{ij}$. Separating the gauge field $\hat a_i$ into its pure curl and divergence parts
\begin{equation}
	\hat a_i=a_i+\nabla_i\chi,
\end{equation}
such that $\nabla_ia^i=0$, and defining an effective pressure
\begin{equation}
	P=p_e-\frac{1}{2}h_{\tau\tau}+\frac{\partial\chi}{\partial\tau},
\end{equation}
the spatial components yield 
\begin{equation}
\begin{split}
	\nabla_a T^{ai}=&\epsilon^3(\rho_0+p_0) [\nabla_i P+\partial_\tau v_i +v^j\nabla_j v_i-\nu(\nabla^j\nabla_j v_i+R_{ij}v^j)\\
	&+\partial_\tau a_i+v^j f_{ji}]+\mathcal{O}(\epsilon^5),
\end{split}
\end{equation}
where
\begin{equation}
	\nu=\eta/(\rho_0+p_0)
\end{equation}
is the kinematical viscosity, $f_{ij}\equiv\partial_ia_j-\partial_ja_i$ is the field strength of $a_i$ (equivalently $\hat a_i$) and $R_{ij}$ is the Ricci tensor of $g_{ij}$. If a fluid carries conserved charges, equation \eqref{fluid charge current conservation} will also yield incompressibility \eqref{incompressibility scaling}. Taking the hydrodynamic limit $\epsilon\rightarrow0$, these give the non-relativistic incompressible Navier-Stokes (INS)\footnote{INS will refer to the non-relativistic incompressible Navier-Stokes equations \emph{without} forcing terms, i.e.\ when $a_i = 0$ in \eqref{E}.} equations with a forcing function due to an external electromagnetic field,
\begin{subequations}
\begin{gather}
\partial_\tau v_i-\nu(\partial^2 v_i+R_{ij}v^j)+\partial_i P+v^j \partial_j v_i=-\partial_\tau a_i-v^j f_{ji} \label{Ei}\\
\partial_iv^i=0. \label{E0}
\end{gather}
\label{E}%
\end{subequations}
Ideal fluids are described by Euler's equations, obtained by setting $\nu=0$. We will mostly be concerned with fluid flow on flat space ($R_{ij}=0$) in the absence of external forces, where $H_{ab}$, thus $a_i$, are zero.

\subsection{Open problems in the Navier-Stokes equations}

Despite extensive and successful application in physics and engineering, and two hundred years of intense mathematical study, the Navier-Stokes equations still obscure a great number of secrets. In particular, the questions of existence, uniqueness and regularity, see \cite{Constantin2001}.

Consider a solution $v^j(\tau,x^i),P(\tau,x^i)\in C^\infty(\mathbb{R}^d)$ to the incompressible Navier-Stokes equations in some domain $\Gamma\in\mathbb{R}^d$, initially at time $\tau=\tau_0$. Existence refers to the condition that the kinetic energy $v^2/2$ remains bounded for all time. In fact, we expect for finite mean dissipation, solutions to decay to equilibrium for $\tau\rightarrow\infty$. Uniqueness refers to the question of whether $v^j(\tau,x^i)$ is uniquely determined by $v^j(\tau_0,x^i)$ for all $\tau\geq \tau_0$, and regularity is concerned with, for an isolated system, the most general conditions for smooth solutions to exist, which may be expressed in terms of the mean-square of the vorticity $\omega_{ij}=\partial_iv_j-\partial_jv_i$, as whether
\begin{equation}
\int_{\tau_0}^{\tau_1\geq\tau_0} \ud \tau \left[\int_\Gamma |\omega(\tau,x^i)|^2\right]^2<\infty. \label{regularity}
\end{equation}
For $d=2$, the vorticity is self-parallel, and resulting conservation laws fix the vorticity to remain finite, so solutions are smooth. Moreover, it is known that solutions are unique \cite{ladyzhenskaya1969mathematical}.

\section{The fluid/gravity correspondence}
\label{sec:The fluid/gravity correspondence}

In 1974, Damour \cite{Damour}, and later in 1986, Thorne et al. \cite{Thorne1986}, considered an observer outside a black hole, interacting with (perturbing) the event horizon. Surprisingly, they found that the observer will experience perturbations of the ``stretched" horizon (no observer can exist at the horizon itself) described by modes of a viscous fluid possessing electric charge and conductivity. While this model is just that, a model, it did inspire investigation into whether a more concrete correspondence could be found. This began when Policastro et al.\ \cite{Policastro:2001yc} related the shear viscosity of $\mathcal{N}=4$ super Yang-Mills theory to the absorption of energy by a black brane (a further set of coefficients were subsequently similarly derived in \cite{Baier:2007ix}). This was carried out in the context of the anti-de Sitter/conformal field theory correspondence, an equivalence between a theory of gravity in anti-de Sitter space and a conformal field theory in one dimension less \cite{Maldacena:1997re}.

This signalled a major emergence in holography, the correspondence between gravitational theories on $d$-dimensional manifolds and $(d-1)$-dimensional quantum field theories dual to the the dynamics on a hypersurface within. The correspondence holds independent of the coupling of the QFT, though in the strong coupling limit, the gravitational dual becomes classical. Motion of the hypersurface along the bulk dimension parametrizes the renormalisation group (RG) flow, or energy scaling of the dual QFT.

What was now understood was that this holographic correspondence can be viewed in the long-wavelength limit of the dual QFT, where one recovers a dual hydrodynamics. In particular, the long-wavelength limit can be expressed as the set of hydrodynamic solutions varying slowly on a scale set by the extrinsic curvature of the hypersurface. This motivated what is now known as the fluid/gravity correspondence. The gravitational dual to the thermal state is given by the structure of the spacetime, for example in AdS space by an AdS black hole. As one expects from the UV/IR connection \cite{susskind1998holographic}, where infrared effects in the bulk emerge as ultraviolet effects in the boundary theory, the relativistic hydrodynamics of perturbations about this thermal state determine the bulk gravity solution in a gradient expansion off the hypersurface. The corresponding metric construction was developed initially in \cite{Bhattacharyya:2008jc}.

\subsection{The Navier-Stokes fluid on a Rindler boundary}
\label{sec:The Navier-Stokes fluid on a Rindler boundary}

A metric dual to the INS equations was first developed in \cite{Bredberg:2011jq} on the Rindler wedge, up to third order in the non-relativistic, small amplitude expansion detailed later in this section. An algorithm for generalising this metric to all orders was subsequently developed in \cite{Compere:2011dx}, though terms calculated beyond third order are not universal. They receive corrections from quadratic curvature in Gauss-Bonnet gravity \cite{Cai:2011xv}. The construction of general metrics dual to relativistic fluids is set up in \cite{Pinzani-Fokeeva:2014cka}. We summarise the construction in \cite{Compere:2011dx} here.

\begin{SCfigure}\centering
\caption[The Rindler spacetime dual to a thermal state in global equilibrium. Long-wavelength perturbations of the spacetime are dual at the hypersurface $\Sigma_c$ to a codimension one fluid]{The past $\mathcal{H}^-$ and future $\mathcal{H}^+$ horizons define the boundary of the Rindler wedge. Grey lines demonstrate lines of constant $r$ (curved) and $\tau$ (straight). Long-wavelength perturbations of the hypersurface $\Sigma_c$ are described by the equations of hydrodynamics (image from \cite{Compere:2012mt}).\newline\newline\newline\newline}
\includegraphics[scale=0.3]{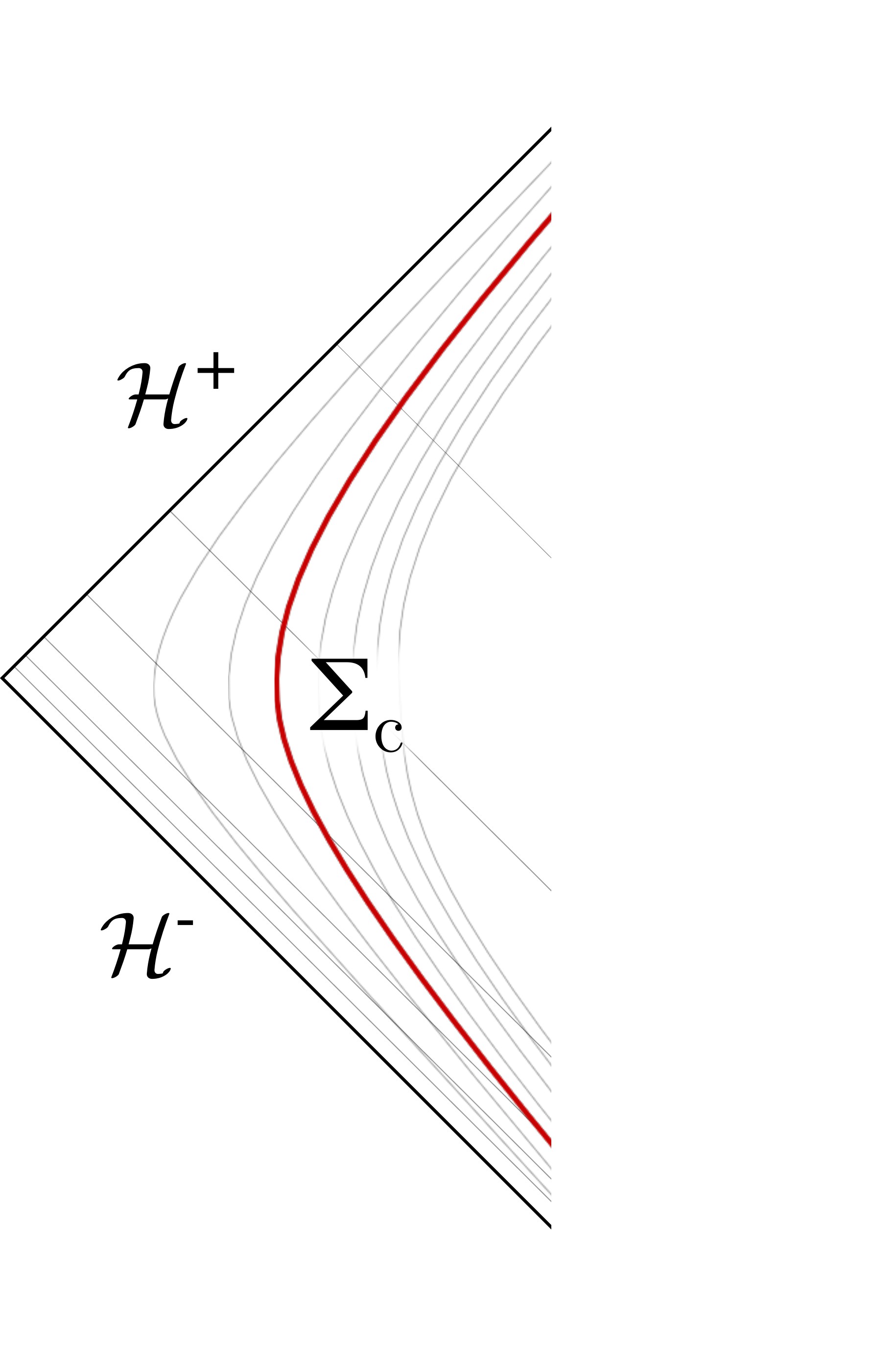}\label{fig:Rindler space}
\end{SCfigure}

Consider the $(d-1)$-dimensional surface $\Sigma_c$ with induced metric
\begin{equation}
y_{ab} \ud x^a \ud x^b		=		-r_c \ud \tau^2		+		\ud x_i \ud x^i,
\label{metric on Sigma_c}
\end{equation}
where the parameter $r_c$ is an arbitrary constant. One metric embedding this surface is
\begin{equation}
\ud s^2		=		- r \ud \tau^2		+		2 \ud\tau \ud r		+		\ud x_i \ud x^i,
\end{equation}
which describes flat space (fig. \ref{fig:Rindler space})
in ingoing Rindler coordinates $x^\mu=(\tau,x^i,r)$, defined in terms of the Cartesian chart $(t,x^i,z)$ by
\begin{equation}
z^2 - t^2 = 4r
\qquad
z + t = e^{\tau/2}.
\label{rindler coordinates in terms of cartesian coordinates}
\end{equation}
The hypersurface $\Sigma_c$ is defined by  $r = r_c$ where $r$ is the coordinate into the bulk. Allowing for a family of equilibrium configurations, consider diffeomorphisms satisfying the three conditions
\begin{itemize}
\item[i)] The induced metric on the hypersurface $\Sigma_c$ takes the form \eqref{metric on Sigma_c}.
\item[ii)] The stress tensor on $\Sigma_c$ describes a perfect fluid.
\item[iii)] Diffeomorphisms return metrics stationary and homogeneous in $(\tau,x^i)$.
\end{itemize}
The allowed set is given by the following boost, shift and rescaling of $x^\mu$. First, a constant boost $\beta_i$,
\begin{equation}
\sqrt{r_c} \tau		\rightarrow		\gamma(\sqrt{r_c}\tau - \beta_i x^i),
\qquad
x^i		\rightarrow		x^i		-		\gamma \beta^i \sqrt{r_c} \tau		+		(\gamma-1)\frac{\beta^i\beta_j}{\beta^2}x^j,
\label{rindler diffeomorphism: boost wrt beta}
\end{equation}
where $\gamma=(1-\beta^2)^{-1/2}$ and $\beta_i\equiv {r_c}^{-1/2}v_i$. Second, a shift in $r$ and a rescaling of $\tau$,
\begin{equation}
r\rightarrow r-r_h,
\qquad
\tau\rightarrow(1-r_h/r_c)^{-1/2}\tau.
\label{rindler diffeomorphism: shift in r and rescaling of tau}
\end{equation}
These yield the flat space metric in rather complicated coordinates,
\begin{equation}
\ud s^2		=
\begin{aligned}[t]
	&	\frac{\ud \tau^2}{1-v^2/r_c}
		\left(
			v^2-\frac{r-r_h}{1-r_h/r_c}
		\right)
		- \frac{2\gamma}{\sqrt{1-r_h/r_c}}\ud\tau\ud r
		- \frac{2\gamma v_i}{r_c\sqrt{1-r_h/r_c}}\ud x^i\ud r
	\\
	&	+ \frac{2 v_i}{1-v^2/r_c}
		\left(
			\frac{r-r_c}{r_c-r_h}
		\right)\ud x^i \ud\tau
		+ \left(
			\delta_{ij}-\frac{v_i v_j}{r_c^2(1-v^2/r_c)}
			\left(
				\frac{r-r_c}{1-r_h/r_c}
			\right)
		\right)
		\ud x^i\ud x^j.
\end{aligned}
\label{flat space metric after diffeomorphism}
\end{equation}

The Brown-York stress tensor on $\Sigma_c$ (in units where $16\pi G=1$) is given by
\begin{equation}
T_{ab}=2(Ky_{ab}-K_{ab}),
\end{equation}
where
\begin{equation}
K_{ab}=\dfrac{1}{2}(\mathcal{L}_n y)_{ab},\qquad K=K^a{}_a,
\end{equation}
are the extrinsic curvature and its mean, and $n^\mu$ is the spacelike unit normal to the hypersurface.

By imposing that the Brown-York stress tensor on $\Sigma_c$ gives that of the stress-energy tensor of a fluid we can identify the parameters of the metric \eqref{flat space metric after diffeomorphism} with the density $\rho$, pressure $p$ and four-velocity $u^a$ of a fluid, as follows:
\begin{equation}
\rho=0, \qquad p=\frac{1}{\sqrt{r_c-r_h}}, \qquad u^a=\frac{1}{\sqrt{r_c-v^2}}(1,v_i).
\end{equation}
The Hamiltonian constraint
\begin{equation}
R_{\mu\nu}n^\mu n^\nu=0
\end{equation}
on $\Sigma_c$ yields a constraint on the Brown-York stress tensor
\begin{equation}
(d+2)T_{ab}T^{ab}=(T^a_a)^2.
\end{equation}
When this constraint is applied to the equilibrium configurations described above, one finds the equation of state is $\rho=0$ (as above), or $\rho=-2(d+2)(d+1)p$ which occurs for a fluid on the Taub geometry \cite{Eling:2012ni}. 

Promoting $v_i$ and $p$ to slowly varying functions of the coordinates $x^a$, and regarding $v^i(x^a)$ and $p=r_c^{-1/2}+r_c^{-3/2}P(x^a)$ as small perturbations scaling as
\begin{equation}
v_i\sim\epsilon,
\qquad
P\sim\epsilon^2,
\label{hydrodynamic scaling of v^i and P}
\end{equation}
about equilibrium, yields the metric
\begin{equation}
\begin{split}
	\ud s^2 =& -r\ud\tau^2 + 2\ud\tau \ud r + \ud x^i \ud x_i \\
	&- 2\left(1-\frac{r}{r_c}\right)v_i \ud x^i \ud\tau - 2\frac{v_i}{r_c}\ud x^i \ud r \\
	& +\left(1-\frac{r}{r_c}\right)\left[(v^2+2P)\ud\tau^2 + \frac{v_i v_j}{r_c}\ud x^i \ud x^j\right] + \left(\frac{v^2+2P}{r_c}\right)\ud\tau \ud r +\mathcal{O}(\epsilon^3)
\end{split}
\end{equation}
which satisfies the Einstein's equations to $\mathcal{O}(\epsilon^2)$ if $v_i$ satisfies incompressibility, $\partial_i v^i=\mathcal{O}(\epsilon^3)$. Corrections appear in powers of $\epsilon^2$, so this is the complete metric to second order. 

The metric may now be built up order by order in the hydrodynamic scaling. Assume one has the metric at order $\epsilon^{n-1}$, where the first non-vanishing component $\hat R_{\mu\nu}^{(n)}$ of the Ricci tensor appears at order $n$. By adding a correction term $g_{\mu\nu}^{(n)}$ to the metric at order $n$, resulting in a shift in the Ricci tensor $\delta R_{\mu\nu}^{(n)}$, we can ensure the vanishing of the Ricci tensor is guaranteed to order $n$ if
\begin{equation}
\hat R_{\mu\nu}^{(n)} + \delta R_{\mu\nu}^{(n)}		=		0	. \label{Ricci(n) + delta Ricci(n) = 0}
\end{equation}
Recalling that, in the hydrodynamic scaling, derivatives scale as
\begin{equation}
\partial_r\sim\epsilon^0,
\qquad
\partial_i\sim\epsilon^1,
\qquad
\partial_\tau\sim\epsilon^2,
\label{hydrodynamic expansion scaling of bulk coordinates}
\end{equation}
one sees that corrections $\delta R_{\mu\nu}^{(n)}$ at order $n$ will appear only as $r$ derivatives of $g_{\mu\nu}^{(n)}$. It is shown in \cite{Compere:2011dx} that, using the Bianchi identity and the Gauss-Codacci relations, integrability of the set of differential equations \eqref{Ricci(n) + delta Ricci(n) = 0} defining $\delta R_{\mu\nu}^{(n)}$ in terms of $g_{\mu\nu}^{(n)}$ is given by imposing the momentum constraint, equivalent to the conservation of the stress tensor on $\Sigma_c$,
\begin{equation}
R_{a\mu}n^\mu=\nabla_a T^{ab}|^{(n)}_{\Sigma_c}=0,
\end{equation}
which is precisely the fluid equations of motion, to order $n$.

The perturbation scheme contains several degrees of freedom. The gauge freedom of the infinitesimal perturbations
\begin{equation}
g_{\mu\nu}^{(n)}\rightarrow g_{\mu\nu}^{(n)}+\partial_\mu \varphi_\nu^{(n)}+\partial_\nu \varphi_\mu^{(n)}
\end{equation}
for some arbitrary vector $\varphi^{\mu(n)}(\tau,\vec x,r)$ at order $\epsilon^n$, which may be fixed by demanding that $g_{r\mu}$ is that of the seed metric to all orders in $\epsilon$. The $x^a$-dependent functions of integration from equation \eqref{Ricci(n) + delta Ricci(n) = 0} may be fixed by imposing the boundary form \eqref{metric on Sigma_c} of the metric on $\Sigma_c$, and also requiring regularity of the metric at $r=0$, which in this construction translates to the absence of logarithmic terms in $r$. Corrections to the bulk metric under these conditions then become
\begin{equation}\begin{split}
g_{r\mu}^{(n)}=& 0\\
g_{\tau\tau}^{(n)}=& (1-r/r_c)F_\tau^{(n)}(x^a)+\int_r^{r_c}\,\ud r' \int_{r'}^{r_c}\,\ud r'' (\hat R_{ii}^{(n)}-r\hat R_{rr}^{(n)}-2\hat R_{r\tau}^{(n)})\\
g_{\tau i}^{(n)}=& (1-r/r_c)F_i^{(n)}(x^a)-2\int_r^{r_c}\,\ud r'\int_{r'}^{r_c}\,\ud r'' \hat R_{ri}^{(n)}\\
g_{ij}^{(n)}=& -2\int_r^{r_c}\,\ud r' \frac{1}{r'}\int_0^{r'} \,\ud r''\hat R_{ij}^{(n)},
\end{split}\end{equation}
where the $F_a^{(n)}(x^b)$ comprise of the remaining integration functions, and the final degree of freedom; field redefinitions $\delta v_i^{(n)}$ and $\delta P^{(n)}$ at order $\epsilon^n$. $F_i^{(n)}$ is related to redefinitions of the fluid velocity and is fixed by the isotropic gauge condition $P_a^bT_{bc}u^c=0$. $F_\tau^{(n)}$ is related to redefinitions of the pressure and is fixed by defining the isotropic part of $T_{ij}$ to be
\begin{equation}
T_{ij}^\text{isotropic}=\left(\frac{1}{\sqrt{r_c}}+\frac{P}{r_c^{3/2}}\right) \delta_{ij}
\end{equation}
to all orders.

Applying the perturbation scheme to the seed metric yields to third order
\begin{equation}
\begin{split}
\ud s^2 =&  -r\ud\tau^2 +2\ud\tau \ud r + \ud x^i \ud x_i - 2\left(1-\frac{r}{r_c}\right)v_i \ud x^i \ud\tau - 2\frac{v_i}{r_c}\ud x^i \ud r \\
& +\left(1-\frac{r}{r_c}\right)\left[(v^2+2P)\ud\tau^2 + \frac{v_i v_j}{r_c}\ud x^i \ud x^j\right] + \left(\frac{v^2+2P}{r_c}\right)\ud\tau \ud r \\
& -\left[\frac{(r^2-r_c^2)}{r_c}\partial^2 v_i + \left(1-\frac{r}{r_c}\right)\left(\frac{v^2+2P}{r_c}\right)v_i\right]\ud x^i \ud\tau + \mathcal{O}(\epsilon^4),
\end{split}
\label{metric dual to the Navier-stokes fluid to epsilon^3}
\end{equation}
which satisfies the vacuum Einstein equations if
\begin{equation}
r_c^{3/2}\nabla^a T_{ai}|_{\Sigma_c}=\partial_\tau v_i-r_c\partial^2 v_i+\partial_i P+v^j\partial_j v_i=\mathcal{O}(\epsilon^5) ,
\end{equation}
which are the Navier-Stokes equations with kinematical viscosity
\begin{equation}
\nu=r_c.
\end{equation}
We will refer to \eqref{metric dual to the Navier-stokes fluid to epsilon^3} as the \emph{fluid metric}, not to be confused with the metric \eqref{metric on Sigma_c} of the space on which the fluid evolves.

Higher order corrections to the Navier-Stokes equations follow from conservation of the stress tensor on $\Sigma_c$. Vector and scalar quantities are at odd and even orders respectively in the scaling $\epsilon$. Accordingly, corrections to the scalar incompressibility equation appear at even orders, and to the vector Navier-Stokes equations at odd orders.

\section{Duality in the context of holography}
\label{sec:Duality in the context of holography}

The defining equations in general relativity are the Einstein field equations, and in the non-relativistic limit of hydrodynamics, the Navier-Stokes equations \eqref{E}. Each is a set of non-linear partial differential equations whose solutions exhibit fantastically varied phenomenology. When approaching any complex physical system with a view to finding solutions, it is often advantageous to consider the symmetries, intensively studied in both of these systems since their conceptions. Beyond diffeomorphisms, the search in gravity has in general been somewhat limited \cite{Torre:1993jm, Kramer1980}, however in the presence of a spacetime isometry, the symmetry group becomes remarkably large \cite{maison2000duality}, particularly for vacuum spacetimes. For symmetries of the Navier-Stokes equations see \cite{gusyatnikova1989symmetries}, and with regards to the conformal group \cite{Bhattacharyya:2008kq, Horvathy:2009kz}. In light of the fluid/gravity correspondence, one may ask whether the symmetries of these systems are linked.

In \cite{stewart1982generalisation, Berger1987, Stephani1988, Garfinkle:1996ur, Racz:1996ei, Racz:1997wq, Zsigrai01082000, Boonserm:2006vr}, the authors apply known symmetries of the Einstein equations to spacetimes with perfect fluid sources in the presence of one or two Killing vectors, constructing new spacetimes with fluids possessing the same equation of state. Our approach will differ from theirs in that the fluid we will consider is no longer a perfect fluid evolving in the $d$-dimensional bulk spacetime, but is a holographically dual dissipative fluid on a flat $(d-1)$-dimensional hypersurface. Meanwhile, rather than employing the symmetries of Einstein gravity with a perfect fluid source, we will employ invariance transformations of the vacuum Einstein equations and holographically project these to the dual fluid to find invariance transformations of the Navier-Stokes fluid. It is in this way that we look for dualities between solutions to the Navier-Stokes equations arising from dualities between solutions to the vacuum Einstein equations: dual metrics yield dual fluid configurations.

To do this, we will will impose an isometry on the fluid metric, and then act on the metric with the generalised Ehlers group defined in terms of this isometry, which does not in general generate vacuum metrics from vacuum metrics, but does contain such transformations. At the same time, we will preserve certain aspects of the fluid metric \eqref{metric dual to the Navier-stokes fluid to epsilon^3} such as the embedded flat background on which the fluid evolves. Despite the fact that we will not have explicitly demanded that the transformed metric is vacuum, we will find that it is regardless, and that no additional constraints are required for the transformed fluid to satisfy the incompressible Navier-Stokes equations. We will discuss this unexpected result in \S\ref{sec:Are the metric transformations vacuum to vacuum?}.

\section[The generalised Ehlers transformation in the fluid/gravity correspondence]{The generalised Ehlers transformation in the \\ fluid/gravity correspondence}
\label{sec:The generalised Ehlers transformation in the fluid/gravity correspondence}

\subsection*{The generalised Ehlers group}

We will use a solution-generating technique detailed in \cite{Mars:2001gd}, which is similar to that in chapter \ref{ch:Solution-generating transformations in NS-NS supergravity and DFT}, in that it preserves the conformally rescaled reduced metric $F\gamma$ defined in \eqref{metric on V_d-1}, orthogonal to a Killing vector $\xi$. However, it has the crucial difference that it allows the use of null Killing vectors. For Killing vector\footnote{The generalised Ehlers transformation as a group is actually defined in \cite{Mars:2001gd} for any vector $\xi$. However, we will in our analysis restrict to the assumption it is a Killing vector. We discuss this choice in \S\ref{sec:Are the metric transformations vacuum to vacuum?}.} $\xi = \xi^\mu \partial_\mu$ and one-form $W = W_\mu \ud x^\mu$, the generalised Ehlers transformation is given by
\begin{equation}
g_{\mu\nu}
	\rightarrow		h_{\mu \nu} (\xi,W,g)
	=				\Omega^2 g_{\mu\nu}		- 2 \xi_{(\mu} W_{\nu)}		+ \frac{ F }{\Omega^2} W_\mu W_\nu,
\label{generalised Ehlers transformation}
\end{equation}
where $\Omega^2 = 1 + \xi^\mu W_\mu \geq 1 $, and the transformation defines a group. As such, the group does not in general send vacuum metrics to vacuum metrics, but does contain such transformations.

\subsection{Introducing the transformed fluid metric ansatz}
\label{sec:Introducing the transformed fluid metric ansatz}

The Rindler metric \eqref{metric dual to the Navier-stokes fluid to epsilon^3} dual to the incompressible Navier-Stokes fluid is defined in terms of the fluid velocity, pressure, and the hypersurface position within the bulk, $(v_i(x^a), P(x^a), r_c)$. Let us consider those transformed metrics $h(\xi,W,g)$ which have the same functional form as \eqref{metric dual to the Navier-stokes fluid to epsilon^3}, but now in terms of a transformed set $(v'_i(x^a), P'(x^a), r'_c)$. On satisfying the vacuum Einstein equations on $\Sigma'_c$, now at $r = r'_c$ in the new geometry, the metric $h$ will yield the incompressible Navier-Stokes equations in the transformed set
\begin{subequations}
\begin{gather}
\partial_i v'_i=0
\\
\partial_\tau v'_i		+ \partial_i P'		+ v'_k \partial_k v'_i		- r'_c \partial^2 v'_i=0.
\end{gather}
\label{incompressible Navier-stokes equations in v'_i, P', r'_c}%
\end{subequations}
Crucially, if $(v_i,P)$ satisfy the Navier-Stokes equations with viscosity $\nu=r_c$, and if the metric transformation is vacuum to vacuum, the transformed velocity and pressure $(v'_i, P')$ represent a new set of solutions for viscosity $\nu=r'_c$. We will confine our search for vacuum to vacuum metric transformations to those that lie within the generalised Ehlers map. Thus, we look for a subset of the generalised Ehlers transformation acting on the fluid metric \eqref{metric dual to the Navier-stokes fluid to epsilon^3}, obeying some Killing isometry, which corresponds to solution-generating transformations of the velocity and pressure, and RG flow parametrised by $r_c$, of an incompressible Navier-Stokes fluid in arbitrary dimension.

Now the Rindler metric is just one fluid metric supporting flat background geometries on the boundary. Rather than demanding we preserve the \emph{full} form of the fluid metric, we therefore retain only some common features of such metrics; the metric gauge $g_{\mu r}$, and the flat boundary metric of the form \eqref{metric on Sigma_c}. This data is sufficient to define both the fluid to all orders and the background upon which it evolves. The equation we wish to solve is thus
\begin{equation}
g(v_i,P,r_c)_{\mu\nu} \rightarrow h(\xi,W,g)_{\mu \nu} = g'(v'_i, P', r'_c)_{\mu\nu},
\label{g(r_c, v, P) rightarrow h(xi, W, g) = g'(v', P', r'_c)}
\end{equation}
where
\begin{subequations}
\begin{gather}
g'_{\tau r}		=		1		+ \frac{(v')^2 + 2P'}{2r'_c},
\qquad
g'_{ir}		=		- \frac{v'_i}{r'_c},
\qquad
g'_{rr}		=		0,
\\
g'_{ab}|_{r'_c}		=		y'_{ab},
\qquad
\text{where}
\qquad
y'_{\tau\tau}		=		- r'_c,
\qquad
y'_{ai}		=		y_{ai}.
\end{gather}
\label{required properties of transformed metric}%
\end{subequations}

\subsection{Deriving the transformation on the fluid}
\label{sec:Deriving the transformation on the fluid}

Equation \eqref{required properties of transformed metric} provides us with sufficient information to derive the possible fluid transformations via the form of the one-form $W$. The parameters of the transformed fluid are determined, to all orders in $\epsilon$, by $g'_{ar}=g'_{ar}|_{r'_c}$. Consequently, the fluid transformations will be given by the transformation of these components. Evaluation at $r'_c$ is necessary in order to circumvent the ambiguity in the dual metric and provide explicit fluid transformations. Before we specialise to the cases of null and non-null Killing vectors, we note the relations
\begin{align}
\frac{F}{\Omega^2} W_\mu		& =		\xi^\nu(g_{\mu\nu}-g'_{\mu\nu}),
\label{ehler*xi}
\\
F/\Omega^2		& =		\xi^\mu\xi^\nu g'_{\mu\nu},
\label{ehler*xi*xi}
\end{align}
valid for all Killing vectors, found by contraction of \eqref{generalised Ehlers transformation} once and twice respectively with the Killing vector.

Now, for non-null Killing vectors, the transformation \eqref{generalised Ehlers transformation} for the vanishing components of the metric $g'_{rr} = g_{rr} = 0$ fixes
\begin{equation}
W_r		=		2 \alpha \xi_r \Omega^2 / F
\qquad
\text{where}
\quad
\alpha=0,1.
\label{W_r from g'_rr = g_rr = 0}
\end{equation}
Also, one may obtain an expression for $W_a$ by contracting \eqref{generalised Ehlers transformation} with the boundary indices $(a,b,\ldots)$ of the Killing vector\footnote{Here and in what follows $\xi_\mu= g_{\mu \nu} \xi^\nu$, i.e.\ indices are raised and lowered with the metric $g$ and never with $g'$.}:
\begin{equation}
W_a		=		\frac{ \Omega^2 \xi^r (g_{ar} - \xi_a\xi_r/F) + \xi^b g'_{ab}}{-F/\Omega^2+(1-2\alpha)\xi^r\xi_r}		+ \frac{\Omega^2\xi_a}{F}.\label{Wa@wrc}
\end{equation}
Note: this expression is uniquely defined only at the hypersurface $\Sigma'_c$ of the transformed geometry, where we have defined the form of $g'_{ab}$ and $W_a$ becomes independent of the transformed fluid velocity and pressure. We also have
\begin{equation}
\xi^a(g'_{ar}-(1-2\alpha)g_{ar})=0,\label{xidgar}
\end{equation}
derived by comparing \eqref{ehler*xi} and \eqref{W_r from g'_rr = g_rr = 0}. Inserting $W_r$ \eqref{W_r from g'_rr = g_rr = 0} and $W_a$ \eqref{Wa@wrc} into the generalised Ehlers transformation \eqref{generalised Ehlers transformation} and employing \eqref{xidgar} and \eqref{ehler*xi*xi}, one finds, on evaluating at $r'_c$,
\begin{equation}
g'_{ar}=\left[\frac{ F g_{ar}+\xi_r((1-2\alpha)\xi^b y'_{ab}-\xi_a)}{\xi^c\xi^d y'_{cd}+(1-2\alpha)\xi^r\xi_r} \right]_{r'_c}.\label{g_artransformation}
\end{equation}

Meanwhile, for Killing vectors null at $r'_c$: $F|_{r'_c}=0$, we find \eqref{ehler*xi} gives
\begin{equation}
g'_{ar}		=		g_{ar}		+ \left[ \frac{\xi^b(g_{ab}-y'_{ab})}{\xi^r}\right]_{r'_c},
\label{g_artransformation-null}
\end{equation}
alongside the requirement the Killing vector remains null in the transformed geometry $\xi^\mu\xi^\nu g'_{\mu\nu}=0$.

\subsection{Energy scaling invariance from an isometry into the bulk}
\label{sec:Energy scaling invariance from an isometry into the bulk}

We begin with an example of a (null) Killing vector into the bulk, $\xi=\xi^r(x^\mu)\partial_r$. The Killing equations require the integrability condition 
\begin{equation}
\partial_{[i}v_{j]}		=		\mathcal{O}(\epsilon^4),
\label{integrability condition for radial killing vector}
\end{equation}
while transformation \eqref{g_artransformation-null} yields $g'_{ar}=g_{ar}$, or
\begin{equation}
v'_i		=		\frac{r'_c}{r_c} v_i
\qquad
\frac{1}{2} (v')^2		+ P'		=		\frac{r'_c}{r_c} \left( \frac{1}{2} v^2		+ P \right)		,
\label{SGT from radial Killing vector}
\end{equation}
which is exact to all orders in $\epsilon$. Since rescaling $r_c$ allows for renormalisation group flow of the hypersurface position into the bulk, it is unsurprising that this results in a rescaling of the fluid energy $v^2/2 + P$. That this rescaling is exact to all order appears however to be non-trivial. If one were to calculate the integrability condition for the Killing vector order by order in $\epsilon$, then solutions to these equations along with the incompressible Navier-Stokes equations to the corresponding order, will satisfy the exact rescaling invariance \eqref{SGT from radial Killing vector} to all orders.

For example, at leading order, the general solution to the integrability condition and INS equations is potential flow defined in terms of a potential $q(x^a)$,
\begin{equation}
v_i		=		\partial_i q,
\qquad
P		=		- \partial_\tau q		- \frac{1}{2} (\partial_i q) (\partial_i q)		+ P_0(\tau),
\end{equation}
where $P_0(\tau)$ is an arbitrary function of $\tau$, and the INS equations reduce to
\begin{equation}
\partial^2 q		=		0	.
\label{Laplace equation in q}
\end{equation}
Meanwhile, the fluid transformation \eqref{SGT from radial Killing vector} becomes
\begin{equation}
q		\rightarrow		\frac{r'_c}{r_c} q
\qquad
P_0		\rightarrow		\frac{r'_c}{r_c} P_0,
\label{SGT from radial Killing vector: the potential flow case}
\end{equation}
and thus it is trivial that the pair $(v'_i, P')$ satisfy the fluid equations with if $(v_i,P)$ do so due to the scale invariance of the Laplace equation \eqref{Laplace equation in q} in $q$. It is interesting to consider the problems of existence, uniqueness and regularity of the Navier-Stokes in this case. Vanishing mean square vorticity ensures the class of solutions $(v_i,P)$ generated by \eqref{SGT from radial Killing vector: the potential flow case} are regular. With respect to existence, the kinetic energy $v^2$ scales by a factor $(r'_c/r_c)^2$ and thus is bounded for finite $r'_c$ and smooth $q$.

\subsection*{The timelike Killing vector}

One might expect, in the presence of a timelike Killing vector $\xi=\partial_\tau$ (it is sufficient for this discussion to consider stationary solutions) a duality of the form
\begin{equation}
v'_i	=	- v_i	,
\qquad
P'		=	P		,
\qquad
r'_c	=	-r_c
\end{equation}
enacting time-reversal of the fluid, but this is not the case. This is explained by noting that time-reversal is enacted by redefining the viscosity by $\nu=\pm r_c$, see \cite{Bredberg:2011jq}, rather than by changing $r_c$ itself. This is because sending $r_c\rightarrow -r_c$ brings the fluid outside the causal region of the spacetime.

\subsection{Fixed viscosity \texorpdfstring{$\mathbb{Z}_2$}{Z2} transformations}
\label{sec:Fixed viscosity mathbb Z_2 transformations}

We turn to fixed boundary (viscosity) transformations, where $r'_c = r_c$. If $\alpha=0$, or for Killing vectors null at the dual boundary, one recovers the identity. For non-null Killing vectors with $\alpha=1$, one finds
\begin{equation}
g'_{ar}		=		g_{ar}		- 2\xi_r \left[ \frac{ \xi^b y_{ab} - \xi^r g_{ar}}{ \xi^c\xi^d y_{cd} - \xi^r\xi_r} \right]_{r_c},
\label{transformation of g_ar for fixed rc}
\end{equation}
which defines a $\mathbb{Z}_2$ group.

\subsubsection{Spacelike Killing vectors} \label{sec:spacelike}

Consider a generic space-like Killing vector $\xi=\xi^k\partial_k$. Under \eqref{transformation of g_ar for fixed rc}, the pressure is preserved, while the velocity transforms as
\begin{equation}
v'_i		=		v_i - 2\left[ \xi_i \frac{\sum_k\xi^k v_k}{\sum_j(\xi^j)^2} \right]_{r_c},
\label{transformation of v_i for xi = xi^k partial_k}
\end{equation}
which is a reflection in the hyperplane normal to the Killing vector (evaluated at the hypersurface) and containing the point $x^i$ at which the velocity is defined.

\subsubsection*{Translational isometry}

Consider $\xi = c_k\partial_k$ where the constants $c_k$ are normalised to $\sum_k c_k^2=1$. The corresponding isometries are $c_k\partial_k v_i	=0$ and $c_k\partial_k P=0$. The pressure is preserved, while the velocity transforms as
\begin{equation}
v'_i		=		v_i		- 2c_i c_k v_k.
\label{reflection transformation of v_i for constant xi = xi^k partial_k}
\end{equation}
The incompressibility condition
\begin{align}
\partial_i v'_i		=		\partial_i v_i		- 2c_i c_k \partial_i v_k		=		0
\label{incompressibility for v' w xi = partial_1}
\end{align}
and Navier-Stokes equations
\begin{equation}
\begin{split}
	\partial_\tau v'_i		+ \partial_i P'		+ v'_k \partial_k v'_i		- r_c \partial^2 v'_i
		=	&	(\delta_{ik}-2c_i c_k)(\partial_\tau v_k		+ \partial_k P		+ v^j\partial_j v_k		- r_c\partial^2 v_k)
		\\
			&	+ 2c_ic_k\partial_k P		- 2c_jv_jc_k\partial_k(v_i		- 2c_ic_lv_l)		=		0
\end{split}
\label{NS for v' and P' w xi = partial_1}
\end{equation}
are satisfied by the INS equations in the original fluid parameters along with the isometries.

\subsubsection*{Rotational isometry}

Consider a Killing vector $\xi=-x_2\partial_1+x_1\partial_2$ corresponding to a rotational isometry in the fluid.  In cylindrical coordinates
\begin{equation}
x_1=\rho\cos\theta,
\qquad
x_2=\rho\sin\theta
\qquad
x_k=x_k
	\quad \forall k>2,
\end{equation}
where the Killing vector becomes $\xi=\partial_\theta$, the isometries are
\begin{equation}
\partial_\theta v_1=-v_2
\qquad
\partial_\theta v_2=v_1
\qquad
\partial_\theta v_k=0
	\quad \forall k>2
\qquad
\partial_\theta P=0. \label{angularisometries}
\end{equation}
Solutions satisfy
\begin{equation}
v_1=x_1\mu-x_2\eta
\qquad
v_2=x_2\mu+x_1\eta		,
\label{angularvelocities}
\end{equation}
where $\mu=\mu(\tau,x_{k>2},\rho)$, $\eta=\eta(\tau,x_{k>2},\rho)$, with transformed velocities
\begin{equation}
v'_1		=		x_1\mu		+ x_2\eta
\qquad
v'_2		=		x_2\mu		- x_1\eta		.
\end{equation}
That is, the transformation sends $\eta\rightarrow-\eta$ (equivalently $\theta\rightarrow-\theta$). The incompressible Navier-Stokes equations for the original fluid may be expressed as
\begin{subequations}
\begin{align}
0	& =	(2+\rho\partial_\rho)\mu+ \sum_{k>2} \partial_k v_k\\
0	& =	\begin{aligned}[t]
			\partial_\tau\mu+\rho^{-1}\partial_\rho P+\mu^2-\eta^2 +\mu\rho\partial_\rho\mu &+ \sum_{k>2} v_k \partial_k \mu\\
			&-r_c\left(3\rho^{-1}\partial_\rho\mu+\partial_\rho^2\mu+ \sum_{k>2} \partial^k \partial_k \mu\right)
		\end{aligned}\\
0	& =	\partial_\tau\eta+2\mu\eta+\mu\rho\partial_\rho\eta+ \sum_{k>2} v_k \partial_k \eta -r_c\left(3\rho^{-1}\partial_\rho\eta+\partial_\rho^2\eta+ \sum_{k>2} \partial^k\partial_k\eta\right)\\
0	& =	\partial_\tau v_j		+ \partial_j P		+ \mu\rho\partial_\rho v_j		+ \sum_{k>2} v_k\partial_k v_j		- r_c\rho^{-1}\partial_\rho(\rho\partial_\rho v_j)
	\qquad
	\forall j>2,
\end{align}
\end{subequations}
It is clear from the parity of these equations in $\eta$ that if there exists a fluid solution defined in terms of a pair $(\mu,\eta)$ by \eqref{angularvelocities}, then there also exists a solution parametrized by the pair $(\mu,-\eta)$. That is, the dual fluid satisfies the INS equations.

We provide an example with the three-dimensional fluid solution
\begin{subequations}
\begin{gather}
v_1=A\left(x_1-x_2e^{-2A(\tau-\tau_0)}\right)\qquad v_2=A\left(x_2+x_1e^{-2A(\tau-\tau_0)}\right)\\
v_3=B\exp\left(4A(\tau-\tau_0)+\frac{A\rho^2}{2r_c}\right)-e^{2A\tau}\int^\tau\ud\tau'\, q(\tau')e^{-2A\tau'}-2A x_3\\
P=\frac{1}{2}A^2\rho^2(e^{-4A(\tau-\tau_0)}-1)+q(\tau)x_3-2A^2 x_3^2
\end{gather}
\end{subequations}
which satisfies the isometries \eqref{angularisometries} (note that this is in general an unphysical solution, presented primarily for illustrative purposes). Here, $A$, $B$ and $\tau_0$ are arbitrary non-vanishing constants and $q(\tau)$ is an arbitrary function of time. The duality is equivalent to sending $\tau_0\rightarrow\tau_0+i\pi/2A$.

\subsection{Are the metric transformations vacuum to vacuum?}
\label{sec:Are the metric transformations vacuum to vacuum?}

We have found solution-generating transformations of an INS fluid by applying the generalised Ehlers transformation with respect to a Killing vector, while requiring a particular ansatz for the the transformed metric at the hypersurface. However, that this procedure has provided us with solution-generating transformations of the dual fluid is somewhat unexpected. After all, we have not yet demanded that the metric transformation takes vacuum metrics to vacuum metrics. Here, we review the metric transformations and whether they are in fact vacuum to vacuum. We do this in the case of four bulk dimensions by considering the potential formalism in \S\ref{ch:Solution-generating transformations in NS-NS supergravity and DFT} (in particular \S\ref{sec:An example: (3+1)-d vacuum Einstein gravity with an isometry}).

However, before we do this, we recall that the generalised Ehlers transformation \eqref{generalised Ehlers transformation} was defined in the literature for arbitrary $\xi$, in particular that it need not be a Killing vector. We briefly discuss whether $\xi$ need be a Killing vector for our fluid transformations to yield valid INS fluid solutions.

\subsubsection*{Spacelike Killing vector $\xi^i \partial_i$}

For the case $\xi = c_k\partial_k$, we can determine whether $\xi$ need be a Killing vector from the equations of motion \eqref{incompressibility for v' w xi = partial_1} and \eqref{NS for v' and P' w xi = partial_1} in the transformed variables. A simple calculation in the frame where $c_k = \delta_k^1$ determines that the INS in the transformed variables are satisfied only if $v_i$ and $P$ are independent of $x^1$. Since there is no other dependence of the fluid metric on $x^1$, this implies that $\partial_1$ must be a Killing vector in the bulk geometry. We have not completed the corresponding calculation for the rotational vector $\xi=-x_2\partial_1+x_1\partial_2$, since our construction of fluid solutions, along with our presentation of the INS equations, relies on $-x_2\partial_1+x_1\partial_2$ being a Killing vector. However, we expect an equivalent result: that the vector $\xi$ must be a Killing vector in the bulk geometry for the transformed fluids to satisfy the INS equations. We now discuss whether the transformed metric is vacuum.

In this case, the norm $F$ and twist $\omega$ are not defined unambiguously for the transformed metric, and so we cannot determine whether the transformations are vacuum to vacuum using these. However, the fluid transformations acts as a reflection in the isometry direction---e.g.\ $\xi = \partial_1$ sends $v_1 \rightarrow -v_1$. It is quite reasonable to expect that this is also the case for the metric---simply a coordinate reflection and therefore a (somewhat trivial) symmetry of the vacuum Einstein equations.

\subsubsection*{A Killing vector into the bulk $\xi = \xi^r\partial_r$}

Again, we first determine whether $\xi$ need be a Killing vector for \eqref{SGT from radial Killing vector} to be a valid fluid transformation. One finds that to leading order in $\epsilon$, the transformed fluid satisfies the INS equations only if $\partial_{[i}v_{j]}\partial_{[i}v_{j]}$ vanishes (assuming finite $r_c$). This is true if and only if $\partial_{[i}v_{j]}$ vanishes, which is the integrability condition for $\xi$ being a Killing vector. That is, the transformed fluid is once again a fluid solution only if $\xi$ is a Killing vector. We now discuss whether the transformed metric is vacuum.

The transformed potentials are in this case determined unambiguously for the transformed metric. The Killing vector is null to all orders for both the metric and the transformed metric by the gauge condition $g_{rr} = 0$. The twist can also be determined by noting
\begin{equation}
	0		= \xi^\mu \omega_\mu		= \xi^\mu \partial_\mu \chi		= \xi^r \partial_r \chi
	\qquad
	\Rightarrow
	\qquad
	\chi = \chi|_{r_c}
\end{equation}
and similarly $\chi' = \chi'|_{r'_c}$. To leading order, a direct calculation yields $\partial_\mu \chi = \partial_\mu \chi' = 0$. Meanwhile, the metric transformation \eqref{generalised Ehlers transformation} preserves the conformally rescaled reduced metric $\hat\gamma = F\gamma$, which is here
\begin{equation}
	\hat\gamma_{\mu\nu}		=		- g_{\mu r} g_{\nu r}		.
\end{equation}
Collecting these results, the transformation we encounter in \S\ref{sec:Energy scaling invariance from an isometry into the bulk}, corresponds to\footnote{One may ask why we have not uncovered the map $g'_{\mu r} = - g_{\mu r}$. This is simply because this map would change the metric signature, while the generalised Ehlers transformation is shown in \cite{Mars:2001gd} to preserve metric signature.}
\begin{equation}
g'_{\mu r} = g_{\mu r}		\qquad		F' = F		\qquad 		\partial_\mu \chi' = \partial_\mu \chi,
\end{equation}
which corresponds to a trivial invariance transformation of the vacuum field equations. However, we saw in \S\ref{sec:Reduction of Einstein gravity with respect to an isometry} that for non-null Killing vectors, trivial invariance transformations correspond to redefinitions of the coordinate defined by the Killing vector, e.g.\ $\tau$ for $\xi = \partial_\tau$. We therefore expect that, under the constraint \eqref{integrability condition for radial killing vector}, the transformation \eqref{SGT from radial Killing vector} applied to the fluid metric is equivalent to a redefinition of the radial coordinate $r$, though we do not offer a proof here. Whether these results hold true in greater than four bulk dimensions has not been determined.

Our brief analysis therefore suggests that all the fluid transformations we have encountered, even if non-trivial from the fluid perspective, are trivial from the gravity perspective.

\subsection{Extension to magnetohydrodynamics}
\label{sec:Extension to magnetohydrodynamics}

The fluid/gravity correspondence has been extended to include various fields and matter content in the bulk spacetime. We look now to applying the solution-generating technique of this section to Einstein-Maxwell theory dual to magnetohydrodynamics (MHD). We use the metric as given in \cite{Lysov:2013jsa}, where they use units with $8\pi G = 1$. The bulk dynamics are governed by the Einstein-Maxwell equations of motion
\begin{equation}
R_{\mu\nu}		=		\frac{1}{4\pi} \left( F_{\mu\rho} F_\nu{}^\rho		-		\frac{1}{2(d-2)} g_{\mu\nu} F_{\sigma\rho} F^{\sigma\rho} \right)
\qquad
\nabla^\mu F_{\mu\nu}		=		0		.
\label{Einstein-Maxwell eoms}
\end{equation}
The field strength additionally satisfies the Bianchi identity
\begin{equation}
\ud F	=		0		.
\end{equation}

Meanwhile, the MHD equations are given, for magnetic conductivity $1/4\pi r_c$ and viscosity $r_c$, by
\begin{equation}
\begin{gathered}
\partial_\tau v_i		+ v^j\partial_j v_i		+ \partial_i \left( P - \frac{1}{16\pi}\frac{d}{ d-2} f^2 \right)		- r_c \partial^2 v_i		=		- \partial^j \pi_{ji}
\\
\pi_{ji}		= \frac{1}{4\pi} \left( f_{jl} f_{il}		- \frac{1}{4} f^2 \delta_{ji} \right)
\qquad
f_{\tau i}		=		- r_c \partial_j f_{ij}		-		v^j f_{ji}
\qquad
\partial_{[a} f_{bc]}		=		0
\qquad
\partial_i v_i		= 0 .
\end{gathered}
\label{MHD equations}
\end{equation}

A similar metric expansion to that outlined in \S\ref{sec:The Navier-Stokes fluid on a Rindler boundary} is employed here, which uses the invariance of the equations under the hydrodynamic scaling \eqref{hydrodynamic expansion scaling of bulk coordinates}, \eqref{hydrodynamic scaling of v^i and P} with
\begin{equation}
f_{ij}		\sim		\epsilon
\qquad
f_{\tau i}		\sim	\epsilon^2,
\end{equation}
and the resulting metric and field strength are given by
\begin{align}
g_{\mu\nu} \ud x^\mu \ud x^\nu	& =
\begin{aligned}[t]
&
	-	r \ud \tau^2
	+	2\ud\tau \ud r
	+	\ud x^i \ud x_i
	-	2 \left( 1- \frac{r}{r_c} \right) v_i\ud x^i \ud \tau
	-	2 \frac{v_i}{r_c} \ud x^i \ud r
\\
&	+	\left( 1 - \frac{r}{r_c} \right)
		\left[
			(v^2+2P) \ud\tau^2
			+	\frac{v_iv_j}{r_c} \ud x^i \ud x^j
		\right]
	+	\left(
			\frac{v^2}{r_c} + \frac{2P}{r_c}
		\right)
		\ud\tau \ud r
\\
&		
	-	\frac{1}{16\pi (d-2)} \left( 1 - \frac{r}{r_c} \right)^2 f^2 \ud \tau^2
	+	\frac{1}{2\pi r_c} \left( 1 - \frac{r}{r_c} \right) \pi_{ij} \ud x^i \ud x^j
\\
&
	-	\frac{(r^2 - r_c^2)}{r_c} \partial^2 v_i \ud x^i \ud \tau
	+	\mathcal O(\epsilon^3)
\end{aligned}
\\
r_c F		& =		\frac{1}{2} f_{ij} \ud x^i \wedge \ud x^j		+ f_{i\tau} \ud x^i \wedge \ud \tau		- \partial_j f_{ij} \ud x^i \wedge \ud r		+ \mathcal O(\epsilon^3),
\end{align}
which satisfy the Einstein-Maxwell equations to order $\mathcal O(\epsilon^3)$ provided \eqref{MHD equations} hold.

\subsubsection*{Bulk Killing vector}

The Killing equations for a Killing vector $\xi^r\partial_r$ are highly restrictive. In particular, they demand that $f_{ij}$ vanishes. The MHD equations then implies that $f_{\tau i}$ vanishes. We therefore recover the unforced INS equations, and the scaling invariance of \S\ref{sec:Energy scaling invariance from an isometry into the bulk}.

\subsubsection*{Fixed viscosity $\mathbb Z_2$ transformation}

Transformation \eqref{transformation of g_ar for fixed rc} with the choice $\xi = \partial_1$ again yields the simple reflection of the velocity \eqref{reflection transformation of v_i for constant xi = xi^k partial_k} with $c_k = \delta_k^1$:
\begin{equation}
v_1 \rightarrow - v_1		\qquad		v_{\hat\imath} \rightarrow v_{\hat\imath}		,
\label{reflection of v_1}
\end{equation}
where hatted indices omit 1. The pressure is preserved. Rather than looking to determine the bulk or boundary field strength transformations explicitly, we can look to the MHD equations which can be expanded as
\begin{gather}
\partial_{[\hat c} f_{\hat a\hat b]}		=		0
\qquad
\partial_{[\hat a} f_{\hat b]1}		=		0
\qquad
\partial_{\hat \imath}v^{\hat \imath}		=		0
\\
f_{\tau 1}	= - r_c \partial_{\hat \jmath} f_{1 \hat \jmath} - v^{\hat \jmath} f_{\hat \jmath 1}
\qquad
f_{\tau \hat \imath}	= - r_c  \partial_{\hat \jmath} f_{\hat \imath \hat \jmath} - v^{\hat \jmath} f_{\hat \jmath \hat \imath} - v^1 f_{1 \hat \imath} 
\label{MHD with partial_1 isometry (E=J+v x B)}
\\
\partial_\tau v_1		+ v^{\hat \jmath}\partial_{\hat \jmath} v_1		- r_c \partial^{\hat \jmath} \partial_{\hat \jmath} v_1		=		- \frac{1}{4\pi} \partial^{\hat \jmath} \left( f_{\hat \jmath \hat k} f_{1\hat k}	 \right)
\\
\partial_\tau v_{\hat \imath}		+ v^{\hat \jmath}\partial_{\hat \jmath} v_{\hat \imath}		- \partial_{\hat \imath} \left( P - \frac{1}{16\pi}\frac{d}{d-2} f_{\hat l\hat k} f_{\hat l \hat k} \right)		- r_c \partial^{\hat \jmath} \partial_{\hat \jmath} v_{\hat \imath}		=		- \frac{1}{4\pi} \partial^{\hat \jmath} \left( f_{\hat \jmath \hat k} f_{\hat \imath \hat k}		- f_{\hat l\hat k} f_{\hat l \hat k} \delta_{\hat \jmath \hat \imath} \right).
\end{gather}
We see that a reflection \eqref{reflection of v_1} with \eqref{MHD with partial_1 isometry (E=J+v x B)} requires a reflection
\begin{equation}
f_{\hat a 1} \rightarrow 		- f_{\hat a 1}		\qquad		f_{\hat a \hat b} \rightarrow f_{\hat a \hat b}
\label{reflection of f_(hat a 1)}
\end{equation}
while all other equations are invariant under \eqref{reflection of v_1} and \eqref{reflection of f_(hat a 1)}. This transformation is simply equivalent to reflection of the coordinate $x^1 \rightarrow - x^1$. One can find transformations allowed by other spatial Killing vectors by suitable rotations of the $x^i$ coordinates.

On projecting solution-generating transformations to the dual fluid, we had hoped to arrive at some non-trivial fluid transformations deriving from non-trivial symmetries in the vacuum field equations, such as the $F \rightarrow 1/F$ of Buchdahl. For example, perfect fluids with an isometry along $x^1$ are invariant under $v_1\rightarrow 1/v_1$. However, we have arrived, particularly for spacelike Killing vectors, at rather trivial transformations. It would be interesting to see if extensions of this work could uncover further non-trivial physics, and if not, whether this trivial result was an inevitable product of some unknown property of the fluid/gravity correspondence.

\chapter{Conclusion and outlook}
\label{ch:Conclusion and outlook}
\chead{\textsc{conclusion and outlook}}

\section[Double field theory and U-duality-invariant M-theory]{Double field theory and U-duality-invariant \\M-theory}

We have presented new derivations of the equations of motion for the generalised coset metrics of DFT and $SL(5)$-invariant extended geometry from their respective actions. In the DFT case, we offered a pp-wave solution in the doubled geometry, which reduces under two different solutions to the section condition to the F1 string and the spacetime pp-wave. The Goldstone modes of the DFT pp-wave are found, in the string case, to be governed by the equations of motion of the duality-symmetric string of Duff \cite{Duff:1989tf} and Tseytlin \cite{Tseytlin:1990nb, Tseytlin:1990va}. In the $SL(5)$ geometry, we offer a pp-wave generalised metric solution which reduces in the supergravity frame to the M2-brane.

We have looked at the $O(d,d,\mathbb R)$-covariant string (gauged) $\sigma$-model actions of \cite{Tseytlin:1990nb, Tseytlin:1990va, Hull:2004in, Lee:2013hma}, and applied a Buscher-type procedure with both one and two Lagrange multipliers. With one Lagrange multiplier, the resulting action is not of the form of a DFT $\sigma$-model. For two Lagrange multipliers, we find the expected $O(1,1,\mathbb Z)$ T-duality transformation.

There are plenty of avenues of further research in the duality-invariant geometries. U-duality-invariant theories have been constructed in all cases where the dimension of the duality group is finite, i.e.\ $3\leq d\leq 8$. However, the full equations of motion have been constructed only for $d = 4$, and it would be relatively simple to extend our chain rule approach to higher dimensional geometries, where the Lagrangians, vielbein, and Y-tensors are known \cite{Berman:2011jh, Berman:2012vc}. However, we note that one can find \emph{some} solutions in these geometries purely by variation of the Lagrangian with respect to the generalised metric treated as an arbitrary symmetric object. For example, in the $SL(5)$ geometry, solving $\mathcal K = 0$. Indeed, the $SL(5)$ pp-wave generalised metric does this. That the generalised metric takes the appropriate coset form only projects out, via the appropriate projector $P$, some components of $\mathcal K$---that is while all solutions satisfy $P\mathcal K =0$, some also satisfy $\mathcal K = 0$. A related example of note is the self-dual solution in $E_7$ exceptional field theory constructed in \cite{Berman:2014hna}, which reproduces the complete $1/2$-BPS spectrum of ten- and eleven-dimensional supergravity. However, determining the full equations of motion would be required in, for example, a Goldstone analysis of these solutions.

\section[Solution-generating techniques in supergravity, double field theory and beyond]{Solution-generating techniques in supergravity, \\double field theory and beyond}

In our study of symmetries of NS-NS supergravity with one non-null isometry, we decompose all field content into fields living on a codimension one submanifold defined by the Killing vector. Specifically, we define a set of $p$-form potentials and a submanifold reduced metric. We restrict to the cases where these potentials are scalars, and determine the equations of motion of these potentials and the reduced metric. Via the corresponding effective actions, we determine a set of symmetry groups which preserve the submanifold metric. In static Einstein-dilaton theory we find a non-trivial $\mathbb Z_2\times\mathbb Z_2$ symmetry group, and in five-dimensional static magnetic supergravity, a non-trivial $SL(2,\mathbb R)\times \mathbb Z_2$ group.

We looked at how similar techniques might be applied to DFT with an isometry. We discussed which components of the generalised metric to transform, and which to keep fixed when looking for symmetry groups, and looked at a specific case corresponding in the supergravity frame to static Einstein-dilaton gravity. In this case, we recovered only the $\mathbb Z_2$ T-duality we encountered in Einstein-dilaton theory, which is of course manifest in DFT. We would like to see how the remaining $\mathbb Z_2$ appears in the DFT context. Moreover, it would be interesting to see this extended to the full DFT case, especially in a covariant form. One could perhaps decompose the DFT fields in to potentials analogous to those for Einstein gravity in \S\ref{sec:Reduction of Einstein gravity with respect to an isometry}, by employing the studies of DFT geometry in \cite{Hohm:2010xe, Jeon:2010rw, Hohm:2011si, Hohm:2012mf, Berman:2013uda, Cederwall:2014kxa}, where analogues of the covariant derivative and curvature tensors are constructed.

For example, in \cite{Pinkstone:1995wk} the author constructed an effective action for stationary four-dimensional NS-NS supergravity containing two sigma models, each with $SL(2,\mathbb R)$ invariance. One group corresponded to S-duality and the other to an Ehlers-type symmetry. Moreover, they point out that these sigma models can be mapped into each other by an $O(1,1,\mathbb Z)$ T-duality transformation. That is, in that case, the Ehlers symmetry is T-dual to S-duality. It would be interesting to see how this would play out within DFT. Since the theory is T-duality invariant in the presence of an isometry, we ask if Ehlers-type transformations would appear on the same footing as S-duality, where one would rotate between the two for different solutions to the section condition. Perhaps this is one way of recovering U-duality from DFT.

Returning to solution-generating transformations in Einstein gravity coupled to form fields, a much greater goal would be to be able to find the symmetry groups for any spacetime dimension and with any form-field content (it would of course be ideal to verify if these methods are valid for null Killing vectors, not only spacelike/timelike). The potential space line element method in \S\ref{sec:Solution-generating symmetries from an effective action} clearly would not suffice, since the potentials are no longer scalars, though it would be interesting to see if one can construct appropriate generalisations of this method.

One could of course introduce more isometries, e.g.\ stationary-axisymmetric systems relevant to stars and black holes. This includes the generalisation of the Ehlers symmetry to the $SL(d-2,\mathbb R)$ invariance of $d$-dimensional vacuum Einstein gravity with $(d-3)$ isometries. One could also consider solution-generating isometries in other gravitational theories; the presence of matter, for example a perfect fluid \cite{stewart1982generalisation, Berger1987, Stephani1988, Garfinkle:1996ur, Racz:1996ei, Racz:1997wq, Zsigrai01082000, Boonserm:2006vr}; form fields with Lagrangian contributions beyond the quadratic term, such as the Chern-Simons term appearing in eleven-dimensional supergravity \eqref{11-d sugra action}; and Lanczos-Lovelock gravity \cite{lanczos1938remarkable, lovelock1971einstein}, where the Ricci curvature in the Lagrangian is generalised to
\begin{equation}
L		=	\sum_{k=0}^{k_\text{max}} a_k \delta_{[\mu_1}^{\rho_1} \delta_{\nu_1}^{\sigma_1} \ldots \delta_{\mu_k}^{\rho_k} \delta_{\nu_k]}^{\sigma_k} \prod_{l=1}^k R^{\mu_l \nu_l}{}_{\nu_l \sigma_l}		,
\end{equation}
where $k_\text{max} \leq (d-1)/2$ and $a_k$ are dimensionful constants. It was shown by Brustein and Medved \cite{Brustein:2012uu} that Lovelock gravity is equivalent to Einstein gravity coupled to forms fields. It would be interesting to see how this works with regards to the potentials which appear in the presence of a Killing vector. After all, our work in \S\ref{sec:Solution-generating symmetries from an effective action} shows how the metric and Riemann tensor can, via \eqref{Riemann on V_d-1 in terms of Riemann on V_d} and \eqref{derivative of lowered killing vector in terms of F and omega} be decomposed in terms of the twist \eqref{twist: general definition}, norm and reduced metric \eqref{metric on V_d-1}. On the other hand, equation \eqref{H in terms of alpha and beta} illustrates how form fields decompose into two potentials each. Brustein and Medved's work then suggests that the potentials encoding the metric in Lovelock gravity in some way correspond to the potentials encoding form fields in Einstein gravity.

\section[Solution-generating transformations in the fluid/gravity correspondence]{Solution-generating transformations in the \\ fluid/gravity correspondence}

We have explored how solution-generating transformations of the Einstein equations in the presence of a Killing vector may be mapped to the codimension one hydrodynamics holographically dual under the fluid/gravity correspondence. Our focus has been on the incompressible Navier-Stokes fluid dual to vacuum Rindler spacetime, where we have uncovered a selection of fluid transformations.

Firstly, we discovered a linear scaling of the fluid velocity and total energy with the viscosity, corresponding in the gravity dual to renormalisation group flow of the hypersurface through the bulk. This scaling invariance is exact, in that it receives no corrections at higher orders in the hydrodynamic expansion. It would be relatively simple to solve the Killing equations for the fluid metric at higher orders and check the validity of the scaling invariance at the corresponding order in the INS equations. Higher order corrections for the metric and INS equations are given in \cite{Compere:2011dx}. Secondly, we find that spacelike isometries in the fluid yield somewhat trivial $\mathbb{Z}_2$ transformations of the fluid velocity (it would be interesting to see if other Killing vectors, such as those with components along both radial and hypersurface directions, might yield more non-trivial results). Explicit examples are given of reflection-like symmetries for translational and rotational isometries in the fluid. Finally, we determine that all of our transformations are trivial from the point of view of the bulk spacetime.

There are plenty of immediate extensions to this work from the a large literature on gravitational systems dual to hydrodynamics, such as: fluid flow on a sphere dual to the Schwarzschild black hole \cite{Bredberg:2011xw}; cyclonic flow dual to the Kerr-Newman AdS$_4$ black hole \cite{Leigh:2012jv}; (non-)conformal relativistic hydrodynamics dual to vacuum Einstein gravity \cite{Bhattacharyya:2008jc, Bhattacharyya:2008mz, Eling:2012ni, Compere:2012mt, Pinzani-Fokeeva:2014cka}; forced fluids dual to non-vacuum gravity, such as bulk ideal fluids \cite{Wu:2013mda}. In particular, we ask if it is possible to transform between fluid solutions on flat backgrounds and those on curved backgrounds. Not only would this allow us to find fluid flows on curved backgrounds, where the equations are often more difficult to solve than the INS equations, one could possibly generate non-trivial fluid flows on flat space from trivial flows on curved space.

Extensions further afield include condensed matter systems dual to gravity (see \cite{hartnoll2011horizons} for a review). One could even look to solution-generating techniques in holography beyond the hydrodynamic limit. Indeed, in the epilogue of \cite{Hubeny:2011hd}, the authors consider the following. If the Einstein equations in the long-wavelength limit are dual to fluid dynamics, then perhaps the there may be a similar connection between the full Einstein equations and a strong coupling analogue of the Boltzmann transport equations, describing the thermodynamics of a system out of thermal equilibrium. Essentially, we wish to open up the use of gravitational solution-generating symmetries in holography.

\appendix

\chapter{Conventions and useful formulae}
\label{app:Conventions and useful formulae}
\chead{\textsc{appendix: conventions and useful formulae}}

\paragraph{Metric signature} Lorentzian: $(-++\ldots+)$, Riemannian: $(++\ldots+)$

\paragraph{(Anti)symmetrization}
\begin{align}
A_{(\mu_1 \ldots	\mu_p)}		& =		\frac{1}{p!} \sum_\pi A_{\mu_{\pi(1)} \ldots \mu_{\pi(p)}}
\\
A_{[\mu_1 \ldots	\mu_p]}		& =		\frac{1}{p!} \sum_\pi \delta_\pi A_{\mu_{\pi(1)} \ldots \mu_{\pi(p)}}		,
\end{align}
where the sum is taken over all permutations, $\pi$, of 1, \ldots , $p$ and $\delta_\pi$ is $+1$ for even permutations and $-1$ for odd permutations.

\paragraph{Alternating symbol and tensor}
\label{app:Alternating symbol and tensor}

For totally antisymmetric \emph{symbol}
\begin{equation}
\epsilon^{\mu_1 \ldots \mu_d} = \epsilon_{\mu_1 \ldots \mu_d}		 =		\epsilon_{[\mu_1 \ldots \mu_d]}	,	\qquad		\epsilon_{012 \ldots d} = 1 ,
\end{equation}
on a $d$-dimensional space, we define the totally antisymmetric \emph{tensor} on \emph{Lorentzian space} with metric $g$ by
\begin{equation}
\hat \epsilon_{\mu_1 \ldots \mu_d}		=		\sqrt{ - \det(g) } \, \epsilon_{\mu_1 \ldots \mu_d},
\end{equation}
whose indices can be raised and lowered with $g$, which gives
\begin{equation}
\hat \epsilon^{\mu_1 \ldots \mu_d}		=		\frac{ - 1 }{ \sqrt{ - \det(g) } } \, \epsilon^{\mu_1 \ldots \mu_d}		.
\end{equation}
The antisymmetric tensor is annihilated by the covariant derivative. It contracts as
\begin{equation}
\hat \epsilon^{\mu_1 \ldots \mu_r \nu_{r+1} \ldots \nu_d} \hat \epsilon_{\rho_1 \ldots \rho_r \nu_{r+1} \ldots \nu_d}		=		- r! (d-r)! \delta^{\mu_1}_{ [\rho_1 } \ldots \delta^{\mu_r}_{ \rho_r] }		.
\end{equation}

\section{Differential forms}
\label{app:Differential forms}

The results in this section are valid for $d$-dimensional Lorentzian spacetimes with metric $g$, for arbitrary forms $A$, $B$, $C$ (where lowercase Roman subscripts indicate rank), and vectors $\upsilon$.

\subsection*{Definitions}

\begin{align}
(\star A)_{\mu_1 \ldots \mu_{d-p}}									&=		\frac{1}{p!} \hat \epsilon_{\mu_1 \ldots \mu_{d-p}}{}^{\nu_1 \ldots \nu_p} (A_p)_{\nu_1 \ldots \nu_p}
\\
(A_p\wedge B_q)_{\mu_1 \ldots \mu_p \nu_1 \ldots \nu_q}	& =		\frac{(p+q)!}{p!q!} (A_p)_{[\mu_1 \ldots \mu_p} (B_q)_{\nu_1 \ldots \nu_q]}
\\
(\iota_\upsilon A_p)_{\mu_2 \ldots \mu_p}							& =		\upsilon^{\mu_1} (A_p)_{\mu_1 \mu_2 \ldots \mu_p}
\\
(\ud A_p)_{\mu_1 \ldots \mu_{p+1}}									& =		(p+1) \partial_{[\mu_1} (A_p)_{\mu_2 \ldots \mu_{p+1}]}
\end{align}
Musical notation is defined for vectors $\upsilon$ and one-forms $\omega$ as
\begin{gather}
(\upsilon_\flat)_\mu = g_{\mu\nu} \upsilon^\mu		\qquad		(\omega^\sharp)^\mu = g^{\mu\nu} \omega_\nu		.
\end{gather}

\subsection*{Useful formulae}

For the following identities, references are given in the second column in certain cases.
\begin{align}
		\mathcal L_\upsilon A					& =		( \iota_\upsilon \ud 		+		\ud \iota_\upsilon) A
				& \parbox{0.2\linewidth}{\cite[p.~202]{Nakahara2003} }
\\		\star^2 A_p								& =		(-1)^{1+p(d-p)} A_p
				& \parbox{0.2\linewidth}{\cite[p.~291]{Nakahara2003} }
\\		(A \wedge B) \wedge C				& = 		A \wedge (B \wedge C)
				& \parbox{0.2\linewidth}{\cite[p.~198]{Nakahara2003} }
\\		A_p \wedge B_q							& =		(-1)^{pq} B_q \wedge A_p
				& \parbox{0.2\linewidth}{\cite[p.~198]{Nakahara2003} }
\\		A_p \wedge \star B_p					& = 		B_p \wedge \star A_p
				& \parbox{0.2\linewidth}{ \cite[p.~292]{Nakahara2003} }
\\		\ud(A_p\wedge B_q)					& =		\ud A_p \wedge B_q		+		(-1)^p A_p\wedge \ud B_q
				& \parbox{0.2\linewidth}{ \cite[p.~199]{Nakahara2003} }
\\		\iota_\upsilon (A_p\wedge B_q)	& =		\iota_\upsilon A_p \wedge B_q		+ (-1)^p A_p \wedge \iota_\upsilon B_q
				& \parbox{0.2\linewidth}{ \cite[p.~202]{Nakahara2003} }
\\		\mathcal L_\upsilon (A\wedge B)	& =		(\mathcal L_\upsilon A)\wedge B		+		A\wedge \mathcal L_\upsilon B
				&
\\		(\star \ud \star A_p)_{\mu_1 \ldots \mu_{p-1}}		& =		(-1)^{pd} \nabla^\sigma (A_p)_{\sigma \mu_1 \ldots \mu_{p-1}}
				&
\\		\star (A_p \wedge \star A_p)		& =		\frac{1}{p!} (-1)^{1+p(d-p)} A_{\mu_1 \ldots \mu_p} A^{\mu_1 \ldots \mu_p}
				& 
\\		\upsilon_\flat \wedge \star A_p		& =	(-1)^{d-p} \star \iota_\upsilon A_p
				& 
\\		\star (\upsilon_\flat \wedge A_p)	& =	(-1)^{d-p+1} \iota_\upsilon \star A_p
				&
\\		(-1)^{pd+1} (i_\upsilon \upsilon_\flat) A_p		& =		\left[ (i_\upsilon \star)^2 		+		(-1)^d (\star i_\upsilon)^2 \right] A_p
\end{align}

\newpage
\chead{\textsc{bibliography}}

\bibliographystyle{ieeetr}
\bibliography{ref_stringandM,ref_fluidgravity,ref_gravityandsolgen}
\end{document}